\documentclass[a4paper]{article}
\pdfoutput=1
\usepackage{pdfpages}

\usepackage{icrc2013}
\usepackage{amsmath}
\usepackage[english]{babel}
\usepackage{url}
\usepackage{enumitem}
\usepackage{overpic}
\usepackage{hyperref}
\hypersetup{
  colorlinks=true,
  linkcolor=blue,
  citecolor=blue,
  urlcolor=blue,
}

\newcommand{\nosection}[1]{%
  \refstepcounter{section}%
  \addcontentsline{toc}{section}{\protect\numberline{\thesection}#1}%
  \markright{#1}}

\title{The Sensitivity of HAWC to Steady and Transient Sources of Gamma Rays: Contributions to ICRC 2013}

\authors{
{\bf The HAWC Collaboration:}\\
A.~U.~Abeysekara$^{a}$,
R.~Alfaro$^{b}$,
C.~Alvarez$^{c}$,
J.~D.~{\'A}lvarez$^{d}$,
R.~Arceo$^{c}$,
J.~C.~Arteaga-Vel{\'a}zquez$^{d}$,
H.~A.~Ayala Solares$^{e}$,
A.~S.~Barber$^{f}$,
B.~M.~Baughman$^{g}$,
N.~Bautista-Elivar$^{h}$
% ...
}

\afiliations{
{\center
$^{a}$ Department of Physics and Astronomy, Michigan State University, East Lansing, MI, USA\\
$^{b}$ Instituto de F{\'\i}sica, Universidad Nacional Aut{\'o}noma de M{\'e}xico, Mexico D.F., Mexico\\
$^{c}$ CEFyMAP, Universidad Aut{\'o}noma de Chiapas, Tuxtla Guti{\'e}rrez, Chiapas, Mexico\\
$^{d}$ Universidad Michoacana de San Nicol{\'a}s de Hidalgo, Morelia, Mexico\\
$^{e}$ Department of Physics, Michigan Technological University, Houghton, MI, USA\\
$^{f}$ Department of Physics and Astronomy, University of Utah, Salt Lake City, UT, USA\\
$^{g}$ Department of Physics, University of Maryland, College Park, MD, USA\\
$^{h}$ Universidad Politecnica de Pachuca, Pachuca, Hgo, Mexico\\
% ...
}
}

\abstract{
  The High-Altitude Water Cherenkov (HAWC) Gamma-Ray Observatory is designed to
  record air showers produced by cosmic rays and gamma rays between 100 GeV and
  100 TeV.  Because of its large field of view and high livetime, HAWC is
  well-suited to measure gamma rays from extended sources, diffuse emission,
  and transient sources.  We describe the sensitivity of HAWC to emission from
  the extended Cygnus region as well as other types of galactic diffuse
  emission;  searches for flares from gamma-ray bursts and active galactic
  nuclei; and the first measurement of the Crab Nebula with HAWC-30.
}


\begin{document}
\maketitle

\pagestyle{plain}
\pagenumbering{arabic}

\clearpage

\newpage
\onecolumn{
  \tableofcontents
}

\newpage
\setcounter{section}{0}
\nosection{Studying the Cygnus region with the HAWC Observatory\\
{\footnotesize\sc Chiumun Michelle Hui}}
\setcounter{section}{0}
\setcounter{figure}{0}
\setcounter{table}{0}
\setcounter{equation}{0}
%
%
\title{Studying the Cygnus region with the HAWC Observatory}


\authors{
C. M. Hui$^{1}$
for the HAWC Collaboration$^{2}$
}

\afiliations{
$^1$ Michigan Technological University, Department of Physics,
Houghton, MI 49931, USA\\
$^2$ For a complete author list, see the special section of these proceedings
}

\email{cmhui@mtu.edu}

\abstract{
The Cygnus region is an extremely active star forming region with a
wealth of gamma-ray sources such as pulsar wind nebulae, young star
clusters, and binary systems.  In the TeV regime, the Milagro observatory
has detected two extended sources, MGRO\,J2019+37 and J2031+41, along
with hints of correlated GeV emission in the region.  The VERITAS
observatory has observed MGRO\,J2019+37 with an angular
resolution of $0.1^\circ$ at 1\,TeV, and found a region of extended
emission at energies greater than 600\,GeV that is spatially consistent with
the Milagro detection and associated with multiple sources at lower
energies.  The High Altitude Water Cherenkov (HAWC) Observatory is
ideal for studying the morphology and emission origin of the Cygnus
region.  It is a surveying instrument with an energy range between
100\,GeV and 100\,TeV, an angular resolution of $<0.2^\circ$ for
energies above 10\,TeV, a field of view of 2\,sr, a duty cycle greater
than 90\%, and an area of 22,000\,m$^2$.  Currently the HAWC
Observatory is under construction at Sierra Negra in the state of
Puebla, Mexico, and one third of the full array will be operational by
Summer 2013.  I will present the sensitivity of the full array to the
Cygnus region.}


\maketitle

\section*{The Cygnus Region}
The Cygnus region is an active, gas-rich region and a potential
area for cosmic-ray acceleration.  Observations from radio to TeV
gamma rays have revealed diffuse emission and a variety of objects
such as molecular clouds, star clusters, and pulsars.  The Fermi
Large Area Telescope (LAT) has detected over a dozen gamma-ray
sources \cite{2FGL}, along with a cavity of enhanced gamma-ray
emission with a hard spectrum called the "cocoon".  Particles seem to
be accelerated and confined within the cocoon, and the
cocoon gamma-ray emission appears to be consistent with the Milagro
observation of the region \cite{cocoon,MGRO07}. 

\begin{figure*}
   \centering
   \includegraphics[width=0.95\textwidth]{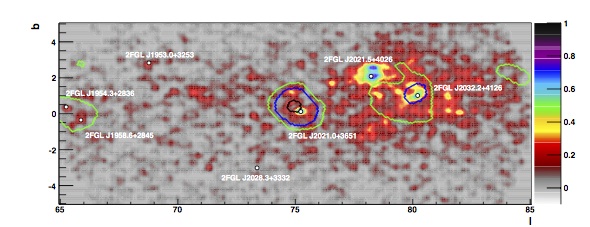}
   \caption{Fermi LAT counts map for photons above 10\,GeV of the Cygnus region from 40 months of data taken between 2008 and 2011.  The color contours indicate significance of Milagro data collected over the last 3 years of operation from 2005 to 2008 (black=$11\,\sigma$, blue=$5\,\sigma$, green=$3\,\sigma$) \cite{MGRO12}.  HAWC will be $\sim15$ times more sensitive than Milagro and be able to reproduce the Cygnus region map with $\sim1$ week of data.}
   \label{map}
 \end{figure*}

At TeV energies, only a handful of sources have been identified in
the Cygnus region.  The Milagro collaboration reported extended TeV
emission from MGRO\,J2019+37 and J2031+41 \cite{MGRO07, MGRO12}
(see Figure \ref{map}).  These two gamma-ray sources have also been
seen by other ground-based gamma-ray observatories with various source
type associations.  In addition, the Milagro data also show a few
$>3\,\sigma$ hotspots that are spatially correlated with sources from
the Fermi Bright Source List \cite{MGRO-BSL}.  

Diffuse excess in the Cygnus region at 15\,TeV is observed by Milagro
after subtracting known sources \cite{MGRO-diffuse}.  This excess is
likely due to interaction of newly accelerated cosmic rays with the
local environment.  \cite{Fermi-diffuse} reported that the whole Cygnus
region on average appears to have a cosmic ray population similar to
that of the local interstellar space from GeV observations.  Despite
the correlations between GeV and TeV observations for point and extended
sources in the Cygnus region, \cite{Fermi-diffuse} reported no GeV
counterpart for the broad diffuse excess seen by Milagro.  HAWC
sensitivity to diffuse emission will be discussed in a separate
contribution in these proceedings \cite{HAWC-diffuse}. 



\section*{HAWC}
\begin{figure}
  \centering
  \includegraphics[width=0.45\textwidth]{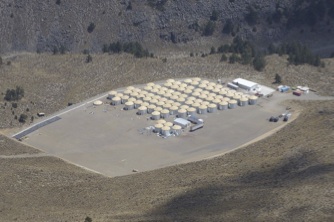}
  \caption{Aerial view of the HAWC array taken in April 2013.}
  \label{HAWC-array}
 \end{figure}

The High Altitude Water Cherenkov (HAWC) Observatory is a second
generation TeV gamma-ray detector based on the water Cherenkov
technique.  It is currently under construction at Sierra Negra,
Mexico, at an elevation of 4100\,m.  When completed, the array will
consist of 300 water Cherenkov detectors that are 7.3\,m in diameter
and 4.5\,m in depth, covering an area of 22,000\,m$^2$.  The array
will be sensitive to energies between 100\,GeV and 100\,TeV and will have a
2\,sr field of view (FOV) and an angular resolution of $<0.2^\circ$ for
energies above 10\,TeV. Since September 2012, the HAWC observatory has
been collecting data with a partial array.  Figure \ref{HAWC-array}
shows a recent picture of the HAWC observatory.  The performance of
the partial array and details of the HAWC observatory are discussed in
\cite{HAWC-HL,HAWC-status}.

\section*{MGRO\,J2019+37}
\begin{figure}[h]
  \centering
  \includegraphics[width=0.45\textwidth]{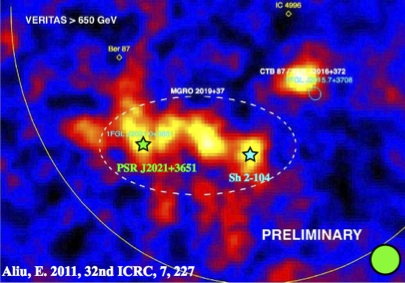}
  \caption{VERITAS excess photon count map with the MGRO\,J2019+37 best
    fit ellipse contour shown in a white dashed line \cite{Aliu-ICRC11}.  The
    green star indicates the pulsar PSR\,J2021+3651 and the blue star
    indicates the HII region Sh\,2-104 \cite{radio}. The green circle
    in the lower right corner shows the angular resolution of
    HAWC at 10\,TeV.}
  \label{2019map}
 \end{figure}

MGRO\,J2019+37 is the brightest source in the Milagro Cygnus map, with
an extent of $0.7^\circ \pm 0.1^\circ$ obtained by fitting a 2D
Gaussian function to the excess observed by Milagro \cite{MGRO12}.  It 
appears to be elliptical in shape, and overlaps two Fermi sources, one
of which is a known pulsar PSR\,J2021+3651 \cite{2FGL}.  Recently
VERITAS reported a $9.5\,\sigma$ detection at energies $>600$\,GeV 
that is spatially compatible with the Milagro detection but with a
smaller extent (see Figure \ref{2019map}).  Their preliminary
spectrum is compatible with the Milagro spectrum and upper limits from
ARGO \cite{Aliu-Gamma12}.  With a PSF of $0.09^\circ$, the VERITAS data
show an elongated region centered on the Milagro detection that
appears to be a combination of multiple hotspots, including a 2FGL
pulsar and the HII region Sh\,2-104.  Observations from Fermi LAT at
above 10\,GeV so far have revealed strong emission from the pulsar
within the VERITAS detection but only marginal emission at the
location of the HII region, and nothing is seen between the two positions
where VERITAS detects strong emission at energies $>600$\,GeV (see Figure
\ref{map} and \ref{2019map}).  \cite{radio} and \cite{xmm} presented
radio, infrared, X-ray, and gamma-ray observations of the region and
showed that the extended TeV emission is unlikely powered by a single
source, i.e. PSR\,J2021+3651.  HAWC, with an angular resolution
of $0.16^\circ$ at 10\,TeV, in comparison to $0.5^\circ-0.8^\circ$ for
Milagro at similar energies, will be able to resolve TeV emission from
this region to study the morphology of this source.

\section*{MGRO\,J2031+41}
\begin{figure}
  \centering
  \includegraphics[width=0.35\textwidth]{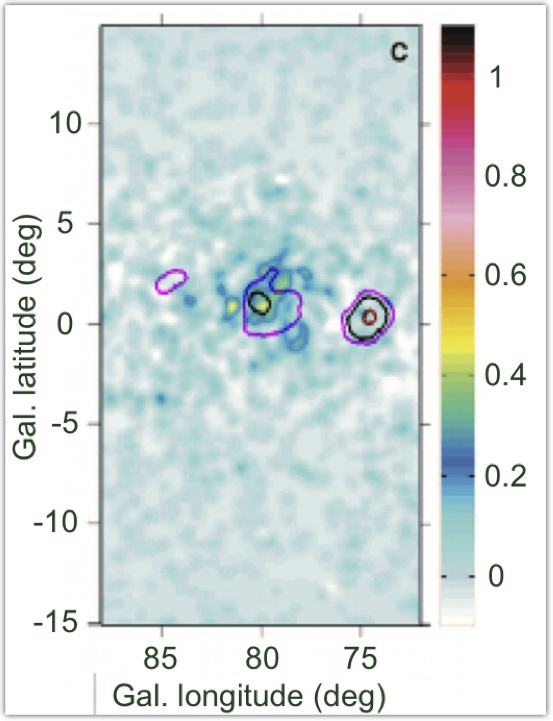}
  \caption{Milagro contours (red=$11\,\sigma$, black=$5\,\sigma$,
    pink=$3\,\sigma$) overlaid on the Fermi 10--100\,GeV photon count map
    of the cocoon \cite{cocoon, MGRO12}.  The Milagro detection likely
    encompasses both the pulsar wind nebula and diffuse emission in
    the region.} 
  \label{cocoon-overlaid}
 \end{figure}

\begin{figure}
  \centering
  \includegraphics[width=0.45\textwidth]{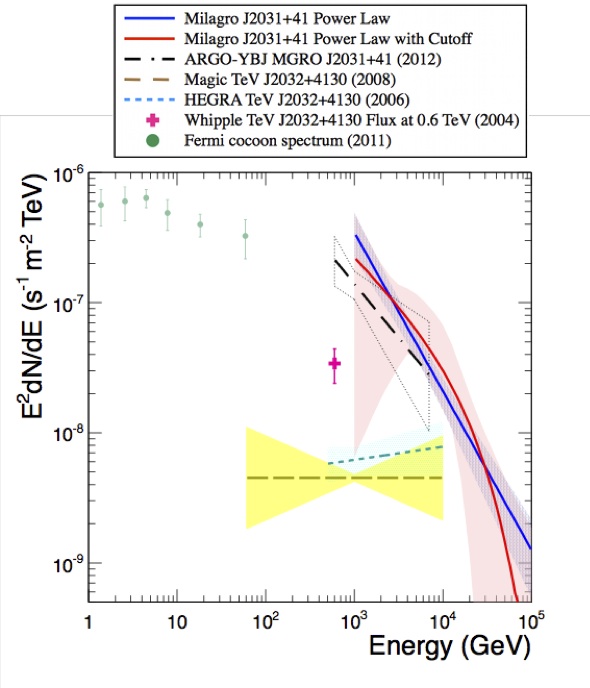}
  \caption{Spectral measurements of the region around MGRO\,J2031+41
    from various instruments \cite{cocoon, MGRO12}.  The IACT spectra
    are approximately an order of magnitude lower than those by
    survey-type instruments.  This can be attributed to the smaller FOVs
    of the IACTs in comparison to Milagro and ARGO. } 
  \label{J2031spec}
 \end{figure}

The Milagro detection of MGRO\,J2031+41 is asymmetric in shape and
encompasses the Fermi pulsar 2FGL\,J2032.2+4126 and the cocoon region
(see Figure \ref{map} and \ref{cocoon-overlaid}).  At the pulsar
position, the imaging atmospheric Cherenkov telescope (IACT) Whipple reported
a point-like emission while IACTs, HEGRA and MAGIC, reported an extended
emission with radius of $\sim0.1^\circ$ \cite{Whipple, HEGRA, MAGIC}.
These are later suggested to be associated with wind from the pulsar
PSR\,J2032+4127 based on radio and X-ray observations
\cite{radioLAT,suzaku}.  Figure \ref{J2031spec} shows the spectra of
the region reported by the various experiments.  The spectra measured
by the IACTs are an order of magnitude less than the Milagro spectrum
and have a very different shape, while the Milagro spectrum appears to
be compatible with an extrapolation of the Fermi cocoon spectrum.
The angular resolution of Milagro is a few times larger than that of the
IACTs; it is likely that the flux measured by Milagro includes both the
pulsar wind nebula emission and the diffuse photons from the cocoon
region, while the measurements from IACTs, with their much smaller FOVs,
may have subtracted out the diffuse emission as background.

Currently, a cosmic-ray acceleration model is favored for the diffuse
gamma-ray emission in the Fermi cocoon region.  The hardness of the
spectrum suggests particles are freshly accelerated and efficiently
confined.  \cite{cocoon} modeled protons with maximum energy from
80 to 300\,TeV and electrons from 6 to 50\,TeV that can explain the Fermi
LAT measurement and noted that the Milagro measurement cannot be
explained by pure inverse Compton emission alone.  HAWC will be
able to improve spectral measurements at TeV energies and help investigate the
origins of the emission of this region.

\section*{Outlook}
A partial array of HAWC (10\% of the full array) has
been in operation since September 2012 while the construction is
ongoing.  This data is currently still being processed.  Beginning this
Fall, 35\% of the array will be operational and known
sources in the Cygnus region will be detected in approximately a
month.  The full array is scheduled to be completed by Fall 2014.  The
sensitivity of the full array will allow for in depth spectral studies
with a year of data and will be presented at the conference.

\section*{Acknowledgments}

We acknowledge the support from: US National Science Foundation (NSF); US
Department of Energy Office of High-Energy Physics; The Laboratory Directed
Research and Development (LDRD) program of Los Alamos National Laboratory;
Consejo Nacional de Ciencia y Tecnolog\'{\i}a (CONACyT), M\'exico; Red de
F\'{\i}sica de Altas Energ\'{\i}as, M\'exico; DGAPA-UNAM, M\'exico; and the
University of Wisconsin Alumni Research Foundation.

\clearpage


\newpage
\setcounter{section}{1}
\nosection{HAWC Sensitivity to Diffuse Emission\\
{\footnotesize\sc Petra H\"{u}ntemeyer, Hugo~A. Ayala Solares}}
\setcounter{section}{0}
\setcounter{figure}{0}
\setcounter{table}{0}
\setcounter{equation}{0}
%
%
\title{HAWC Sensitivity to Diffuse Emission}

\shorttitle{HAWC Sensitivity to Diffuse Emission}

\authors{
Petra H\"untemeyer$^{1}$,
Hugo Albert Ayala Solares$^{1}$,
for the HAWC Collaboration$^2$ 
}

\afiliations{
$^1$ Michigan Technological University, Houghton, MI 49931, USA \\
$^2$ For a complete author list, see the special section of these proceedings \\
\scriptsize{
}
}

\email{petra@mtu.edu}

\abstract{Very high energy (VHE) diffuse gamma-ray emission measurements are an excellent probe of cosmic-ray acceleration, propagation, and density distribution at different locations within our Galaxy. Theoretical models of diffuse gamma-ray emission at GeV and TeV energies usually assume that the cosmic-ray flux and spectrum measured at Earth are representative of the typical flux and spectrum present throughout our Galaxy. But there has been some evidence that these models do not explain the large scale Galactic diffuse gamma-ray emission measured by several experiments at the highest energies. At TeV energies for example, the Milagro experiment reported a significant enhancement of diffuse emission with respect to models of the Cygnus region (measured emission = 8x predicted) and the inner Galaxy (5x). Observations by the HESS telescope of a molecular cloud near the center of the Galaxy also revealed an enhancement and a harder spectrum than expected if the cosmic ray flux were the same as the flux at Earth. In addition to improving theoretical modeling, measuring the diffuse and extended emission in our Galaxy with better sensitivity will help us better understand these enhancements, put tighter constraints on Galactic cosmic-ray emission, and distinguish between hadronic and leptonic acceleration and propagation models. The HAWC observatory, an all-sky high-altitude water Cherenkov detector array currently being constructed in Mexico, will be 15 times more sensitive than the Milagro detector and will be completed in Fall 2014. The simulated sensitivity of the array to Galactic diffuse gamma-ray emission under different model assumptions will be presented.
}


\maketitle

\section*{Introduction}

The origin and acceleration mechanisms that produce the cosmic rays that fill our Galaxy and are bombarding Earth from space have not yet been determined unambiguously. Favored candidate sources of Galactic cosmic rays (GRCs) are supernova remnants (SNRs) and pulsars. The standard production mechanisms for gamma-ray emission are interactions of cosmic rays (hadrons and electrons) with the matter and radiation fields in the Galaxy. 
Cosmic-ray hadrons interact with matter, producing neutral pions, which in turn decay into gamma rays, while cosmic-ray electrons produce TeV gamma rays by inverse Compton (IC) scattering off the interstellar radiation fields. At lower energies (MeV-GeV) bremsstrahlung from cosmic-ray electrons also contributes to gamma-ray emission. 

The detection of TeV gamma rays and X-rays from the same locations within SNRs provides strong evidence that electrons are accelerated in SNRs \cite{bib:Aharonian2006}. However, no definite evidence for the acceleration of protons and nuclei in SNRs has been found and it is not clear whether the proton and electron accelerators are of a different nature. The direct observation of cosmic rays from the candidate injection sites such as SNRs and pulsars is not possible since cosmic rays escape acceleration sites and eventually propagate into the Galactic magnetic field where they are deflected and subsequently mix with the bulk of cosmic rays, also known as the cosmic-ray background or 'sea' of cosmic rays. By measuring diffuse TeV gamma-ray emission the density and spectra of both the cosmic-ray sea and young cosmic-ray accelerators throughout our Galaxy can be studied. 

In general, the TeV sky appears to show more small scale structures than the sky at MeV-GeV energy (with the exception of the fairly recent discovery of the 'Fermi bubbles' \cite{bib:Su2010}).
Close to acceleration sites with ambient target material, gamma rays are produced and may contribute to the diffuse emission at TeV energies. This emission is unique in that it traces the transitional energy regime between 'sea' and freshly released cosmic rays. The detection of extended/diffuse gamma-ray emission near these sites at close to 100 TeV would be a sign of cosmic-ray acceleration up to $10^{15}$ eV in such objects. \cite{bib:Aharonian1996,bib:Aharonian2000,bib:casanova2010,bib:Gabici2009}.
In addition, it will be interesting to investigate if the Fermi bubble structure extends to TeV energies. 

\begin{figure*}[!t]
  \centering
  \includegraphics[width=\textwidth]{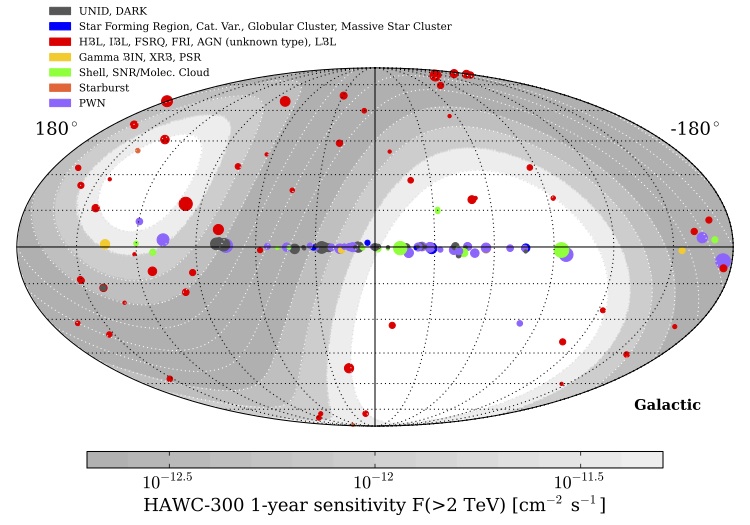}
  \caption{The TeV sky visible to the HAWC observatory in Galactic coordinates.
  The detector sensitivity is declination dependent and darker grey bands
  indicate better sensitivity. Source classes of very-high gamma-ray emission
  as obtained with TeVCat~\cite{bib:TeVCat} are overlaid.}
  \label{HAWCsky_fig}
 \end{figure*}

As a wide field of view instrument, the High Altitude Water Cherenkov (HAWC) observatory is well suited to address the open question of cosmic-ray origins through the measurement of diffuse and extended TeV gamma-ray emission from large-scale areas in our Galaxy. The telescope is capable of providing an unbiased survey of a large portion of the Northern Hemisphere down to regions close to the Galactic center at energies $>$ 100 GeV (see Figure \ref{HAWCsky_fig}). The HAWC data will allow for the study of diffuse emission from large areas along the Galactic plane or from structures such as the Fermi bubbles.

\section*{Previous Measurements}

In diffuse gamma-ray studies the measured spatial and spectral energy distributions of the emission are usually compared to model predictions based on the three dominant standard processes, neutral pion decays, IC scattering, and bremsstrahlung. Thus the relative contribution of hadronic and leptonic gamma-ray production mechanisms is investigated. 

The results of previous studies have been inconclusive.
The H.E.S.S. telescope detected VHE diffuse emission from the Galactic center ridge, which is correlated with giant molecular clouds \cite{bib:Aharonian2006Nature}. The spectrum of this emission is
significantly harder than the spectrum of the diffuse emission predicted with the cosmic-ray spectrum measured at Earth. The resulting enhancement of the measurement with respect to the prediction amounts to a factor of 3-9. This enhancement, similarly visible for
the Sgr B Region, implies that the high-energy cosmic-ray density is much higher than the local value (up to 10 times). In addition, the paper concludes that the cosmic rays producing the gamma rays are likely protons and nuclei rather than electrons because of the measured hardness of the gamma-ray spectrum \cite{bib:Aharonian2006Nature}. 

The first measurement of diffuse gamma-ray emission above 3.5 TeV from a large region of the Galactic plane (40 $<$ Galactic longitude $<$ 100) performed by the Milagro experiment indicated the existence of a TeV excess \cite{bib:Atkins2005}. In addition, the Milagro experiment measured the diffuse emission near 12 and 15 TeV from the Cygnus region of the Galaxy \cite{bib:Abdo2008,bib:Abdo2007} and also found an excess compared to predictions of GALPROP, a numerical model of cosmic-ray propagation in the Galaxy \cite{bib:Strong2000,bib:Strong2004_1,bib:Strong2004_2}. Studies of the latitudinal profiles of the diffuse emission were also performed by the Milagro collaboration. It was found that the measured shape does not agree well with the shape predicted by the GALPROP model that is optimized to explain the GeV gamma-ray excess previously measured by the EGRET experiment. A slight improvement was achieved by increasing the relative contribution from the pion decay channel. But even after this fit to the measured profiles the $\chi^2$value still indicated a poor agreement between the GALPROP model and the Milagro data. 

\begin{figure*}[!t]
  \centering
  \includegraphics[width=\textwidth]{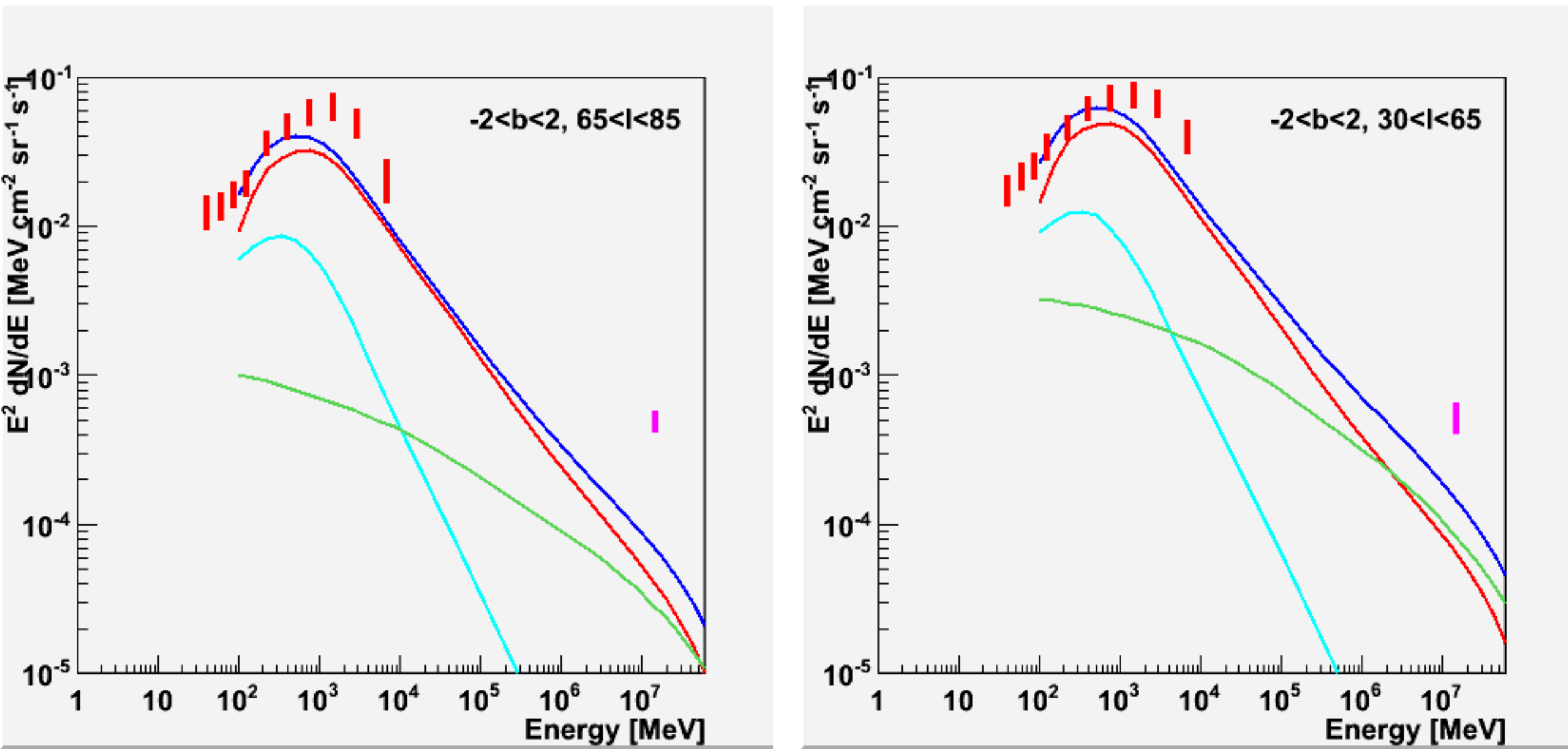}
  \caption{Measured diffuse differential gamma-ray fluxes (EGRET in red, and Milagro in magenta) compared with GALPROP predictions \cite{bib:Abdo2008,bib:Strong2000,bib:Strong2004_1,bib:Strong2004_2}. IC (green), pion (red), and bremsstrahlung (teal) contributions and the total predicted flux (blue) are shown for the Cygnus region and the inner Galaxy.}
  \label{GALPROPMilagro_fig}
 \end{figure*}

The recent results from Milagro and H.E.S.S. support the hypothesis that the cosmic-ray flux is likely to vary throughout the Galaxy. Both excesses have also been studied for an association with the dark matter particles \cite{bib:Bi2009,bib:Crocker2010,bib:Bertone2009,bib:Meade2010}, but for the Milagro result the explanation that the excess is due to unresolved sources is more likely \cite{bib:Casanova2008}. 
In contrast, recent measurements with the Fermi space telescope show no such discrepancy between the conventional assumption of the locally measured cosmic-ray spectrum and measurements of propagated high energy gamma rays \cite{bib:Ackermann2012,bib:Abdo2009}. If the Milagro observations are compared with the conventional GALPROP assumption (see Figure \ref{GALPROPMilagro_fig}) significant excesses are seen both in the Cygnus region (65$^o < $ gal. longitude $< 85^o$, 8x) and in a region outside of Cygnus closer to the Galactic Center (30$^o <$ gal. longitude $< 65^o$, 4.7x) \cite{bib:Sinnis2010}. 

\section*{Expected HAWC Performance}

The Milagro data are not sensitive enough for a spectral energy distribution measurement, so interpretations for the diffuse excess range from leptonic processes - though disfavored by the expected fast cooling of highly energetic electrons - to freshly accelerated cosmic rays that are injected into the interstellar medium. The latter explanation seems simpler in Cygnus, a region that hosts intense star formation activity and is abundant with molecular clouds and candidate cosmic-ray sources. Based on \cite{bib:Gabici2007} it is estimated that more than ten strong young accelerators in the Cygnus region are needed to explain the excess emission, but in order to definitely answer the question if the gamma-ray production is of hadronic or leptonic nature a spectral measurement is necessary. The differential sensitivity of the HAWC observatory is expected to be good enough to perform such a measurement.  In addition, due to its improved point source sensitivity (15x that of the Milagro detector) and angular resolution ($<$ 0.2 deg for energies $>$ 10 TeV), the experiment is expected to detect more point and extended sources with greater accuracy. These will be subtracted from the total flux along the Galactic plane to constrain the truly diffuse emission better. HAWC will also provide an opportunity to perform morphological studies that will reveal the location of cosmic-ray production and acceleration. This will be of particular interest in the Cygnus region that has been the subject of numerous studies recently. None of these studies reach the very high gamma-ray energies that can be measured with the HAWC observatory (up to $\sim$100 TeV). Moreover, HAWC will have access to the Galactic Center ridge, a region where the H.E.S.S. experiment has detected gamma-ray signatures that are inconsistent with expectations based on the locally measured cosmic-ray 
spectra \cite{bib:Aharonian2006Nature}. 

\section*{Outlook}
We will present the sensitivity of HAWC to and significance maps of diffuse gamma-ray emission expected after one to five years of operation of the complete array. We will test the detector response to alternative models of diffuse emission such as GALGROP, and a model based on the Milagro result \cite{bib:Abdo2008,bib:Strong2000,bib:Strong2004_1,bib:Strong2004_2}. Figure 3 shows a first simulated map of the number of events that are due to galactic diffuse gamma-ray emission as predicted by GALPROP after one year of HAWC operation. A detailed description of the software package that is used to simulate the detector behavior can be found elsewhere in these proceedings \cite{bib:BenZiv2013}.

\begin{figure*}[!t]
  \centering
  \includegraphics[width=\textwidth]{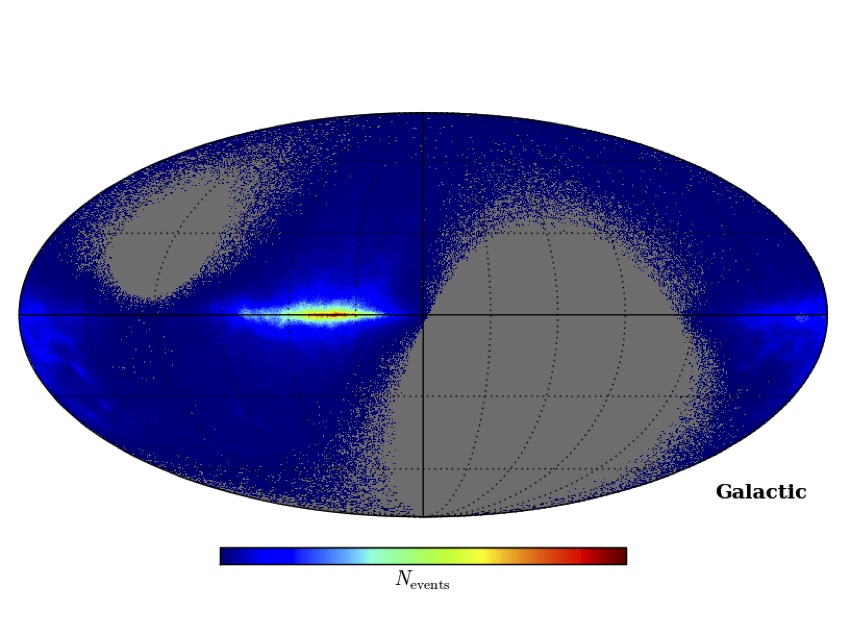}
  \caption{Galactic map of one year of diffuse gamma rays observed with the completed HAWC detector.}
  \label{GALPROPHAWC_fig}
 \end{figure*}

\section*{Acknowledgments}

We acknowledge the support from: US National Science Foundation (NSF); US
Department of Energy Office of High-Energy Physics; The Laboratory Directed
Research and Development (LDRD) program of Los Alamos National Laboratory;
Consejo Nacional de Ciencia y Tecnolog\'{\i}a (CONACyT), M\'exico; Red de
F\'{\i}sica de Altas Energ\'{\i}as, M\'exico; DGAPA-UNAM, M\'exico; and the
University of Wisconsin Alumni Research Foundation.

\clearpage


\newpage
\setcounter{section}{2}
\nosection{Sensitivity and Status of HAWC\\
{\footnotesize\sc Jordan Goodman, John Pretz}}
\setcounter{section}{0}
\setcounter{figure}{0}
\setcounter{table}{0}
\setcounter{equation}{0}
%
%

\title{Sensitivity and Status of HAWC}

\shorttitle{Sensitivity and Status of HAWC}

\authors{
Jordan Goodman$^{1}$,
John Pretz$^{2}$,
for the HAWC Collaboration$^{3}$.
}

\afiliations{
$^1$ Department of Physics, University of Maryland College Park \\
$^2$ Physics Division, Los Alamos National Lab \\
$^3$ For a complete author list, see the special section of these proceedings\\
}

\email{goodman@umdgrb.umd.edu}

\abstract{The High Altitude Water Cherenkov (HAWC) Observatory, 
under construction at Sierra Negra, Mexico will detect energetic air
showers from primary hadrons and gamma rays with energies from 100 GeV
to 100 TeV. The first stage 
of the instrument, HAWC-30, with 10\% of the channels deployed has been 
completed and is performing as expected. We anticipate HAWC-100 will be 
operational by summer 2013 with the full HAWC Observatory (with 300 
detectors) being completed in 2014. HAWC complements existing Imaging 
Atmospheric Cherenkov Telescopes and the space-based gamma-ray telescopes 
with its extreme high-energy reach and its large field-of-view ($\sim$2sr). The full 
HAWC instrument will be used to study particle acceleration in Pulsar 
Wind Nebulae, Supernova Remnants, Active Galactic Nuclei and Gamma-ray 
Bursts. 
}


\maketitle

\section*{Introduction}

The High Altitude Water Cherenkov (HAWC) observatory is a multi-TeV photon 
detector under construction at a high-altitude site in Mexico. Eventually 
covering some 20,000 square meters, the instrument is sensitive to 
100 GeV - 100 TeV photon signals from astrophysical sources. It 
will be used to study high-energy emission from typical high-energy 
photon sources: Galactic pulsar wind nebulae (PWN) \cite{Jim}, supernova remnants 
(SNR), diffuse emission in the galactic plane \cite{Petra} , as 
well as active galactic nuclei (AGN)\cite{Asif} and gamma-ray bursts (GRBs) \cite{Sparks}. 
We will also use it to search for TeV photons from 
dark matter annihilation in our Galaxy \cite{Brian} as well as constrain the 
evaporation of primordial black holes \cite{Tilan}. Here we describe the simulation 
of the instrument and determination of the sensitivity to steady 
high-energy sources.

\section*{HAWC Instrument}

\begin{figure}[t]
  \centering
  \includegraphics[width=0.5\textwidth]{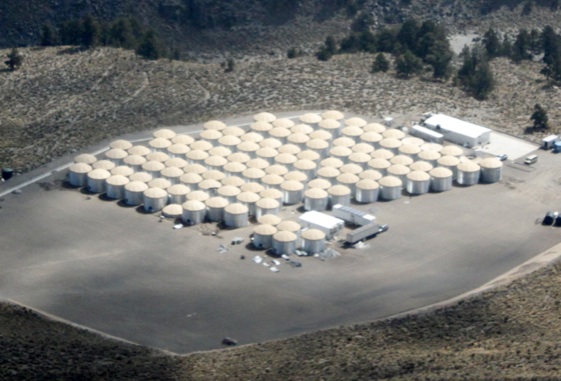}
  \caption{
May 16, 2013 picture from the HAWC site. Steel from 
approximately 
30\% of the final WCDs are visible. The cleared area indicates the size of the 
completed instrument,  $\sim$20000 m$^2$.
}
  \label{detector}
\end{figure}

The instrument consists of 300 water Cherenkov detectors (WCDs): 
4.5-meter tall, 7.3-meter diameter steel water tanks lined with a 
plastic bladder, filled with clear water, and instrumented with four
 photo-multiplier tubes (PMTs) in each. The WCDs are deployed 
close-packed over 20,000 square meters on a 4100-meter plateau near the 
Sierra Negra at $+19^\circ$N in Mexico (Figure \ref{detector}. The WCDs measure the timing and density of air shower 
 particles reaching the ground.
 Custom front-end 
electronics partially re-used from the Milagro experiment are used to 
record the leading-edge time and total charge seen by each PMT during an 
air shower. The particles from an air shower arrive in a thin planar sheet 
propagating at the speed of light which washes over the instrument and the 
arrival time of light is used to determine the direction of the original 
primary particle.

Air showers are modeled using the CORSIKA program developed for the 
KASCADE experiment. The ground detector components are modeled using a
 Geant 4 simulation. The simulation was validated against data from the 
Milagro experiment and comparison to early HAWC data suggests the 
simulation is sufficient to estimate the sensitivity of the whole 
instrument. Reconstruction algorithms developed for Milagro are applied 
to the simulated output.

The HAWC angular resolutions varies 
between $\sim 1^\circ$ to $0.1^\circ$ depending on how many channels are hit 
in the event. The requirement that the core be accurately identified 
practically limits the instrument to observe particles with a core 
within the geometric area of the instrument.

\begin{figure}[t]
  \centering
  \includegraphics[width=0.5\textwidth]{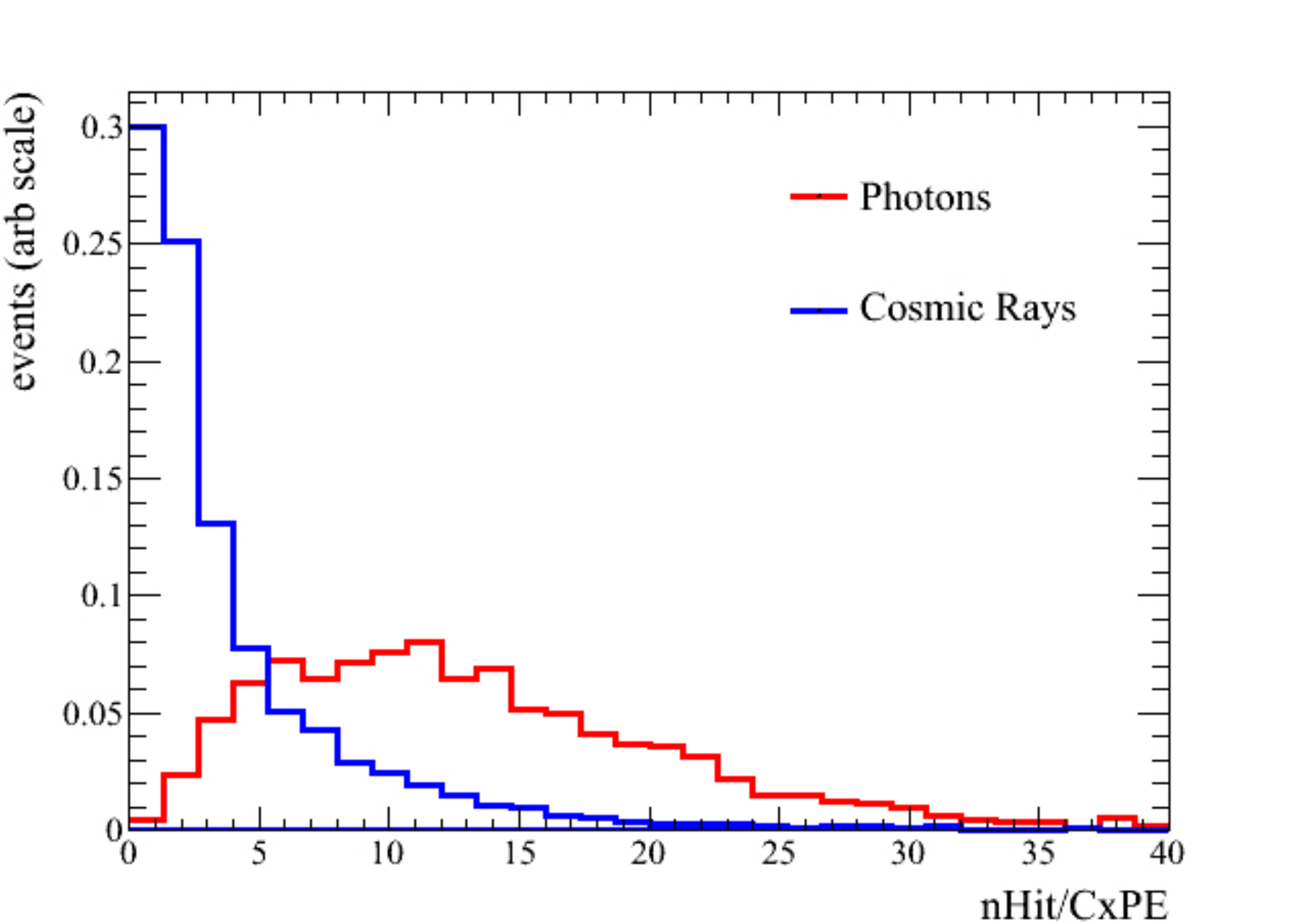}
  \caption{Compactness (nHit/CxPE) distribution for 
photon/hadron discrimination for a photon source at 35 degrees
declination for events with more than 700 PMTs hit. 
CxPE is the number of PEs detected
in the hardest-hit PMT outside of a raidus of 40 meters. The separation,
particularly good for photons over 1 TeV, is evident.
}
  \label{cxpe}
\end{figure}

The challenge in detecting photon sources is the large background of 
hadronic cosmic rays. Due to randomization by Galactic magnetic fields, 
cosmic-ray particles are nearly isotropic where photon sources are 
well-localized. Gamma-ray sources appear as a small bump on the 
nearly-smooth cosmic-ray background. Additionally, cosmic rays produce 
a hadronic air shower where gamma rays produce a nearly pure 
electromagnetic shower. The penetrating particles (primarily muons) 
and clumpy structure (from sub-showers with high transverse momentum) 
of a hadronic shower differs from the smooth distribution produced by 
a pure electromagnetic shower and we exploit this difference to 
suppress the hadronic background. 
Figure \ref{events} 
exhibits some sample simulated signal and background events. 
We utilize a parameter, the compactness, 
defined as nHit/CxPE where nHit is the total number of PMTs 
participating in an event and CxPE is the number of PEs recorded in 
the hardest-hit channel outside of a radius of 40 meters from the shower 
core to distinguish photons from hadrons. The background rejection is 
strongly dependent on the shower energy but rejections of 10$^{-2}$ are 
attainable while maintaining a signal efficiency of ~50\%. Figure 
\ref{cxpe} shows characteristic distributions of CxPE for photon
events and hadron backgrounds.

In order to estimate the sensitivity to a localized source of gamma 
rays at a specified declination, we consider following a small angular 
bin of some angular radius around a potential source as the source 
transits through the HAWC field of view. The simulated output of the 
detector is weighted by the amount of time a source at the supposed 
declination will spend at each zenith angle in the detector. The 
data are divided into bins of nHit and log10(nPE), the number of 
hit PMTs in an event and the number of photo-electrons seen during 
in all PMTs during the event, as a proxy for the primary particle 
energy. Within each bin, we determine optimal cuts on the angular 
bin used and on the compactness.

\section*{Sensitivity Results}
\begin{figure}[t]
  \centering
  \includegraphics[width=0.4\textwidth]{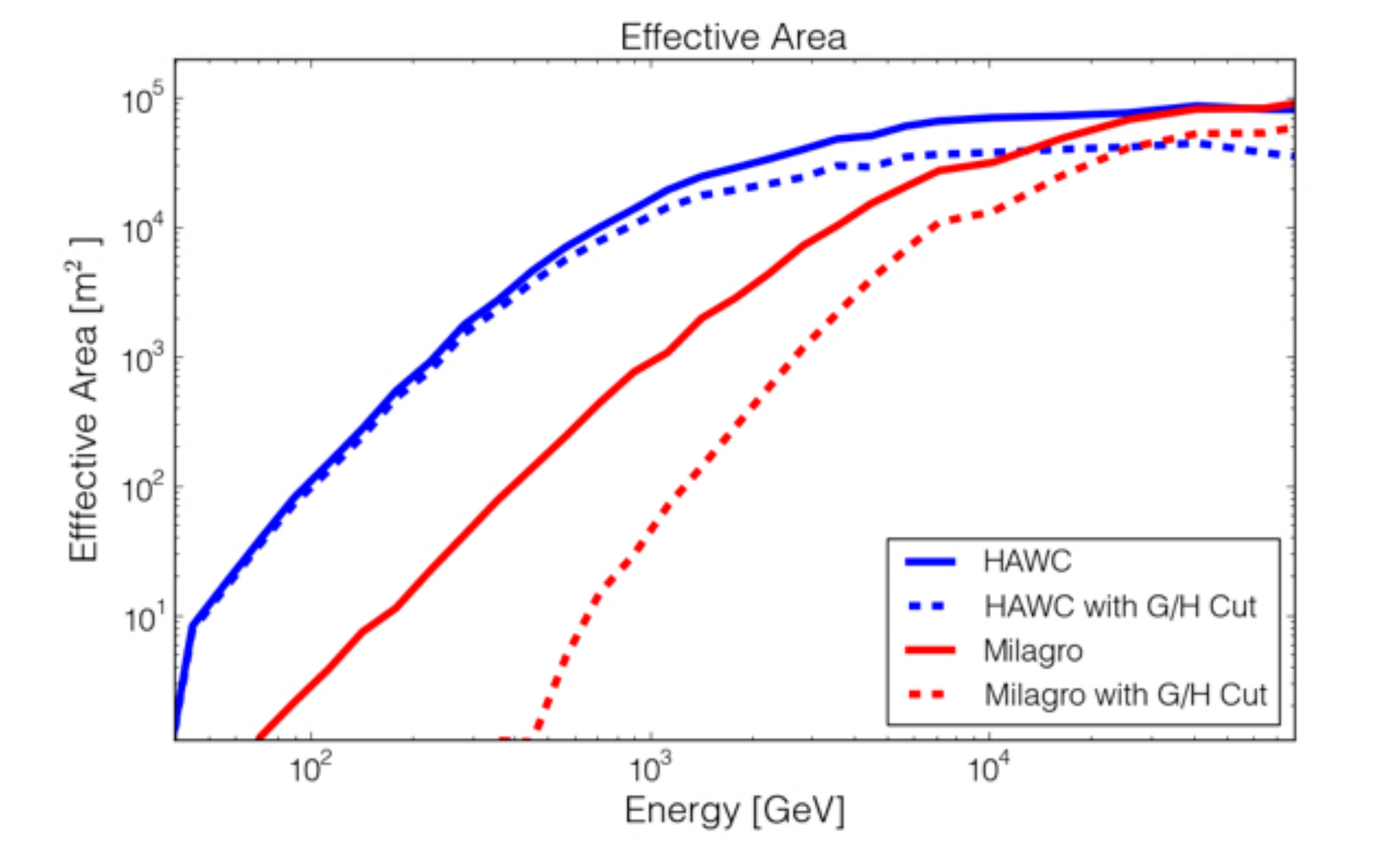}
  \includegraphics[width=0.4\textwidth]{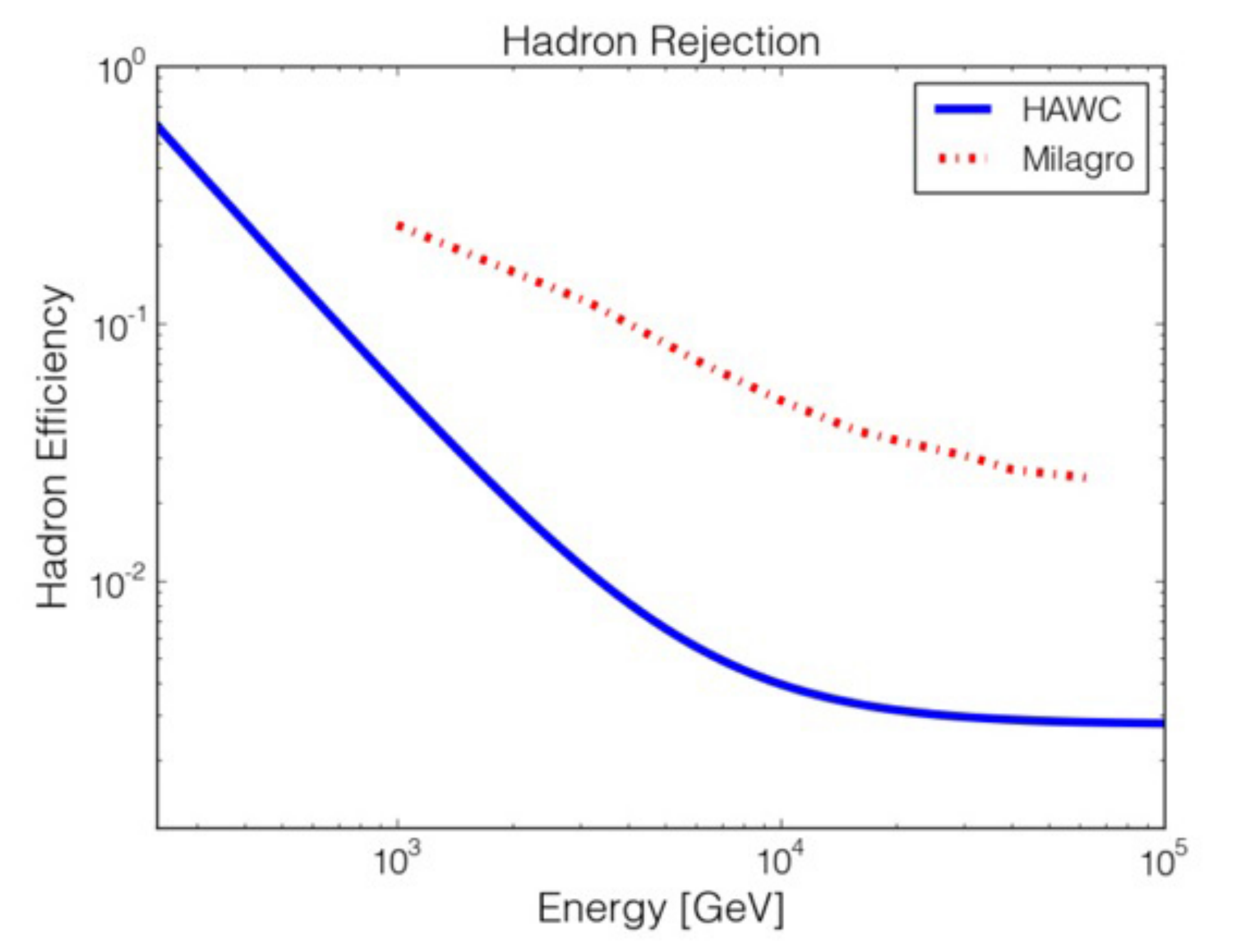}
  \includegraphics[width=0.4\textwidth]{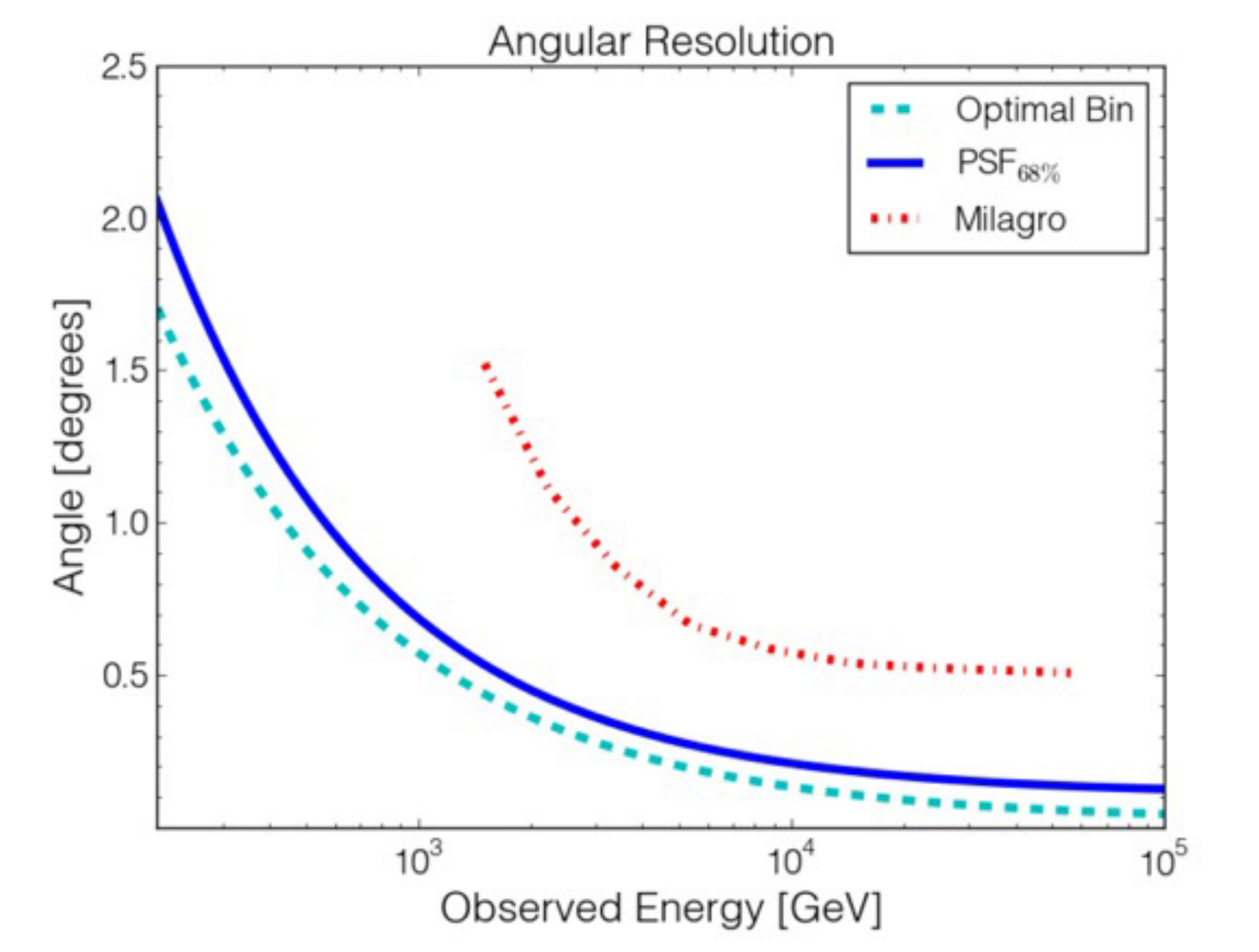}
  \caption{HAWC and Milagro effective area vs energy (top) gamma hadron rejection (middle) and angular resolution (bottom)}
  \label{resolution}
\end{figure}

Figure \ref{resolution} shows the effective area of the HAWC instrument to photons. 
The effective area rises with energy up to 1 TeV. This rise is due 
to the increasing probability for a particle to produce a detectable 
number of energetic particles at the ground. If a photon randomly 
were to interact higher in the atmosphere (since at 4100 meters we 
are lower than the peak particle production in the air shower) the 
resulting shower would be smaller. Up to 1 TeV we detect particles 
which randomly fluctuate to interact deeper in the atmosphere. At 1 
TeV this probability is nearly 1.0 and the effective area plateaus 
at nearly the geometric area of the instrument. This plateau is due 
to the requirement that the air shower core be identified for accurate 
correction of the air shower curvature. When the core lies off the 
detector the core location is ambiguous with the current generation 
of algorithms and hardware. This limits the effective area to the 
geometric area at high energies.

\begin{figure}[t]
  \centering
  \includegraphics[width=0.49\textwidth]{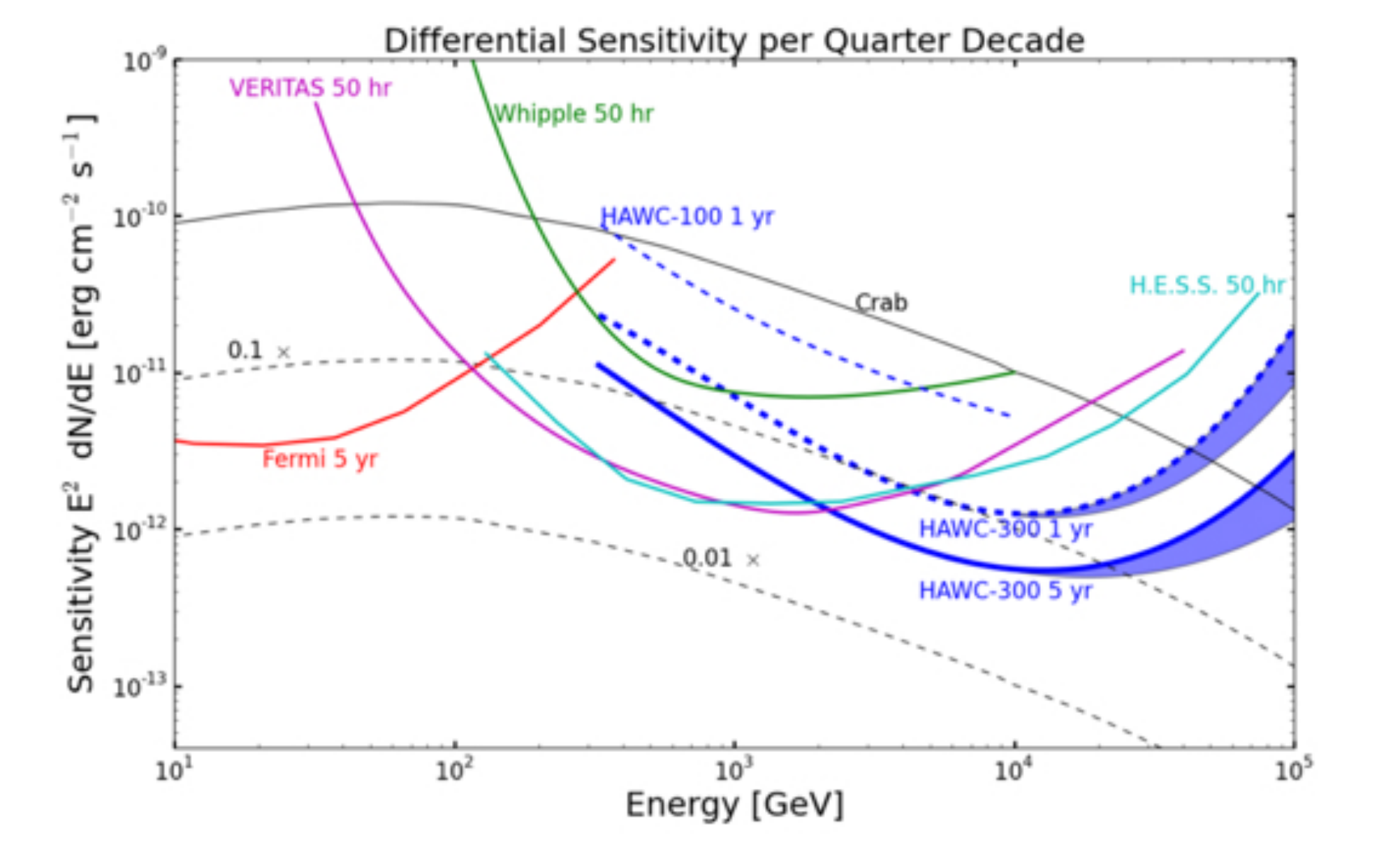}
 \caption{Differential sensitivity per quarter decade of HAWC (for 1
    and 5 years and 1 year of HAWC 100) is shown compared to other
    existing and future IACTs. Note that sensitivity of HAWC and Fermi-LAT is for a sky survey while IACTs are 50 hours on a source.
}
  \label{sensi}
\end{figure}
To characterize the instrument’s sensitivity to sources we consider 
source differential energy spectra of the form 
$\frac{dN}{dE} = \phi_0(E/TeV)^{-\alpha} e^{(E/Ecut)}$ where $\phi_0$ is the source overall flux, 
$\alpha$ is the spectral index and $E_{cut}$ is the exponential cutoff energy.

\begin{figure}[t]
  \centering
  \includegraphics[width=0.5\textwidth]{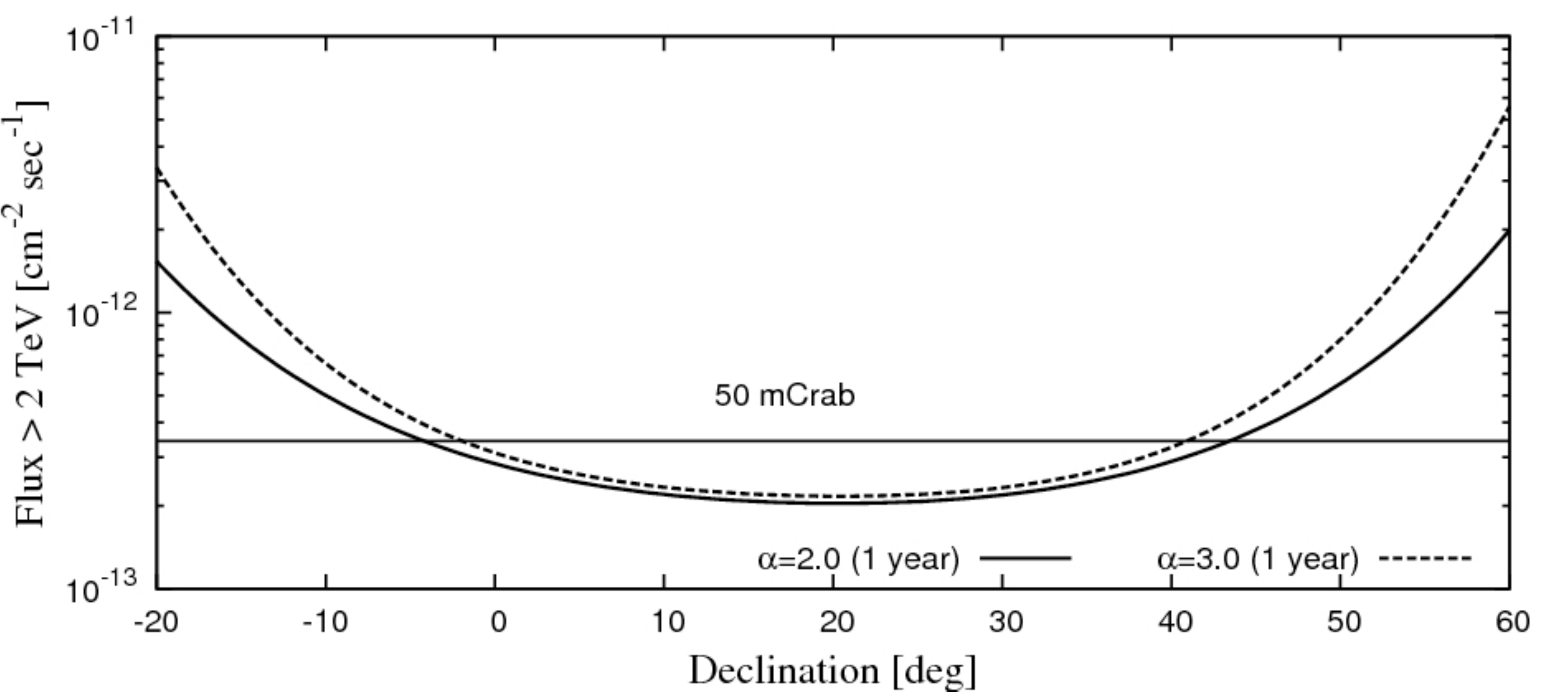}
  \includegraphics[width=0.5\textwidth]{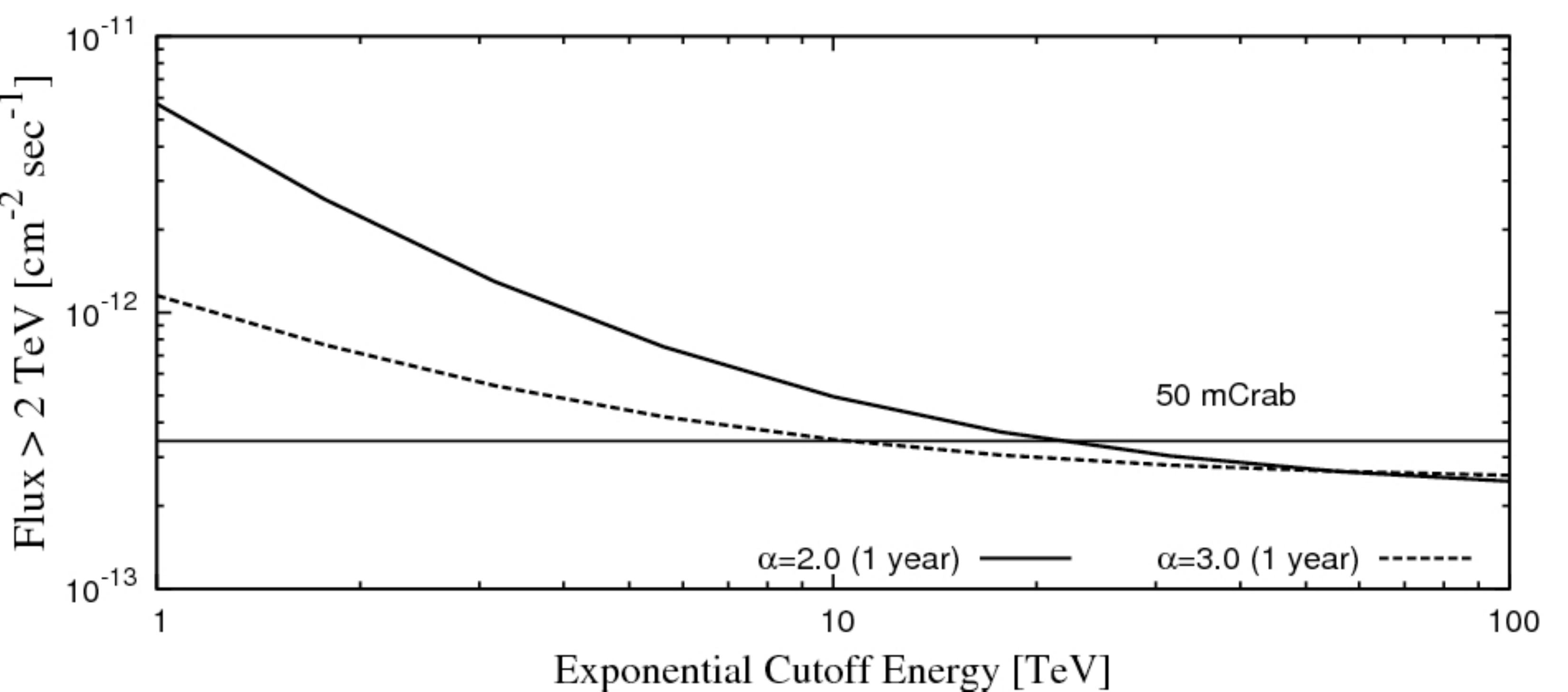}
  \includegraphics[width=0.5\textwidth]{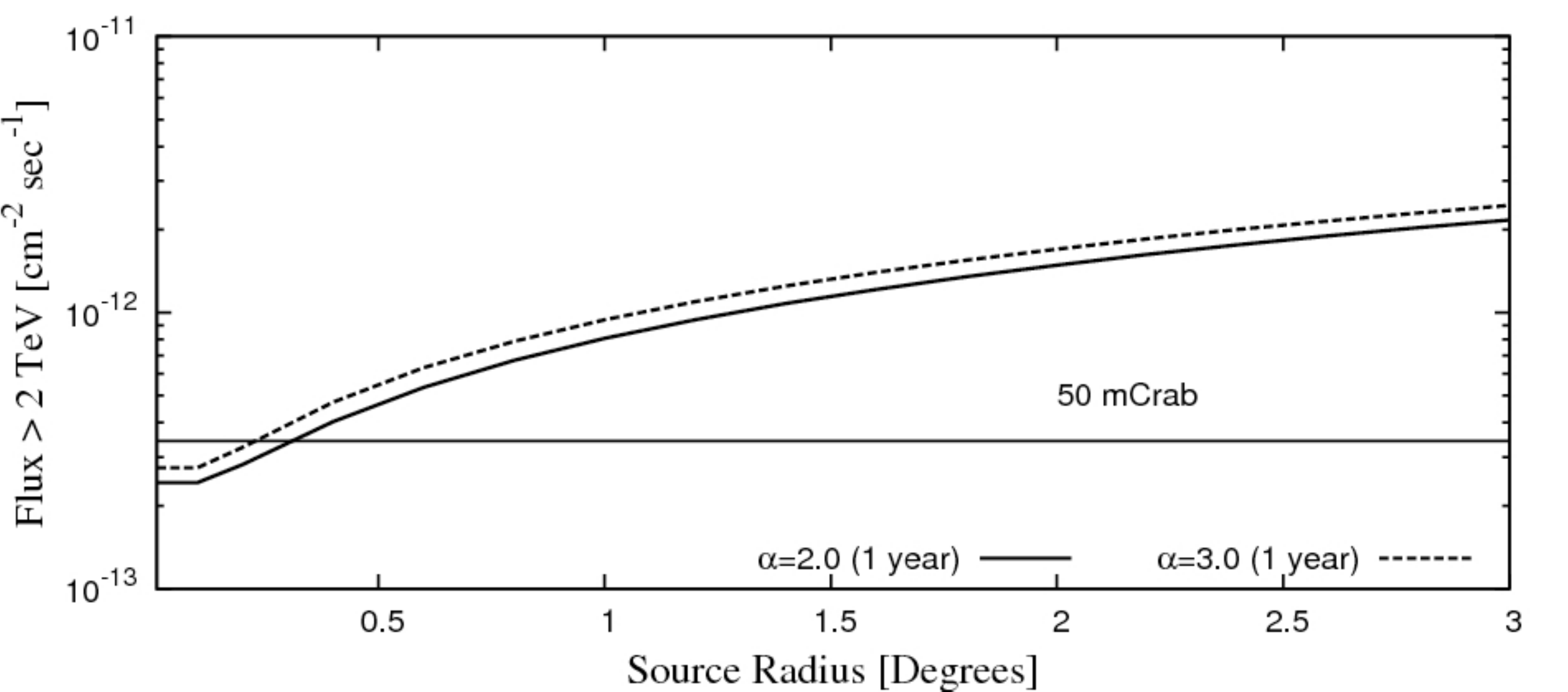}
  \caption{
The sensitivity of HAWC to sources with varying spectral parameters.
The top panel shows the sensitivity of HAWC to sources with pure power-law
spectra as a function of declination. In the center panel, the sensitivity to a
source at $+35^\circ$ declination is shown as a function of spectral cutoff energy. In
the bottom panel, the sensitivity to a source (also at $+35^\circ$) with a pure power
law spectrum is shown as a function of the spatial extent of the source. Note
that one day of data corresponds to one transit of the source, which means the
source spends only a few hours in the field of view of the detector.}
  \label{sensitivity}
\end{figure}

High-energy showers landing off the detector trigger the instrument 
but the core, with the current generation of hardware and algorithms, 
cannot be accurately determined meaning the curvature correction fails. This
results in the uncertainty of our differential sensitivity etimate at high
energies and is seen in the bands in figure \ref {sensi}. Figure
\ref{sensitivity} shows the HAWC sensitivity vs source declination, cutoff energy  and source radius.

\section*{Discussion}

The HAWC instrument is designed to study particle acceleration in Galactic
and extra-Galactic sources as well as the propagation of high-energy particles
through the Galaxy and the extra-Galactic background light (EBL). 

Pulsar Wind Nebulae (PWN) are the most common Galactic source of TeV
gamma rays. The central pulsar drives a flow of energetic electrons
into the surrounding material lighting it up with syncrotron radiation.
Further acceleration is possible in shocks created when the 
flow interacts with surrounding material. Perhaps the most intriguing
current topic in PWN science is the detection of flares from the 
Crab Nebula, challenging our understanding of these objects.
It is currently unknown how high in energy these flares go or 
whether any other PWN flares. 

Additionally, electrons and positrons accelerated in PWN constitute
a background to dark matter searches. The most striking example
of this is the anomalous positron excess discovered by PAMELA.
While most attention has been focused on an interpretation
as positrons from the annihilation of dark matter, interpretation
via conventional pulsar physics is more likely. Developing an unbiased
high-sensitivity survey of PWN is crucial to understanding energetic
particle backgrounds for more exotic searches.

Supernova remnants (SNR) are also known to produce TeV photons. 
SNR accelerate particles at shock boundaries by 
the conventional Fermi mechanism. To date, the strongest 
positive evidence that SNR are responsible for the 
Galactic cosmic-ray population is due to TeV emission 
near SNR coincident with molecular clouds with which the SNR
is interacting. More recently, evidence of 
characteristic spectral features
from pion decay has been seen in these objects, strengthening the 
case that they are hadron accelerators. Nevertheless, 
while we have a few demonstrated examples of hadron accleration
in SNR, we still have not yet seen evidence of PeV hadron accelration
to fully account for the believed Galactic cosmic ray population. 
HAWC's unbiased survey out to 100 TeV will help identify or constrain
instances of PeV hadron accleration. 

Furthermore, HAWC will be used to study Active Galactic 
Nuclei (AGN), one of the leading candidates for UHE 
cosmic-ray acceleration. While AGN are very well-established
TeV sources much remains to be learned. Some are known to flare
by an order of magnitude or more in a matter of hours meaning
very small regions of emission. Furthermore, no long-term monitoring
of AGN at TeV energies is possible because TeV
Cherenkov telescopes are constrained in the number of hours they
can look at a source and may only observe when the source is
up on dark moonless nights. HAWC will conduct an unbiased
survey of the TeV sky every night and can search for transient
and high-energy emission from AGN.

\section*{HAWC-30 Results}

\begin{figure}[t]
  \centering
  \includegraphics[width=0.4\textwidth]{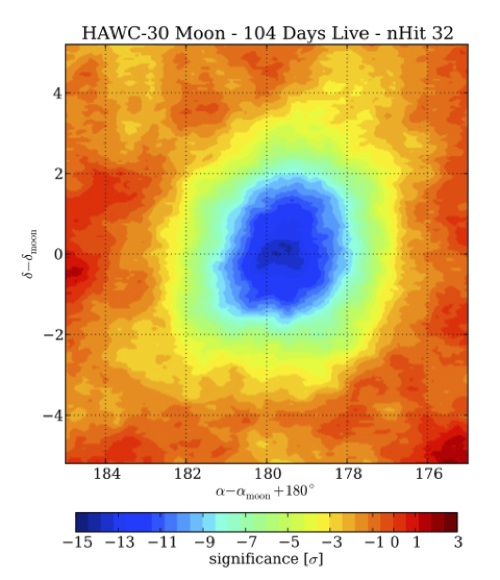}	
 \caption{104-days of the HAWC-30 Moon map.}
  \label{moon}
\end{figure}

\begin{figure}[t]
  \centering
  \includegraphics[width=0.4\textwidth]{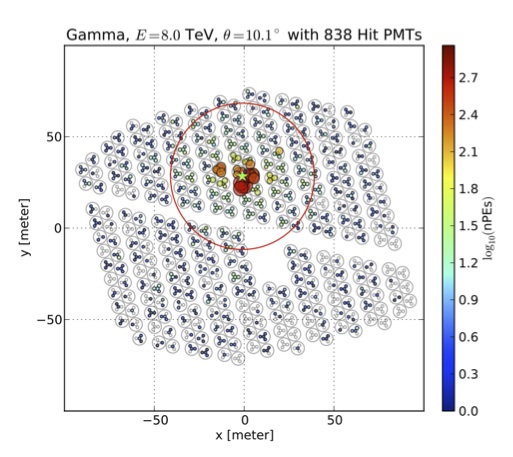}
  \includegraphics[width=0.4\textwidth]{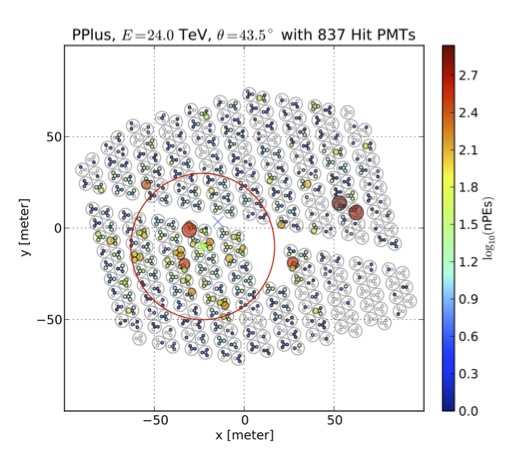}
  \caption{
Two simulated air showers in HAWC. The color and size
scale show the number of PEs detected in each PMT. The photon 
event (top panel) has most of the high light deposition in the
central core of the air shower whereas the proton event (bottom panel)
shows significant localized charge deposition outside of the shower
core region.}
  \label{events}
\end{figure}

The modular design of HAWC allows us to take data even as counters are still being deployed. Starting in October 2012 we began operating the HAWC detector with 30 WCDs.  An observation of the cosmic-ray Moon shadow with HAWC-30 serves as a diagnostic test of the resolution and pointing in the detector’s angular reconstruction. (see Figure \ref {moon}). We observed the Moon shadow from 2012 October 22 to 2013 March 08 and accumulated 35 billion cosmic ray events over 104 days of livetime \cite{moonpaper}. Using the data quality cut of nHit≥32, seven billion events survive. The peak significance is -14.1 and is centered at ($179.6^\circ \pm 0.1^\circ, 0.1^\circ \pm 0.1^\circ$ )

 \section*{HAWC Construction and Schedule}
HAWC construction began in February 2011. As of May 22, 2013, 106 WCD Tanks have been built, with 77 WCDs and 283 PMTs in the data stream. Although we have been operating the detector more or less continuously since HAWC-30 in October 2012, we will begin physics operations of HAWC-100 (more likely HAWC-110) in August 2013. HAWC-250 is on schedule for completion in August 2014 and HAWC-300 should be completed by the end of 2014.

\section*{Acknowledgments}

We acknowledge the support from: US National Science Foundation (NSF); US
Department of Energy Office of High-Energy Physics; The Laboratory Directed
Research and Development (LDRD) program of Los Alamos National Laboratory;
Consejo Nacional de Ciencia y Tecnolog{\'\i}a (CONACyT), M{\'e}xico; Red de
F{\'\i}sica de Altas Energ{\'\i}as, M{\'e}xico; DGAPA-UNAM, M{\'e}xico; the
University of Wisconsin Alumni Research Foundation and the Institute of
Geophysics, Planetary Physics and Signatures at Los Alamos National Lab.

\clearpage


\newpage
\setcounter{section}{3}
\nosection{Search for High-Energy Emission from GRBs with the HAWC
Observatory\\
{\footnotesize\sc Kathryne Sparks}}
\setcounter{section}{0}
\setcounter{figure}{0}
\setcounter{table}{0}
\setcounter{equation}{0}
%
%
%
%
\title{Search for high-energy emission from GRBs with the HAWC Observatory}

\shorttitle{Search for GRBs with HAWC}

\authors{
K. Sparks$^{1}$,
for the HAWC Collaboration.
}

\afiliations{
$^1$ Department of Physics, Pennsylvania State University, University Park, PA, USA \\
}

\email{kjs361@psu.edu}

\abstract{A second generation water Cherenkov detector, the High Altitude Water Cherenkov (HAWC) Observatory is currently being constructed in Sierra Negra, Mexico at an altitude of 4100~m asl. With higher altitude than its predecessor Milagro, HAWC will be almost two orders of magnitude more sensitive to GRBs at 100 GeV. Due to its wide instantaneous field of view ($\sim2$~sr) and long duty cycle, this Extensive Air Shower detector can observe the beginning of the prompt phase of GRBs without needing to slew. HAWC is sensitive to showers in the sub-TeV to TeV energy range and will be able to help constrain the shape and cutoff of high-energy GRB spectra, especially in conjunction with observations from other detectors such as Fermi. Data taking with a partially built array began in 2012. With only 10\% of the array completed, HAWC already provides a substantial improvement over Milagro's sensitivity to GRBs. We present the results of a search for high energy emission from GRBs detected by other instruments using HAWC data.}


\maketitle

\section*{Introduction}

Gamma-ray bursts (GRBs) are extremely powerful transient events that occur at cosmic distances. 
The exact origins are still unknown, but GRBs are thought to occur during neutron star-neutron star or neutron star-black hole mergers \cite{bib:ruffert,bib:rosswog} or the core collapse of massive stars \cite{bib:woosley,bib:macfadyen}.
A jetted, highly relativistic fireball interacting with itself or the surrounding interstellar matter, forming internal and external shocks in which Fermi-acceleration takes place, delivers a plausible explanation of the non-thermal spectrum of GRBs \cite{bib:meszaros,bib:meszaros2,bib:piran}.
The emission mechanism for the high-energy gamma rays is not yet completely understood.

Observations of energy spectra of GRBs can provide information about the intervening space between the burst and Earth as well as about the source itself.
As gamma rays propagate through the interstellar media, they interact with the extra-galactic background light (EBL), which can cause attenuation via pair-production \cite{bib:gilmore}. 
The density of the EBL can consequently be probed with the observation of a high-energy cutoff.
GRB prompt emission is typically described by the Band function \cite{bib:band}, two power laws joined by an exponential cutoff.
However, recent observations have shown that a Band function alone cannot sufficiently describe many high-energy gamma-ray bursts \cite{bib:grb090510,bib:grb090902b}.

The release of the \emph{Fermi} LAT Gamma-Ray Burst Catalog \cite{bib:fermilatcatalog} earlier this year summarized knowledge of the high-energy component of LAT GRBs.
The \emph{Fermi} Gamma Ray Space Telescope consists of two detectors, the Large Area Telescope (LAT), operating at energies between $\sim$20~MeV and more than 300~GeV, and the Gamma-ray Burst Monitor (GBM), whose energy range is lower at 8~keV to 40~MeV.
Thirty-five GRBs were detected by LAT in a three-year period starting in August 2008.

From these detections, we garner more information about the characteristics of high-energy GRBs. 
LAT-detected emission is commonly delayed with respect to the lower energy counterpart detected by GBM.
Additionally, a temporally extended phase during which LAT flux decays following a single or broken power law with index close to $F_{\nu} \propto t^{-1}$ is observed.
This extended emission is consistent with that expected from forward shock emission from a relativistic blast wave and favors an adiabatic fireball model over a radiative one.
Finally, joint GBM-LAT spectral fits to all LAT GRBs require a power-law component in addition to the Band function to fit the bursts.

As a space-based instrument, \emph{Fermi} LAT's effective area is limited by the size of the satellite. Due to the paucity of the flux, observation of the highest energy gamma rays requires a larger effective area.
To do this, two different classes of ground-based detectors exist: Imaging Atmospheric Cherenkov Telescopes (IACTs) \cite{bib:hinton} and Extensive Air Shower (EAS) particle detector arrays \cite{bib:sinnis}. 
IACTs have very good angular and energy resolution enabling a high degree of sensitivity. 
However, they can only observe on clear, moonless nights ($\sim$10\% duty cycle) and have a small field of view ($<5^{\circ}$). The prompt phase of a GRB is often missed by IACTs due to slewing.
EAS detectors, such as the High Altitude Water Cherenkov (HAWC) Observatory, have a nearly 100\% duty cycle and a large field of view ($\sim2$~sr), thus allowing for easier detection of the prompt phase of a GRB.

\section*{The HAWC observatory}

Located at 4100~m in Sierra Negra, Mexico, HAWC will consist of 300 7.3~m wide and 4.5~m deep water tanks when completed.
There are three 20~cm photomultiplier tubes (PMTs) and one 25~cm high quantum efficiency PMT at the bottom of each tank.
Charged particles from an extensive air shower are detected when they pass through each tank and emit Cherenkov radiation. HAWC data is collected by two data acquisition systems (DAQs). 
The main DAQ measures the arrival time and time over threshold (TOT) of PMT pulses, hence providing information for the reconstruction of the shower core, direction and lateral distribution, which in turn helps to determine the species of primary particle and its energy. 
A secondary DAQ, the scaler system, operates in a PMT pulse counting mode \cite{bib:vernetto} and is sensitive to gamma ray and cosmic ray (i.e. due to solar activity) transient events that produce a sudden increase or decrease in counting rates with respect to those produced by atmospheric showers and noise. GRB results discussed in the following sections are primarily from the main DAQ while results from the scaler DAQ are presented in \cite{bib:icrcscalers}.
For more information on the HAWC observatory see \cite{bib:grbsensitivity} and \cite{bib:icrchawc} in these proceedings.

\section*{VAMOS / HAWC GRB selection}

 \begin{figure}[ht]
  \centering
  \includegraphics[width=0.45\textwidth]{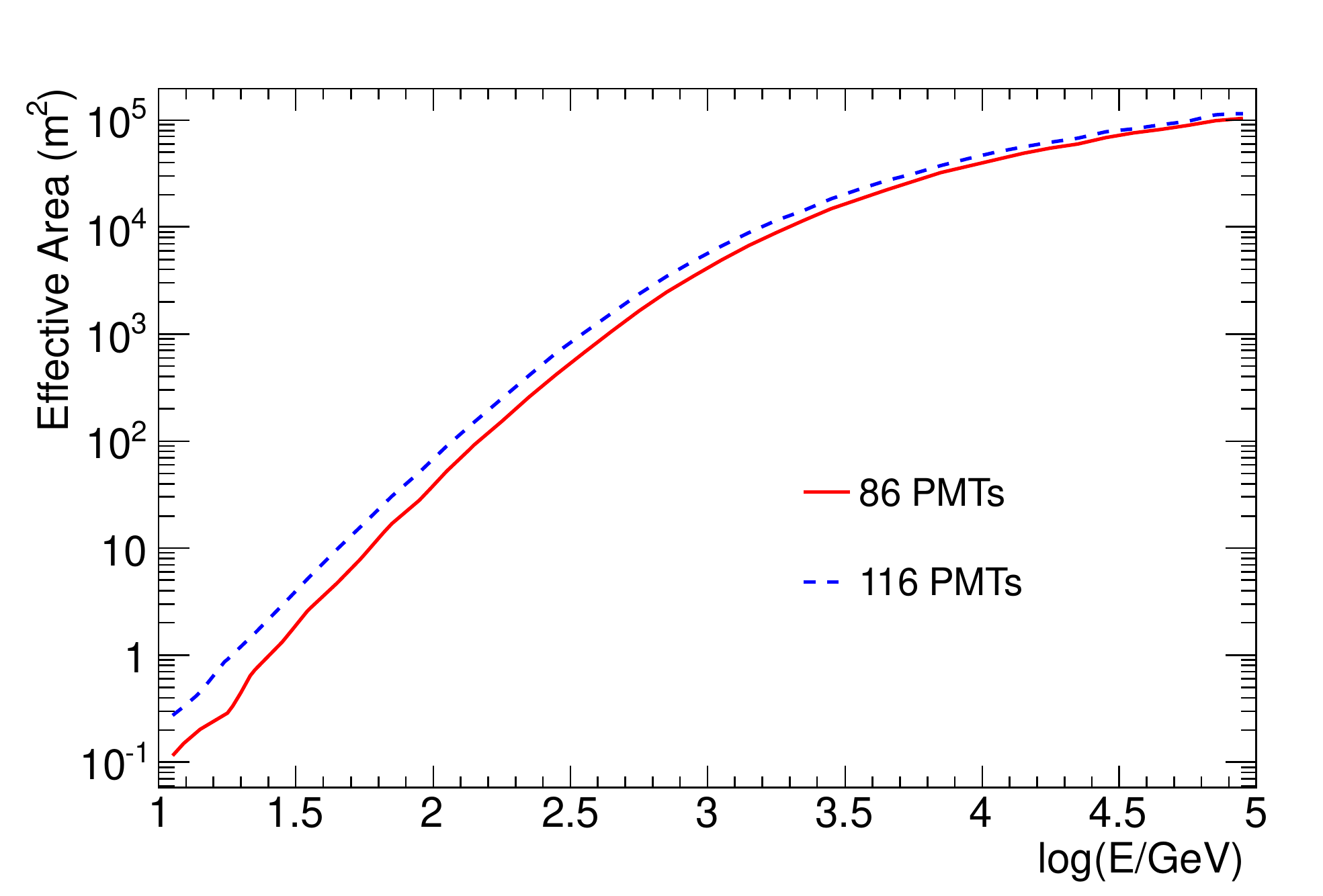}
  \caption{Effective area of HAWC main DAQ as a function of $\gamma$-ray energy for showers $<45^{\circ}$ zenith. A trigger threshold of nHit $\geq$10, corresponding to a rate of $\sim$7 kHz, is assumed. Showers reconstructed with $> 4^\circ$ error are excluded. No gamma-hadron separation cut is applied.}
 \label{fig:effarea}
 \end{figure}

HAWC's Verification and Assessment Measuring of Observatory Subsystem (VAMOS) prototype took data from September 2011 to June 2012 with $\sim$30\% live time. VAMOS consisted of 6 tanks with 31 PMTs. 

Based on GCN circulars\footnote{http://gcn.gsfc.nasa.gov/gcn3 archive.html}, two intense GRBs that occurred within VAMOS' field-of-view (FoV) (zenith angle $< 45^{\circ}$) and uptime were selected for analysis: GRB 111016B and GRB~120328B.
Both bursts were discovered by IPN (GCN circulars 12452 and 13157),
with a gamma ray fluence reported by Konus-Wind in excess of 10$^{-4}$~erg/cm${^2}$ in both cases.
In addition to the low-energy gamma rays seen by GBM, \emph{Fermi} LAT reported a detection of high-energy emission from GRB~120328B (GCN circular 13165).
For VAMOS, GRB~111016B had a zenith angle of 32$^{\circ}$, while GRB~120328B was at a less favorable 41$^{\circ}$.
No redshift information is available for these GRBs. 

Regular data taking with the partially constructed HAWC array began in September 2012. 
Approximately 145 days of live time was accumulated during the first 7 months (September through April) by the growing experiment. 
The dataset used in this analysis contains events with a maximum of 86 to 114 PMTs, triggered at a rate of approximately 7 kHz. 
The effective area for the complete HAWC array was shown in \cite{bib:grbsensitivity} while Figure \ref{fig:effarea} shows the relevant effective
area for the following analysis.

Based on the \emph{Fermi} GBM\footnote{http://heasarc.gsfc.nasa.gov/W3Browse/all/fermigbrst.html} and Swift\footnote{http://heasarc.nasa.gov/docs/swift/archive/grb\_table/} 
GRB catalogs, we selected 12 bursts that were in the FoV of HAWC (zenith angle $< 45^{\circ}$) during its main DAQ uptime and had a localization accuracy better than 5$^\circ$ (for GBM bursts). 
This includes GRB~130504C (GCN circulars 14574, 14583, and 14587), a extremely bright burst with long lasting emission from which LAT detected a photon of $\sim$5~GeV.
The results of this selection process are summarized in Table \ref{tab:results}.

The nearby super-luminous burst GRB~130427A \cite{bib:grb130427a} was at 57$^\circ$ zenith in HAWC's FoV and setting at the time of its GBM trigger. The main DAQ was not taking data at the time, but the scaler DAQ was running. A full analysis of this burst can be found in \cite{bib:icrcscalers}.

\section*{GRB search results}

We first analyzed the main DAQ VAMOS data for GRB 111016B to establish if an intense high-energy emission was present. 
The number of air showers detected during a 155~s time interval around the GRB (including 5~s before T0 = 22:41:40 UT plus the reported GRB duration) and reconstructed within 6$^{\circ}$ from the GRB position was compared to the background estimate based on the event rate in the same angular bin during a 7~hr period including the GRB.

A negative fluctuation of $\approx\,0.6\,\sigma$ was found. 
We then derived a 90\% C.L. upper limit on the number of signal events following the method of Feldman and Cousins \cite{bib:feldman}. 
The limit was then converted to flux units using a Monte Carlo simulation of the detector response.
Assuming a power law spectrum with a cutoff at 100~GeV, the upper limit on E$^2$ dN/dE at 65~GeV is 8.6 $\cdot$ 10$^{-4}$~erg/cm$^2$.
For a spectrum extending up to 316~GeV, the limit on the $>$100~GeV emission is 1.5 $\cdot$ 10$^{-4}$~erg/cm$^2$ at 208~GeV.


A similar analysis was applied to VAMOS data for GRB~120328B. 
Shower events were selected within a 7$^\circ$ radius bin centered at a location corresponding to the center of the
improved IPN error box (RA, Dec = 229.202$^\circ$, +24.818$^\circ$, based on data from \cite{bib:hurley}).
A $+2\sigma$ fluctuation was found in a 30~s time window following GRB onset (06:26:23 UT).
Consequently, and due to a less favorable zenith angle, the obtained limits are weaker than for GRB~111016B:
$3.3 \cdot 10^{-3}$~erg/cm$^2$ at 141~GeV (100 - 200~GeV band) and $1.4 \cdot 10^{-3}$~erg/cm$^2$ at 283~GeV (200 - 400~GeV band).
The limits have been corrected by a factor of 1.6 to account for systematic uncertainties in the signal detection efficiency.
The data complement the spectral measurements made at lower energies by \emph{Fermi} LAT.

\begin{table*}[!t]
\begin{center}
\begin{tabular}{l|c|c|c|c|c|c|c|c}
GRB Name & R.A.               & Dec.               & Inst. & Start & T90 (s) & Zenith 	& Bkg & \# Sig Evts \\
		   & 		          & 		      &  & Time (UT) &   & Angle ($^\circ$)	 & (evts/T90) & for $5\sigma$ det. \\ \hline\hline
121209A & 21:47:8.93 & -8:14:7.1 & Swift & 21:59:11 & 42.7 & 31.1 & 793.4 & 145.3 \\ \hline
121211A & 13:02:7.99 & 30:08:54.9	& Swift & 13:47:02 & 182.0 & 12.2 & 7329.8 & 428.2 \\ \hline
130102A & 20:45:41.63 & 49:49:03.5 & Swift & 18:10:54 & 77.50 & 40.7 & 835.6 & 149.1 \\ \hline
130131511& 12:38:31.2 & -14:28:48 & Fermi & 12:15:17.0 & 147.50 & 43.2 & 1881.1 & 216.9 \\ \hline
130215A &   02:53:56.64 & 13:23:13.2 & Swift & 01:31:27 & 65.70 & 26.8 & 2788.2 & 264.8 \\ \hline
130216A &   04:31:36.24 & 14:40:12 & Swift & 22:15:21 & 6.50 & 42.5 & 95.6 & 53.1 \\ \hline
130219A &   20:14:55.2 & 40:49:48 & Fermi & 18:35:52.4 & 96.10 & 32.3 & 2886.4 & 269.6 \\ \hline
130224370 &   13:43:36.0 & 59:43:12 & Fermi & 08:52:26.5 & 70.90 & 42.2 & 1017.6 & 160.4 \\ \hline
130307126 &   10:23:59.0 & 22:59:53 & Fermi & 03:01:44.4 & 0.40 & 40.1 & 6.99 & 17.7 \\ \hline
130327A &  06:08:9.28 & 55:42:53.3 & Swift & 01:47:30 & 9.00 & 40.4 & 188.0 & 72.7 \\ \hline
130504C & 06:06:31.3 & +03:50:02 & Fermi & 23:29:06.2 & 73.2 & 29.8 & 3205.8 & 283.2 \\ \hline
130507545 & 21:18:57.6 & -20:31:48 & Fermi & 13:04:38.0 & 60.2 & 39.8 & 1317.83 & 182.2 \\ \hline
\end{tabular}
\caption{Results for search for high-energy HAWC GRBs. The first 6 columns show the GRB name, 
  the best available localization (RA, Dec), the relevant instrument (\emph{Swift} or 
  \emph{Fermi}), the start time of the T90 analysis window, and the T90 itself (\emph{Swift} BAT or \emph{Fermi} GBM). 
  The seventh column gives the GRB zenith angle at HAWC.
  The number of events per T90 is shown in the eighth column while the last column gives the number of signal events
  needed for 50\% probability of a 5$\sigma$ detection for the background in the previous column.}
\label{tab:results}
\end{center}
\end{table*}

For GRBs in HAWC's FoV, a circular bin with a radius of 4$^\circ$ was defined for each GRB using its equatorial coordinates. Two time windows were used: one for prompt emission that is equal to the satellite-measured T90, the time over which a burst emits from 5\% of its total measured counts to 95\%, and a second one that is 3$\times$ longer (to cover extended emission). Both time windows used the same start time (see Table \ref{tab:results}). The number of events observed in the search bin defined by the 4$^\circ$ circle and the chosen time window was compared to the background estimated from off-time data.


When measured at a constant location in detector coordinates (zenith and azimuth), HAWC's background level is very stable on time scales relevant for GRBs.
This allows us to predict the background in a circular bin around any source location at any given time.
For a finite time window, the background expectation is the integral of the background rate over the time window duration,
which implies integration along the visible trajectory of the source in the detector field of view.
It is easy to see that shifting the source in time by T and the source RA by the corresponding angle in RA
will place it at the same zenith and azimuth as the original source.
This provides a natural way to estimate the background from off-time data.
In this analysis we use a series of ``test source" locations that covers an 8~hr interval around the GRB.
The background estimates obtained using this method for the T90 time window are shown in Table \ref{tab:results}.

The required number of signal events in the T90 time-window for 50\% Poisson probability of a 5$\sigma$ detection, given the number of background events, is displayed in the ninth column of Table \ref{tab:results}. 
The number of counts in each time-window for each GRB does not exceed the required number of events for a 5$\sigma$ detection. 
The measurements for all selected GRBs are thus consistent, within 5$\sigma$, with statistical fluctuations of the background. 
These results are preliminary. 
To ensure blindness for future analyses, the exact number of events seen in the signal time-window is not given.

\begin{figure}[ht]
  \centering
  \includegraphics[width=0.45\textwidth]{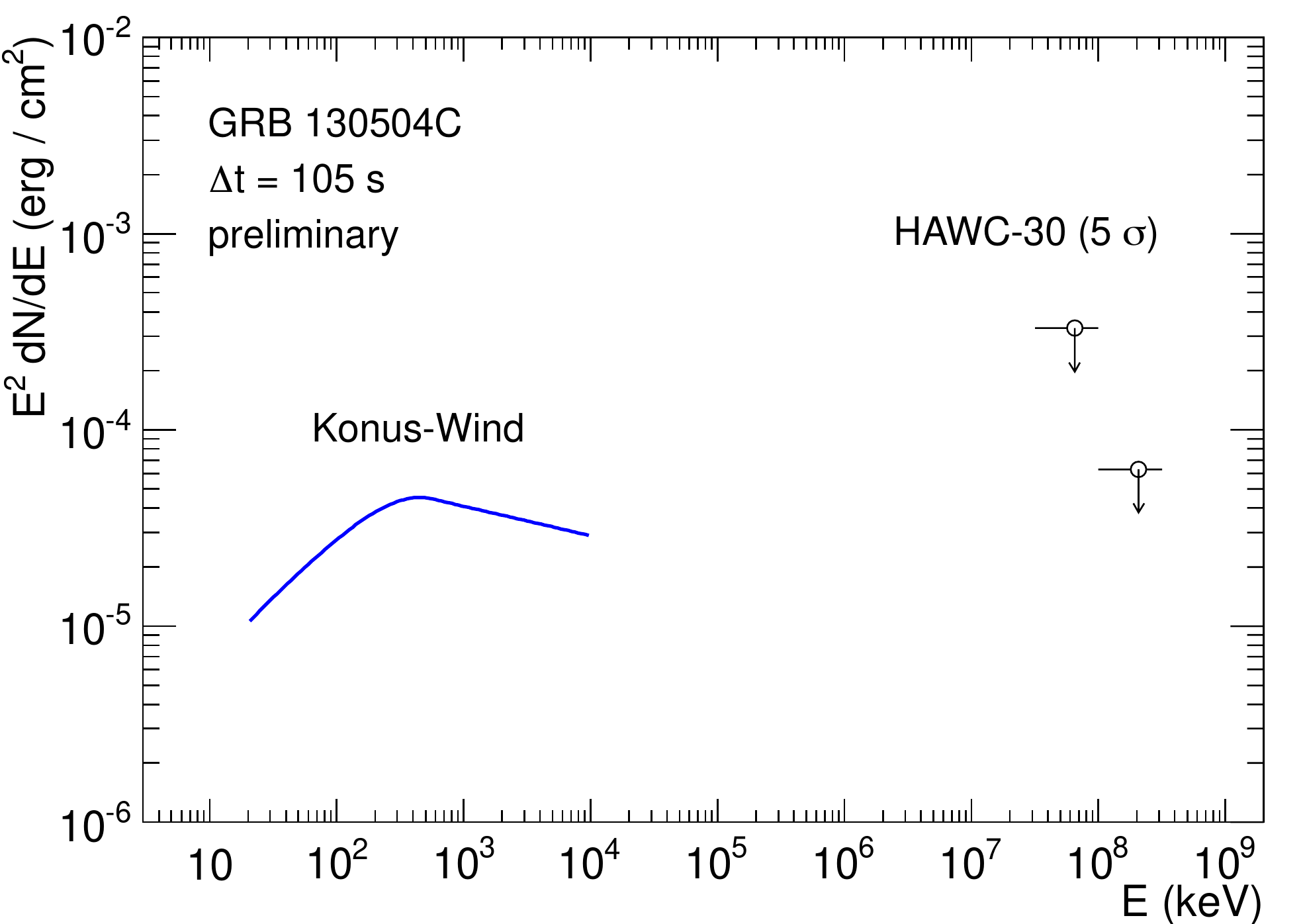}
  \caption{The 5$\sigma$ upper limit on high-energy emission from GRB~130504C imposed by HAWC data. 5$\sigma$ sensitivity is reported rather than 90\% C.L. upper limits to allow the data to remain blind. The spectral fit reported by Konus-Wind (GCN circular 14578) is shown for comparison.}
  \label{fig:grb130504c}
 \end{figure}

For GRB~130504C, a 105~s time window was used in addition to the T90 and 3$\times$T90 windows from the prescription. 
This matches the time of the fluence given by Konus-Wind (GCN circular 14578), which provides a spectral fit for comparison. 
Figure \ref{fig:grb130504c} shows the upper limits set by the 5$\sigma$ discovery requirement of the analysis for GRB~130504C. 
The energy bands are the same as those used for GRB~111016B. 
Analyses that will improve this limit are in progress.


\section*{Future analyses}

In addition to the single time-bin technique described above, two other more complex analyses are being developed for main DAQ data: (1) a model-dependent likelihood analysis and (2) a model-independent scanning time-window analysis. 

The maximum likelihood analysis takes advantage of information from other instruments and theoretical models of GRBs. 
The probability density function (PDF) for the GRB light curve is derived from knowledge of each specific GRB with parameter boundaries appropriately reflecting the high-energy spread seen in the the \emph{Fermi} LAT GRB catalog. 
The background rate is determined with off-time data. 
An extended likelihood function is then constructed from the model of the signal and background PDFs and numerically maximized with respect to the number of signal events using MINUIT\footnote{http://lcgapp.cern.ch/project/cls/work-packages/mathlibs/minuit/doc/doc.html}.

To determine the significance of discovery, we consider the standard likelihood ratio test statistic:
\begin{eqnarray*}
 D = -2 \log \left[\frac{\mathcal{L}(n_{s} = 0)}{\mathcal{L}(\hat{n}_s)}\right]
\end{eqnarray*}
where $\hat{n}_s$ is the best fit value for the number of signal events. 
The maximum likelihood value produced with the model is compared to the null hypothesis, i.e. a model that does not include signal from the GRB. 
This tests how often background mimics signal. 
We use the cumulative distribution of D for simulated background-only light curves to determine the probability that a measured signal under the null hypothesis is simply a statistical fluctuation of the background. 
This analysis is currently underway and will be subject of a future publication.

Unlike the model-dependent likelihood method, the other HAWC technique for GRB detection does not rely on outside sources of information. 
It scans the full sky continuously with a set of time windows that cover the range of known GRB durations looking for upward fluctuations in the expected background rate. 
Trial factors are accounted for using the distribution of probabilities from the full search.

While this technique is intended to work online in the future, it is being run offline on known GRBs at this time. 
Once a GCN notice is released, we take the start time T0 and location of the burst and begin an automated, predefined search. 
Data within the uncertainty of the reported location are tiled for a spatial search with $N$ single- duration time windows shifted from T0 until the GRB is beyond HAWC's FoV. 
The trials-corrected significance of the search result is calculated using simulation.

\section*{Outlook for HAWC GRB sensitivities}

\begin{figure}[ht]
  \centering
  \includegraphics[width=0.45\textwidth]{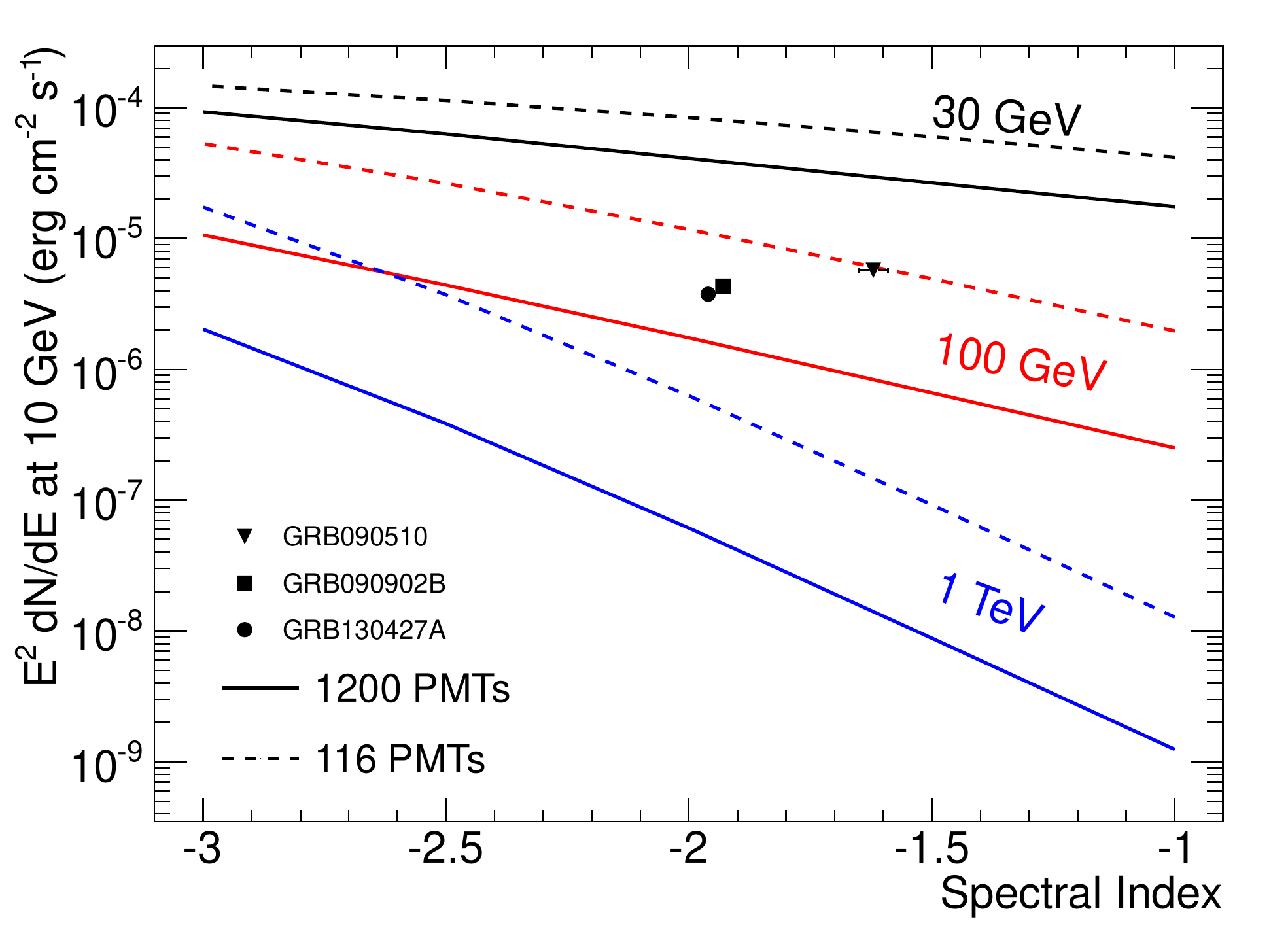}
  \caption{The 5$\sigma$ discovery potential for various values of a sharp high-energy spectral cutoff as a function of spectral index for HAWC with 116 PMTs and 1200 PMTs. The duration of the burst is fixed to 1 s and the zenith angle is fixed to 20$^{\circ}$. Data from 3 different GRBs are corrected for the sensitivity to duration and inserted for comparison \cite{bib:grb090510,bib:grb090902b,bib:zhu}.}
 \label{fig:hawcsensitivity}
 \end{figure}

As the detector array continues to grow, the sensitivity of HAWC increases dramatically. 
Figure \ref{fig:hawcsensitivity} illustrates the effects of different GRB emission spectra on the expected sensitivity of HAWC for several construction phases using the single time-bin analysis. 
We consider a burst at a zenith angle of 20$^\circ$, lasting one second, with a spectrum of the type  dN/dE $\propto E^{-\gamma}$ with a range of indices for 3 different high-energy cutoffs. 
The effect of the EBL is not directly considered because it can be simplistically simulated by the sharp cutoff. 
As an example, for a redshift of $z = 1$, Gilmore et al. \cite{bib:gilmore} predict a cutoff at about 125~GeV. 
Data for GRBs 090510, 090902b, and 130427a, extracted from \cite{bib:grb090510}, \cite{bib:grb090902b}, and \cite{bib:zhu}, are shown for comparison.

HAWC stands an excellent chance of seeing a GRB based on those seen by \emph{Fermi} LAT. 
For example, even if the source spectra or EBL causes the gamma rays to cutoff above 100~GeV, the full HAWC array will still be able to see such bursts as those shown in Figure \ref{fig:hawcsensitivity}.

Physics operations with 440 PMTs is expected to begin in August 2013. 
HAWC-250 (1,000 PMTs) is on schedule for completion in August 2014. 
Even before the array becomes complete at the end of 2014, HAWC is extremely sensitive to high-energy GRBs.

\section*{Acknowledgments}
We acknowledge the support
from: US National Science Foundation (NSF); US Department of Energy Office of
High-Energy Physics; The Laboratory Directed Research and Development (LDRD)
program of Los Alamos National Laboratory; Consejo Nacional de Ciencia y
Tecnolog\'{\i}a (CONACyT), M\'exico; Red de F\'{\i}sica de Altas Energ\'{\i}as,
M\'exico; DGAPA-UNAM, M\'exico; and the University of Wisconsin Alumni Research
Foundation.

\clearpage


\newpage
\setcounter{section}{4}
\nosection{Sensitivity of the HAWC Observatory to Gamma-ray Bursts using the
Scaler System\\
{\footnotesize\sc Dirk Lennarz}}
\setcounter{section}{0}
\setcounter{figure}{0}
\setcounter{table}{0}
\setcounter{equation}{0}
%
%

\title{Sensitivity of the HAWC Observatory to Gamma-ray Bursts Using the Scaler System}

\shorttitle{Sensitivity of HAWC to GRBs Using the Scaler System}

\authors{
Dirk Lennarz$^{1}$
for the HAWC collaboration$^{2}$
}

\afiliations{
$^1$ School of Physics and Center for Relativistic Astrophysics, Georgia Institute of Technology, Atlanta, Georgia, USA \\
$^2$ For a complete author list, see the special section of these proceedings
}

\email{dirk.lennarz@gatech.edu}

\abstract{Gamma-ray bursts (GRBs) are among the most energetic phenomena in the known universe and are predicted to emit very-high-energy (VHE, $>100$~GeV) gamma-ray radiation. The High Altitude Water Cherenkov (HAWC) observatory is a ground based VHE gamma-ray detector currently under construction at Sierra Negra in Mexico at an altitude of 4100 m above sea level. It has two data acquisition (DAQ) systems - one designed to readout full air-shower events (main DAQ) and the other one counting the signals in each photomultiplier tube (scaler DAQ). In this contribution the sensitivity of the scaler DAQ is reviewed, which detects GRBs by a statistical excess over the noise rate. Results of the scalers analysis on selected GRBs are shown.}


\maketitle

\section*{Very-High-Energy Emission from GRBs}
The phenomenon of gamma-ray bursts (GRBs) has been known for almost fifty years now \cite{bib:review_Gehrels}. GRB emission in the keV to MeV energy range is generally well described by Band functions \cite{bib:band_function}. Recently, the Large Area Telescope (LAT) on board the \emph{Fermi Gamma-Ray Space Telescope} (\emph{Fermi}-LAT) has shown that at higher energies (above $\sim 20$~MeV) the brightest bursts inside the LAT field of view require additional spectral components. Most intriguing, and a challenge for GRB modelling, are bursts that exhibit an additional hard power-law (e.g. GRB 090902B and 090510 \cite{bib:Fermi_LAT_GRB090902B,bib:Fermi_LAT_GRB090510}), which may also have a spectral break (GRB~090926A \cite{bib:Fermi_LAT_GRB090926A}). Furthermore, it appears that the LAT emission starts systematically later than the emission at lower energies, e.g. reaching delays of up to 40~s for GRB~090626, and the duration is also longer, reaching up to almost 700~s for GRB~090328 \cite{bib:Fermi_LAT_GRB_catalogue}.

\emph{Fermi}-LAT has observed GRB emission up to 94 GeV (GRB~130427A, see below). Extending the observations beyond that energy is challenging for the LAT because the effective area is approximately constant ($\sim0.6$~$\rm m^2$ at 10~GeV) at these energies and the GRB spectra fall quickly with energy. Such observations would however be desirable because they can help to understand the emission mechanism at work, e.g. by placing a lower limit on the Lorentz boost factor in the GRB jet. Furthermore, they can provide an insight into the extragalactic background light (EBL) and Lorentz invariance violation.

\section*{High Altitude Water Cherenkov Observatory}
HAWC is a very-high-energy (VHE, above $\sim$ 100~GeV) gamma-ray air-shower detector currently under construction at Sierra Negra in Mexico at an altitude of 4100 m above sea level \cite{bib:HAWC} and improves the water Cherenkov technique pioneered by Milagro. VHE photons are detected by measuring Cherenkov light from secondary particles in an extensive air shower. When completed in 2014, HAWC will consist of 300 steel tanks of 7.3~m diameter and 4.5~m depth filled with purified water. The bottom of each tank is outfitted with three $8^{\prime\prime}$ photomultiplier tubes (PMTs) and one $10^{\prime\prime}$ PMT. HAWC will have a substantially larger effective area than Milagro at energies around 100~GeV, primarily because the array is more than three interaction lengths closer to the shower maximum. Furthermore, the optically isolated tanks will improve the hadron rejection efficiency and thereby boost the sensitivity. Compared to Imaging Atmospheric Cherenkov Telescopes, the other type of ground-based instruments sensitive in the VHE regime, HAWC has the advantage of a very large instantaneous field of view ($\sim$2~sr or 16\% of the sky), a duty cycle close to 100\% and no observational delay e.g. due to slewing. It is therefore an ideal detector for studying transient sources like GRBs.

Both HAWC DAQs, the main and the scaler DAQ, have sensitivity to GRBs \cite{bib:HAWC_GRB_sensitivity}. They have different energy sensitivities and therefore complement each other. The main DAQ records the time and charge of individual PMT pulses and the signal arrival time in different tanks. This makes the reconstruction of the incident direction of the shower possible. Its sensitivity to GRBs is reviewed elsewhere \cite{bib:HAWC_GRB_ICRC}.

\section*{Sensitivity of the Scaler System to GRBs}
In the ``single particle technique'' \cite{bib:scaler_method}, a transient flux of gamma rays results in a detector wide increase of the PMT count rates and can be identified on a statistical basis by searching for an excess of the summed count rates over an expected background rate. The PMT count rate is typically dominated by cosmic-ray air showers, natural radioactivity in the tanks and thermal noise of the PMTs and is normally $\approx25$~kHz for the $8^{\prime\prime}$ PMTs and $\approx50$~kHz for the $10^{\prime\prime}$ PMT. Unlike the main DAQ, the scaler system does not measure the energy and arrival direction of the primary gamma rays. However, low-energy gamma rays that cannot be reconstructed by the main DAQ are still observable with the scaler system, thus providing a lower energy threshold which is important for GRB observations due to EBL absorption. In the operation of HAWC, Struck SiS-3820 VME scalers are used and read out in 10~ms time windows. This fine binning will allow the scaler system to produce detailed light curves of the GRB emission at VHE.

 \begin{figure}[t]
  \centering
  \includegraphics[width=0.45\textwidth]{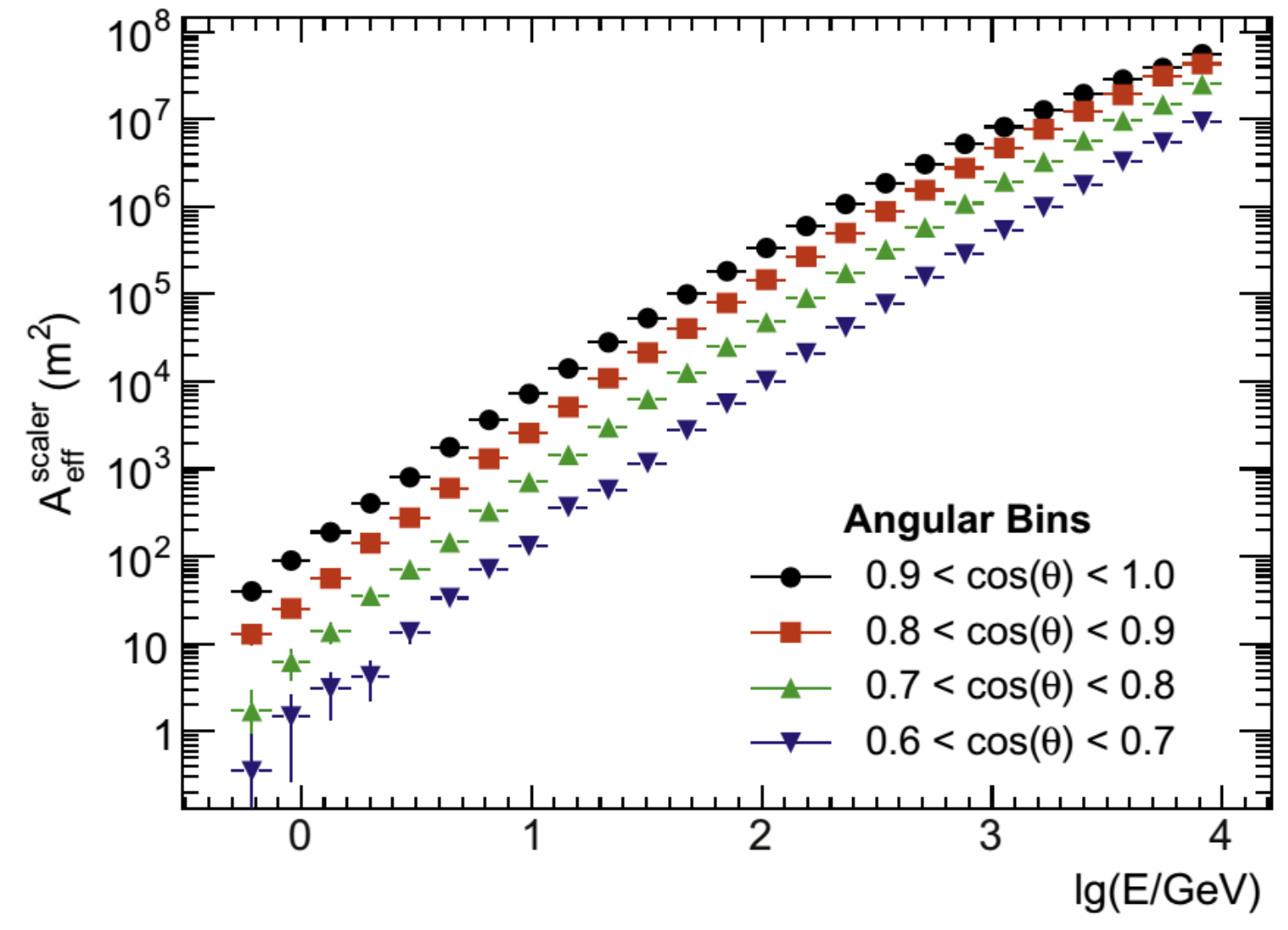}
  \caption{Effective area of the scaler system for the full HAWC detector \cite{bib:HAWC_GRB_sensitivity}. Four different zenith angle ranges are shown.}
  \label{fig:scaler_effective_area}
 \end{figure}

The effective area of the scaler system is derived by means of simulations. Gamma rays are simulated using CORSIKA \cite{bib:CORSIKA} and the detector response is simulated with software using GEANT4 \cite{bib:Geant4}. Fig.~\ref{fig:scaler_effective_area} shows the effective area of the scaler system, $A_{\rm eff}^{\rm scaler}$, for the full HAWC detector as a function of energy for various zenith angle bands. Each shower can create hits in multiple PMTs and therefore the value for the effective area is not restricted to the physical size of the detector. The simulations include only the three $8^{\prime\prime}$ PMTs and adding the $10^{\prime\prime}$ PMTs should result in further improvement.

In order to estimate the sensitivity of the scaler system, an assumption on the background has to be made. The distribution of the total noise rate is not expected to be Poissonian because some sources of noise are correlated. Correlated noise originates from air showers resulting in multiple PMT hits, PMT afterpulsing and Michel electrons created by muons stopping inside the tank. This leads to a background distribution that is wider than a Poissonian and is characterised by the so-called Fano factor $F$. Given a signal rate $S$ and background rate $B$ the significance $\sigma$ of a given observation at zenith angle $\theta$ is:
\begin{equation}\label{eq:significance}
  \sigma = \frac{S\Delta T}{\sqrt{FB\Delta T}} = \sqrt{\frac{\Delta T}{FN_{\rm PMT}R_{\rm PMT}}}\int_{E_{\rm min}}^{E_{\rm max}} dE \frac{dN}{dE} A_{\rm eff}^{\rm scaler}(E,\theta)
\end{equation}
where $N_{\rm PMT}$ is the number of PMTs in the detector, $R_{\rm PMT}$ is the average noise rate and $\Delta T$ the observation window.

Currently, studies are being done to determine the width of the noise distribution and hence the Fano factor from the first experimental data. The sensitivity presented here uses a Fano factor that goes back to a dedicated background simulation. Primary cosmic-rays are simulated with CORSIKA assuming an $E^{-2.7}$ spectrum and normalised to the ATIC measurements at $\approx 100$~GeV \cite{bib:ATIC}. Uncorrelated Gaussian noise (e.g. due to radioactive decay) and PMT afterpulsing is included in the simulation. More information on the simulation can be found elsewhere \cite{bib:HAWC_GRB_sensitivity}. Using the simulation a Fano factor of 17.4 is derived, which reduces the sensitivity of the HAWC scaler system by a factor of 4.2 with respect to purely Poissonian noise.

 \begin{figure}[t]
  \centering
  \includegraphics[width=0.45\textwidth]{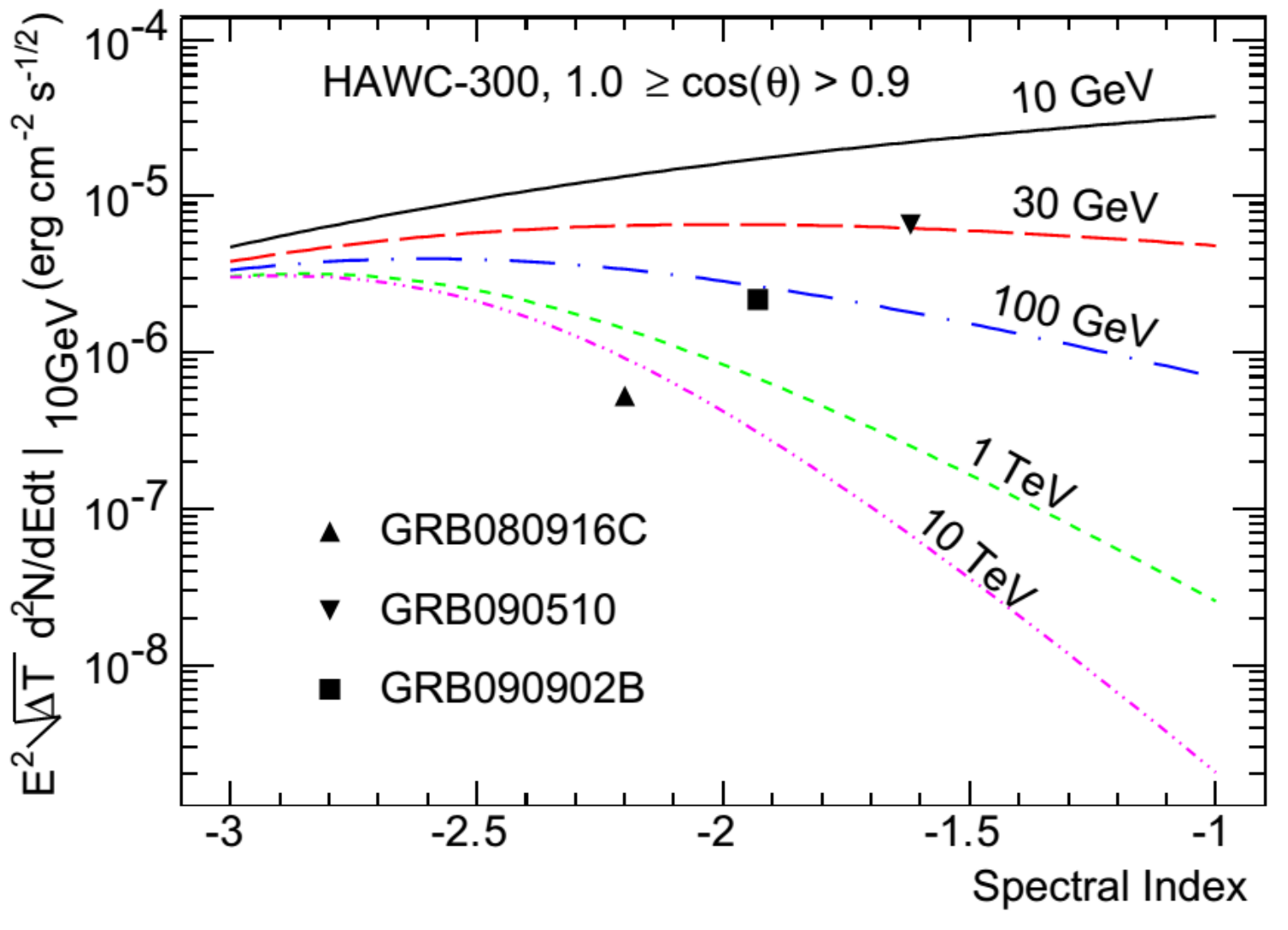}
  \caption{Necessary flux at 10~GeV multiplied by the square root of the GRB duration for a $5~\sigma$ detection \cite{bib:HAWC_GRB_sensitivity}. Shown are various indices of the power-law spectra and sharp high-energy cut-offs. Data from three known GRBs have been inserted for comparison.}
  \label{fig:scaler_sensitivity}
 \end{figure}

The gamma-ray spectra have been re-weighted for power-law spectra of various spectral indices and high-energy cut-offs. Figure~\ref{fig:scaler_sensitivity} shows the flux required for a $5\sigma$ detection. The scaler analysis is always dominated by background, hence the sensitivity is proportional to $1 / \sqrt{\Delta T}$. This means the shorter the burst for a given fluence the better the sensitivity. It can be seen that GRBs like 090510 and 090902B would be significantly detected by the scaler system if they occurred at favourable zenith angles.

A future improvement of the scaler DAQ would be an active veto of air showers. Such a veto could be issued by the main DAQ in case a large amount of light is present in the detector. It would remove some of the correlated noise caused by air showers and thereby decrease the Fano factor. This would improve the sensitivity of the scaler DAQ.

\section*{Analysis of Scaler Data}
The first step is the selection of search windows (regions of interest) to search for an excess in the summed PMT count rate. These windows are chosen using information from observations at lower energies and are different for each analysed GRB. Search strategies include the time of the prompt emission, the peak flux and the time of the highest energy photon. The scaler data is then rebinned to the window size defined by the search windows.

An estimate of the number of background events in the search window is obtained by using a moving average (MA). A symmetric MA is applied to each channel, which means each point $i$ is replaced by the average of the $N$ points before and after (each having $C$ number of counts):
\begin{equation}\label{eq:moving average}
  {\rm MA}(i) = \frac{1}{2N} \sum^{j=i+N}_{j=i-N;j \neq i} C(j)
\end{equation}
The choice of $N$ determines how much data is used in the background estimation. Depending on the binning, it is typically chosen so that the background is derived from $\sim\pm1000$~s of data.

\begin{figure}[t]
 \centering
 \begin{overpic}[width=0.45\textwidth]{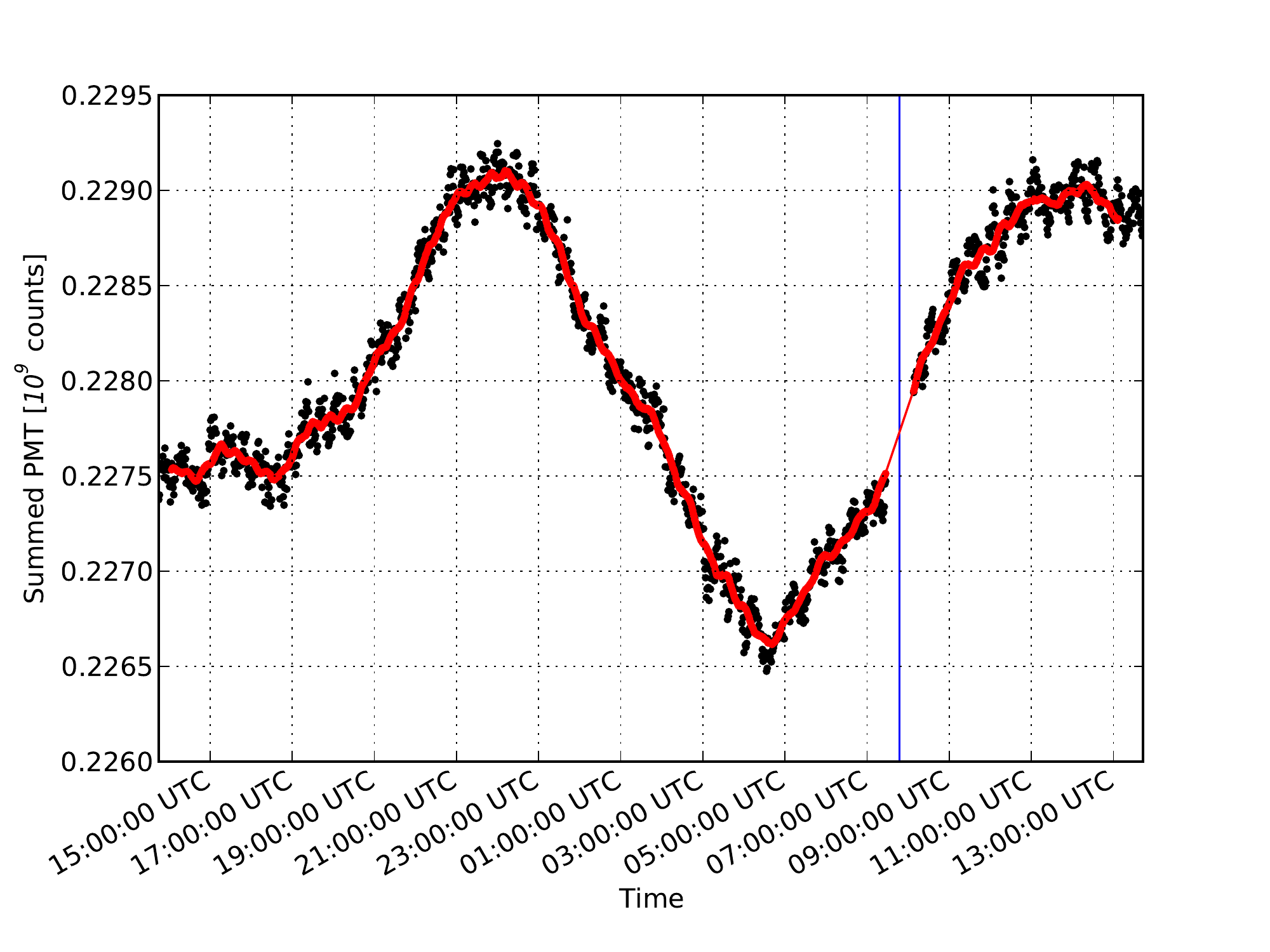}
 \put(15,17){\LARGE Preliminary}
 \end{overpic}
 \caption{The black dots show the summed count rates of all selected PMT channels for GRB~130427A using a 60~s binning. The data is well described by the moving average (red line). A time window between -1200~s and +1200~s around the GRB trigger time (blue line) is kept blind.}
 \label{fig:GRB130427A_data}
\end{figure}

For each PMT the raw data is plotted as a function of time and inspected by eye in order to identify unstable or unreliable channels. Those channels are not included in the summed count rate. Work is currently underway to quantify and select stable and reliable channels automatically.

Figure~\ref{fig:GRB130427A_data} shows an example of the summed count rates of all selected PMT channels for a 60~s binning (black points). The count rates of the PMTs are influenced on the percent level by changes in the atmospheric pressure and temperature due to differences in the shower development and the detector temperature. It can be seen that the count rate has a clear 12~hour modulation. Due to this larger timescale of the variations compared to the period of background estimation it is not necessary to try and correct the data for these environmental and instrumental effects.

Apparently, the data is well described by the MA (red line in Fig.~\ref{fig:GRB130427A_data}). A region between -1200~s and +1200~s around the GRB trigger time (blue line) is kept blind and not plotted to allow for future analysis. As long as, even with the largest search window considered, there is a gap to the start of the earliest search window, binning effects, where data from the blinded region is included in the last re-binned window, are not relevant. The MA is not evaluated inside the blinded region because it would contain information from the search windows. The background estimate for each search window is obtained by averaging the last and first MA outside the blinded region. This value is then compared to the number of counts inside the search window and the number of excess events is calculated.

\begin{figure}[t]
 \centering
 \begin{overpic}[width=0.43\textwidth]{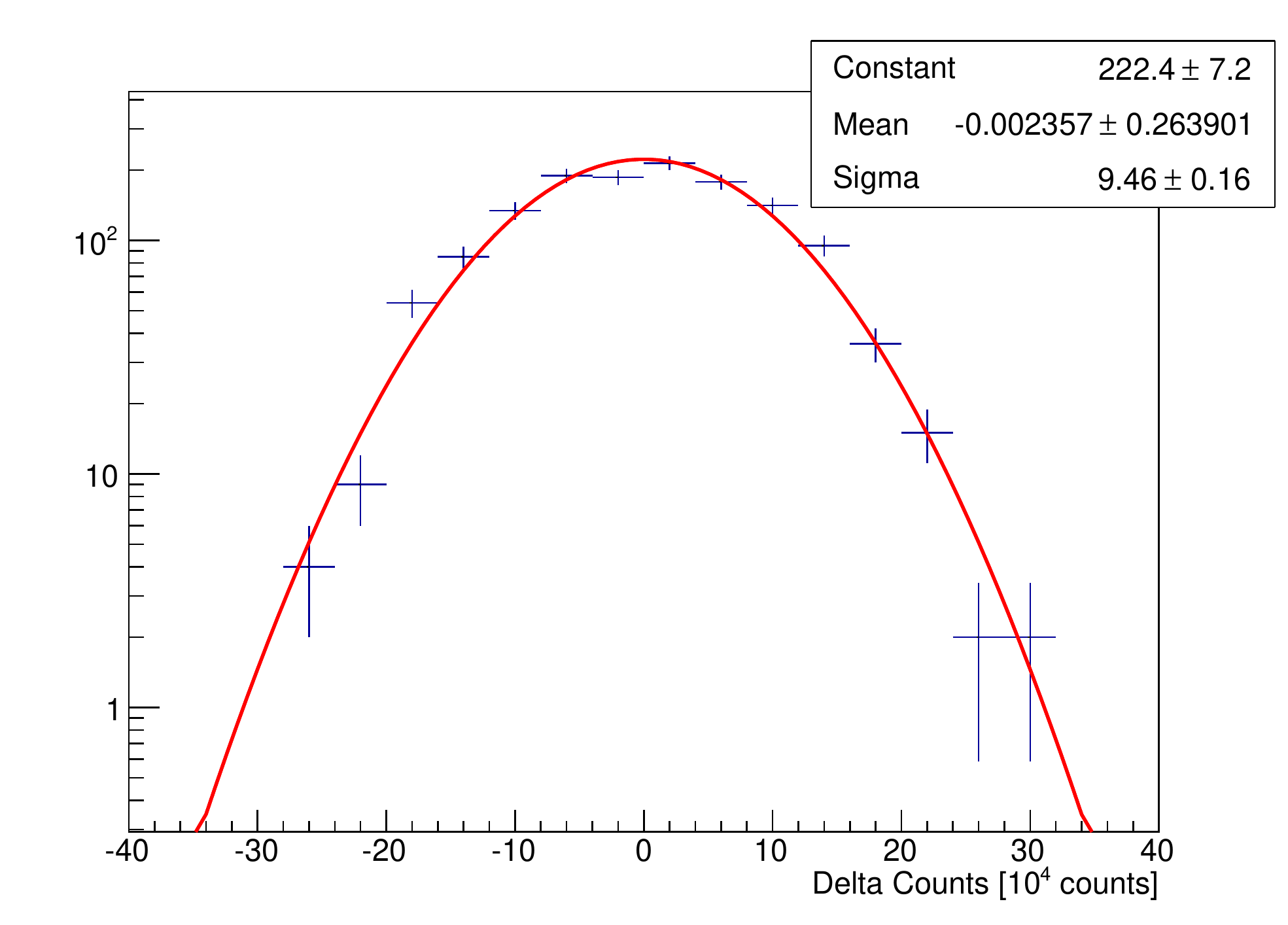}
 \put(37,12){\LARGE Preliminary}
 \end{overpic}
 \caption{A histogram of the residuals between data and the moving average in the 60~s binning for GRB~130427A. There are no significant outliers compared to a Gaussian.}
 \label{fig:GRB130427A_residuals}
\end{figure}

Figure~\ref{fig:GRB130427A_residuals} shows a histogram of the residuals between the data and the MA. The distribution shows no significant outliers when compared to a Gaussian. Using this distribution the p-value of the excess in the search window, which is the probability that the observed excess is caused only by background, can be calculated.

\section*{Analysis of Selected GRBs}
In this contribution the analyses of two GRBs also detected by \emph{Fermi}-LAT, GRB~130427A and GRB~130504C, are presented. At the trigger times, the scaler DAQ comprised 29 operational Water Cherenkov Detectors out of the 300 planned (this stage of the detector being called HAWC-30), thus collecting data from 116 PMTs. Of these channels, two channels were excluded because their rate is either zero or very low and another two PMTs were removed because their rates show an anomalous excess of high count rates.

\subsection*{GRB~130427A}
GRB~130427A was an extremely bright burst, setting the record for the most energetic photon (94 GeV) ever detected from a GRB \cite{bib:LAT_refined}. It is among the handful of bursts that have been detected by \emph{Fermi}-GBM \cite{bib:GBM_detection}, \emph{Fermi}-LAT \cite{bib:LAT_detection} and \emph{Swift} \cite{bib:BAT_detection}. Furthermore, the prompt phase has also been detected by MAXI/GSC \cite{bib:MAXI_detection}, SPI-ACS/INTEGRAL \cite{bib:INTEGRAL_detection}, Konus-Wind \cite{bib:Konus_Wind_detection}, AGILE \cite{bib:AGILE_detection} and RHESSI \cite{bib:RHESSI_detection}. It is the most intense and fluent GRB detected by \emph{Fermi}-GBM so far. The redshift has been ascertained to be $z=0.3399\pm0.0002$ \cite{bib:redshift_gemini,bib:redshift_NOT,bib:redshift_VLT}, which identifies GRB~130427A as one of the rare powerful and nearby bursts.

The \emph{Fermi}-GBM triggered at 07:47:06.42~UT (in the following denoted $T_0$). HAWC data was only collected by the scaler DAQ, as the main DAQ was not operational at that time. The elevation of the burst in HAWC's field of view was only 33 degrees and setting. This results in a sensitivity for HAWC that is more than 2 orders of magnitude poorer than near the zenith. Furthermore, while near zenith the nominal threshold of the scaler system is a few GeV, it is much higher towards the horizon.

\begin{table*}[t!]
\begin{center}
\setlength{\tabcolsep}{4pt}
\begin{tabular}{l|cccccc|cccc}
                 & \multicolumn{6}{|c|}{GRB~130427A}                        & \multicolumn{4}{|c}{GRB~130504C}     \\ \hline
Search Window    & 1      & 2       & 3       &       4 &        5 &      6 &       1 & 2        & 3      & 4      \\
PMT Sum $[10^4]$ & 7594.6 & 22765.2 & 56898.3 & 68307.3 & 113827.1 & 7590.6 & 58062.9 & 116120.5 & 7743.9 & 7738.2 \\
BG Est. $[10^4]$ & 7590.7 & 22773.0 & 56932.1 & 68323.9 & 113879.0 & 7590.7 & 58082.6 & 116149.1 & 7745.1 & 7745.1 \\
Excess $[10^4]$  & +3.9   & -7.8    & -33.8   & -16.6   & -51.9    & -0.1   & -19.7   & -28.6    & -1.2   & -6.9   \\
p-value          & 17\%   & 78\%    & 95\%    & 71\%    & 90\%     & 50\%   & 81\%    & 71\%     & 61\%   & 96\%   \\
\end{tabular}
\caption{For each search window the sum of all PMT counts in the search window is given together with the expectation from the moving average. From these two values the excess is calculated. The p-value is the probability that the observed excess is caused only by the background and is directly derived from the excess distributions like the one shown in Fig.~\ref{fig:GRB130427A_residuals}.}
\label{table:results}
\end{center}
\end{table*}

\emph{Swift}-BAT triggered later on the burst, because it was slewing to a pre-planned target, during which triggering is de-activated. The BAT light curve indicates a first peak between $T_0$ and $T_0+1$ sec and a main, large peak, starting at $T_0+1$, peaking at $T_0+9$ sec and ending at $T_0+41$. At $T_0+51$ a second feature starts, peaking at $T_0+141$ sec and returning to baseline at $\sim T_0+2000$ sec. The BAT $T_{90}$ is $162.83 \pm 1.36$~s (15-350~keV) and the GBM gives a $T_{90}$ of about 138~s (50-300~keV). The LAT emission seems to be correlated with the GBM emission.

Six search windows have been selected based on this light curve:
\begin{enumerate}[noitemsep,nolistsep]
\item  0~s  to  20~s, which covers the bright, structured peak seen by GBM,
\item -5~s  to  55~s, covering the main peak as seen by BAT,
\item -5~s  to 145~s, which is slightly larger than the $T_{90}$ reported by GBM,
\item 120~s to 300~s, covering the main emission around the second peak,
\item -10~s to 290~s, which combines the two peaks,
\item and -10~s to 10~s around the time of the highest energy LAT photon.
\end{enumerate}

For the MA, the choice of $N$ is $N=4$ for the 300~s binning, which means the average is calculated from $\pm1200$~s of data, and $N=6$ for the 180~s binning ($\pm1080$~s of data). Multiples of this are used for the other binnings ($N=8$ for the 150~s binning, $N=18$ for the 60~s binning and $N=54$ for the 20~s binning).

\subsection*{GRB~130504C}
This GRB was detected by \emph{Fermi}-LAT \cite{bib:GRB130504C_LAT_detection}, \emph{Fermi}-GBM \cite{bib:GRB130504C_GBM_detection}, Konus-\emph{Wind} \cite{bib:GRB130504C_Wind_detection} and \emph{Suzaku} \cite{bib:GRB130504C_Suzaku_detection}. The earliest trigger time is $T_0=$23:28:57.518~UT from the GBM. No redshift information is available for this GRB.

The burst was bright enough in the GBM to trigger an autonomous repoint of \emph{Fermi}. The LAT observed it for 2200~s after the trigger and detected long lasting emission with $>70$ photons above 100~MeV out to 1000~s. Multi-peaked emission lasting roughly 40~s is seen at lower energies. The highest energy LAT photon has an energy of $\sim5$~GeV and arrived 251~s after the trigger.

The GBM light curve consists of about 5 peaks associated with the GRB and the $T_{90}$ of the GRB is about 74~s. Konus-\emph{Wind} reports that the light curve shows multiple partly overlapped peaks from $\sim T_0-8$~s to $\sim T_0+112$~s. \emph{Suzaku} saw a multi-peaked structure starting at $T_0-6$~s, ending at $T_0+113$~s with a duration ($T_{90}$) of about 67~s.

Based on this information, the following four search windows have been selected:
\begin{enumerate}[noitemsep,nolistsep]
\item -20~s to 130~s, which covers the whole prompt emission,
\item -10~s to 290~s, which is an extended window,
\item  20~s to  40~s, which covers the peak flux of GBM, Konus-\emph{Wind} and \emph{Suzaku},
\item and -10~s to 10~s around the time of the highest energy LAT photon.
\end{enumerate}

For the MA, $N$ is chosen as for the analysis of GRB~130427A.

\subsection*{Results}
Table~\ref{table:results} shows the results of the analysis of the scaler data for each of the search windows specified earlier. All observations are consistent with the assumption of background only. The only positive excess, in the first time window of GRB~130427A, still has a 17\% chance to be caused by background only. HAWC has reported the results of GRB~130427A in a GCN notice \cite{bib:HAWC_GCN}.

\section*{Conclusions}
This proceeding has presented the scaler DAQ of HAWC as a sensitive instrument and GRBs like 090510 and 090902B would be significantly detected if they occurred at favourable zenith angles. The analysis procedure for scaler data was shown together with first results of selected GRBs. No indication of emission $>10$~GeV is observed. The implications of these non-detections with respect to the VHE fluence of the GRBs will be reported elsewhere.

\section*{Acknowledgments}

We acknowledge the support from: US National Science Foundation (NSF); US
Department of Energy Office of High-Energy Physics; The Laboratory Directed
Research and Development (LDRD) program of Los Alamos National Laboratory;
Consejo Nacional de Ciencia y Tecnolog\'{\i}a (CONACyT), M\'exico; Red de
F\'{\i}sica de Altas Energ\'{\i}as, M\'exico; DGAPA-UNAM, M\'exico; and the
University of Wisconsin Alumni Research Foundation.

\clearpage


\newpage
\setcounter{section}{5}
\nosection{A Real-time AGN Flare Monitor for the HAWC Observatory\\
{\footnotesize\sc Asif Imran, Robert Lauer}}
\setcounter{section}{0}
\setcounter{figure}{0}
\setcounter{table}{0}
\setcounter{equation}{0}
%

\newcommand{\etal}{{et al.}}
\newcommand{\adeg}{\ensuremath{^\circ}}
\newcommand{\asim}{\ensuremath{\sim}}

\title{A Real-time AGN Flare Monitor for the HAWC Observatory}

\shorttitle{AGN Flare Monitor for HAWC}

\authors{
A. Imran$^{1}$,
R. Lauer$^{2}$
for the HAWC Collaboration $^{3}$
}

\afiliations{
$^1$ Department of Physics, University of Wisconsin-Madison, Madison, WI, USA\\
$^2$ Department of Physics, University of New Mexico, Albuquerque, NM, USA\\
$^3$ See M. Mostafa {\it in these proceedings}\\
}

\email{asif.imran@wisc.edu}

\abstract{The High Altitude Water Cherenkov (HAWC) Observatory will commence
  scientific operation with the first 100 of 300 water Cherenkov detectors 
  beginning in August of 2013. The large field of view instrument will
  provide a nearly continuous, unbiased survey of TeV emission from the northern
  hemisphere, making it ideally suited for detecting bright transient events
  such as outbursts from active galactic nuclei (AGN). Observations in the GeV
  and TeV energy regime can be used to probe the structure and emission
  mechanisms in AGN, and to further our understanding of how supermassive black
  holes accrete matter.  Furthermore, gamma-ray observations of AGN are key to
  providing cosmological constraints to the diffuse extragalactic background
  radiation.  The HAWC Collaboration is implementing a comprehensive, online
  flare-monitoring program to promptly detect bright outbursts from
  moderate-redshift AGN. This will provide unique opportunities to trigger
  follow-up observations at complimentary wavelengths (e.g., by Fermi or
  VERITAS). The high duty cycle ($>$90\%) of the HAWC Observatory combined with
  its sensitivity to photon energies up to 100 TeV will allow us to extend the
photon energy spectra of known AGN to the multi-TeV regime. Here, we discuss the
various components of the flare-monitoring system, and present preliminary
results from searches for AGN flares.}


\maketitle

\section*{Introduction} 

AGN are the most prevalent sources of very high energy (VHE; $>$ 100 GeV)
radiation in the universe. The second year Fermi-LAT catalog (hereafter 2FGL)
has revealed over 1000 sources associated with AGN \cite{bib:abdo} . The
predominant non-thermal broadband continuum spectra of AGN contain two distinct,
well separated humps in the $\nu\rm{F}_{\nu}$ representation \cite{bib:wagner}.
The low energy component peaking at radio and X-ray frequencies are generally
attributed to synchrotron emission from highly relativistic electrons in the
jets. Competing models exist to account for the high energy emission at GeV/TeV
energies, but it is widely accepted that the same population of electrons
inverse Compton-scatter soft photons to gamma-ray energies. In this leptonic
scenario, the seed photons are produced either in synchrotron radiation
\cite{bib:maraschi} or possibly in an external source such as CMB photons or
UV/X-ray emission from the accretion disk \cite{bib:dermer}.  In contrast,
hadronic scenarios predict VHE emission as a result of proton-initiated cascades
within the jet or due to synchrotron emission from very energetic protons/ muons
\cite{bib:tavecchio12}. Detailed measurements of the VHE component of the SED are
critical to differentiating competing models of physical processes in the AGN.
The unbroken energy coverage (100 MeV to TeV energies) made possible by Fermi
and VERITAS are already forcing modelers to look beyond the standard one-zone
leptonic models for VHE emission in AGN \cite{bib:aliu}.  Another key feature of
the high-energy emission from AGN, in particular the blazar subclass, is
variability on all timescales from minutes to months. TeV blazars have also been
observed to vary in flux by more than an order of magnitude \cite{bib:punch}.
The origin of flares are not yet fully understood, but observational constraints
from rapid, large scale flux variations can be used to test those models
sensitive to changing source environments or injection parameters (see
e.g.,\cite{bib:tavecchio12}, \cite{bib:cerutti}).  Furthermore, VHE flares have
been recorded from AGN without any visible counterparts at longer wavelengths
(so-called orphan flares) \cite{bib:kraw}. Sparse sampling due to the typical
low-duty cycle of atmospheric Cherenkov telescopes makes it difficult to rule
out lagged counterparts at different frequencies. On the contrary, the large
field of view (FOV) and high duty cycle of HAWC will provide observations of the
Northern sky at very high energies on a daily basis, thus allowing us to
accurately understand and model the emission mechanisms during orphan flares or
any flaring states.

Bright flares also allow us to access distant sources. Not surprisingly, about
40\% of AGN in the 2FGL are found to be variable \cite{bib:ackermann}.
Observations of these distant sources can provide important cosmological
constraints to the primordial radiation field such as the extragalactic
background light (EBL) \cite{bib:ackermann}. Powerful, rapid flares from AGN at
cosmological distances can also be used as probes of exotic physics. For
example, gamma-ray observations from distant sources can be used to test for
Lorentz invariance violation; the effects of quantum gravity may be enhanced by
the large light travel time, resulting in a detectable energy-dependent
dispersion of photon arrival times \cite{bib:abramo13}. In addition, axion-like
particles could potentially increase the expected photon flux from AGN in the
VHE regime. Intense AGN flares coupled with the unique sensitivity of HAWC at
TeV energies will allow us to test new exotic particle models
\cite{bib:abramo12}.  Finally, recent studies have shown that intergalactic
ultra-high energy cosmic ray (UHECR) induced cascade emission (via photohadronic
interactions) can explain the hard but slowly varying TeV gamma-ray spectra
observed from several high-redshift AGN \cite{bib:tavecchio12},
\cite{bib:murase}. Consequently, prominent flares will allow us to test AGN as
possibles sites for the acceleration of UHECR.

Here we report on the planned deployment a real-time, online flare monitoring
system for the HAWC observatory. Preliminary performance of the
monitoring system is evaluated with the HAWC data. Finally, we discuss the
prospect for HAWC to detect powerful AGN outbursts over varying flare duration
and intensity.

\section*{The HAWC Observatory} 
\begin{figure}[t] \centering
   \includegraphics[width=0.5\textwidth]{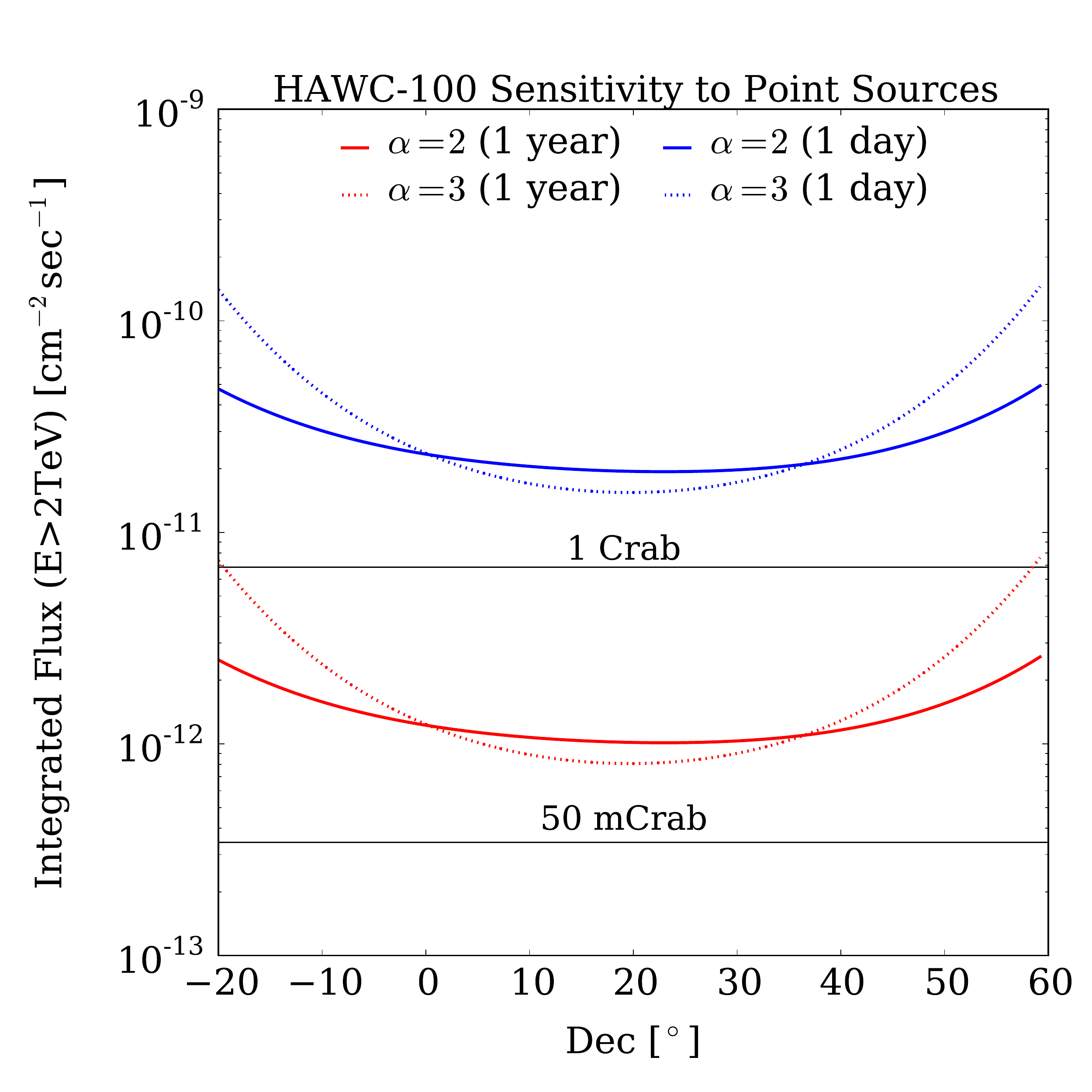} \caption{HAWC-100
     sensitivity to a point-like source in a day (year) as a function of
     declination. We consider a E$^{-\alpha}$ spectrum for $\alpha$ = 2 and 3,
   respectively.}
   \label{hawc100_sensi}
 \end{figure}

HAWC is a next-generation ground-based air shower
array located at a high altitude site in Mexico ($19.0\adeg$N $97.3\adeg$W, 4096
m a.s.l.). It will begin scientific operation in August 2013 once the first 100
water Cherenkov detectors (WCDs) are operational (HAWC-100).  The array will be
completed in August 2014 with 300 WCDs (HAWC-300).  The full array will be
sensitive to photon energies from 50 GeV to 100 TeV with angular resolution of
$0.3\adeg$ at E $>$ \asim1 TeV \cite{bib:abey}.  HAWC-300 will have a
sensitivity of about 5$\sigma$ day$^{-1/2}$ to a Crab-like point source. It will
take \asim7 times longer (i.e. 5$\sigma$ week$^{-1/2}$ ) with HAWC-100 to reach
the same level of sensitivity (Figure~\ref{hawc100_sensi}).  Nevertheless, an
intense outburst such as a 5x Crab flux of any TeV AGN in the detector FOV will
be detected within 6 hours (a full-transit) with HAWC-100. For more details on
the HAWC instrument and analysis techniques, refer to \cite{bib:mostafa}.

Once operational, HAWC will provide a comprehensive survey of the sky at very
high energies, allowing it to carry out non-targeted searches for gamma-ray
sources. Furthermore, HAWC's unique all-sky monitoring capabilities open up
opportunities for real-time detection of transient phenomena at TeV energies,
including gamma-ray bursts and AGN flares. In the event of a possible detection
by HAWC, timely alerts are crucial to increasing the exposure of the source
using follow-up multiwavelength observations by other instruments.

\section*{Analysis} The standard online HAWC analysis chain generates a stream of
reconstructed events from the raw data. At the beginning, these events are
calibrated in multiple stages to remove timing-offsets between photomultiplier
tubes and to apply the charge calibration. This ensures uniform response across
the entire array by minimizing any possible bias across the detector
\cite{bib:lauer}. A quality cut is applied to this dataset to remove poorly
reconstructed events and to ensure enough PMTs participated in the angular
reconstruction fit to reliably reconstruct the direction of the primary
particle. Finally, a set of gamma-hadron separation cuts are applied to the
remaining events to maximize the detection significance for sources with typical
energy spectra. These cuts, derived from monte carlo simulations, are optimized
to remove approximately 99\% of the cosmic-ray background events while
preserving 40-50\% of the simulated gamma-ray events. 

\subsection*{Background Estimation} In order to determine the excess from a
particular region in the sky, we need to estimate the background level from the
remaining events. It is necessary to reject non gamma-ray events whose
reconstructed directions lie close to the source direction. The background
estimation for this analysis uses a modified version of the direct integration
method \cite{bib:atkins} to generate maps of background events in the sky. This
technique assumes that the efficiency for detecting the overwhelming majority of
cosmic-ray background events is a function of the local angles and independent
of the event rate. However, the detector efficiency is affected by daily
variations in the atmospheric conditions along with any changes to the detector
configuration. Consequently, our analysis calculates the background for every
two-hour interval to ensure that the efficiency remains relatively constant over
the duration of the background estimation. For this work, we adopted the HEALPix
library \cite{bib:gorski} to bin the sky into equal-area pixels with an angular
resolution of \asim0.1\adeg.

For a given declination band of width $\rm{d}\delta$, the number of expected
background events from any region in the sky is determined by a convolution of
efficiency at local coordinate bins with the event rate in the detector.

\begin{equation}
  B_{<\theta} (\alpha, \delta) = \int\int \! dh \, d\delta \, d\tau \,\, \epsilon(h,\delta) \, R(\tau)  
  \label{eq:bgr}
\end{equation}

\noindent where $\rm{B}_{<\theta}$ is the number background events within a
circular region (radius $\theta$) centered at celestial coordinate $(\alpha,
\delta)$.  The efficiency map $\epsilon$, is obtained by binning all events in
local hour angle ($h$) and declination ($\delta$) for the entire integration
interval.  This map is scaled into a normalized probability density function for
accepting events within the detector using the total number of events observed
during the 2 hour integration duration. The all-sky background event rate
$R(\tau)$ is calculated by integrating every events in the sky every 24 seconds
(the time it takes for the sky to move by 0.1\adeg or a pixel).

\begin{figure*}[t]
\centering
\includegraphics[width=\textwidth]{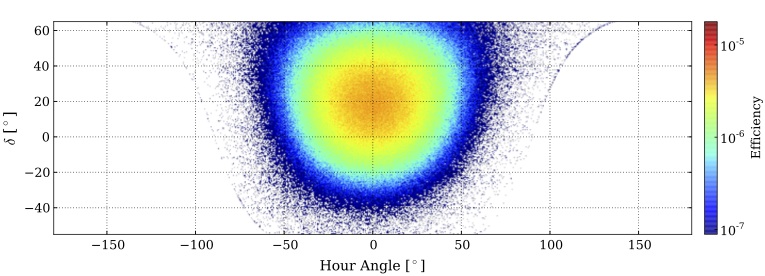}
\caption{Efficiency map, $\epsilon$ for a single 2 hour integration period. It shows the
  distribution of cosmic-ray background distribution in local coordinates,
$(h,\delta)$. Note that the efficiency peaks at zenith or $h = 0\adeg$.}
\label{eff_map}
\end{figure*}

Figure~\ref{eff_map} shows the angular distribution of the arrival direction of
events $\epsilon(h,\delta)$ in local coordinates for a single two-hour
integration period.  Assuming that the arrival directions of the cosmic-ray
events are isotropic, the efficiency distribution peaks overhead (at $h$~=~
0~\adeg) due to the thickness of the atmosphere.

\subsection*{Light Curve} 

For the sidereal time bin $\Delta\tau$ as above, the signal rate S($\alpha,
\delta$) is estimated by integrating every event falling within an optimal
radial search window ($\theta < 2\adeg$) around the putative source location
every 24 seconds. The radial search bin is also used to estimate the background
for the source region using Eq.~(\ref{eq:bgr}). The choice for the optimal bin
size, $\theta = 2\adeg$ is determined from simulation in order to maximize the
detector sensitivity to a point source of gamma rays in the sky.  Once we have
determined both the background and signal, the {\it lightcurve} or the excess
count rate can be calculated as a function of the modified Julian date for any
given location in the sky. In order to accumulate sufficient statistics for the
present background rate, the raw signal and background rates from the previous
section are re-binned over 6-minute intervals or higher to generate the final
lightcurves.

\begin{equation}
  {\rm Excess}_{<\theta} (\alpha, \delta) = S(\alpha,\delta) - B(\alpha,\delta)
\end{equation}

\begin{figure}[t] \centering
   \includegraphics[width=0.5\textwidth]{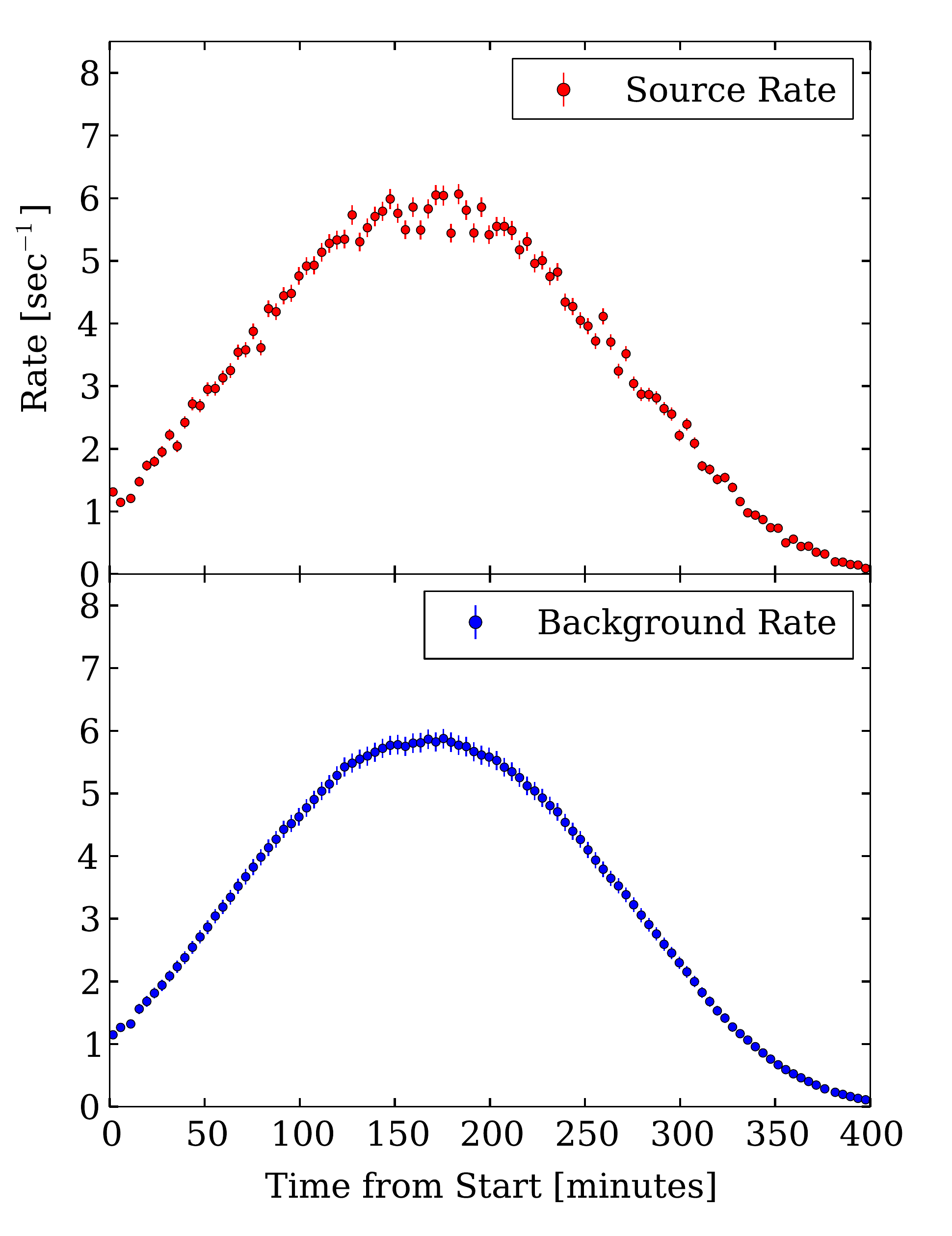} \caption{As an
     example, we show the source rate (Top) and the background rate (Bottom) for
     a specific region calculated from HAWC-43 data. The region was selected at
     a high galactic latitude without any known VHE emitter in the vicinity. We
     used a 2\adeg optimal bin radius for the search window.  The data is binned
     at 6-minute intervals. The background rate is highest when the source is
     directly overhead.} \label{sky_rate} \end{figure}

As an example, Figure~\ref{sky_rate} shows the rate of signal events ({\it Top})
and the rate of background events ({\it Bottom}) from a region of the sky that
is devoid of known VHE emitters. The rates were calculated from a \asim22 hour
run obtained with the HAWC array under 43 WCD configuration. The average all-sky
rate was found to be  \asim1.5 kHz across the sky. No gamma-hadron separation
cuts were applied in this case. The shapes of the curves reflect the region
transiting the sky, as observed by HAWC. Hence, the event rate (predominantly
consisting of cosmic-ray background events) peaks when the region passes
overhead. 

\begin{figure}[t] \centering
   \includegraphics[width=0.5\textwidth]{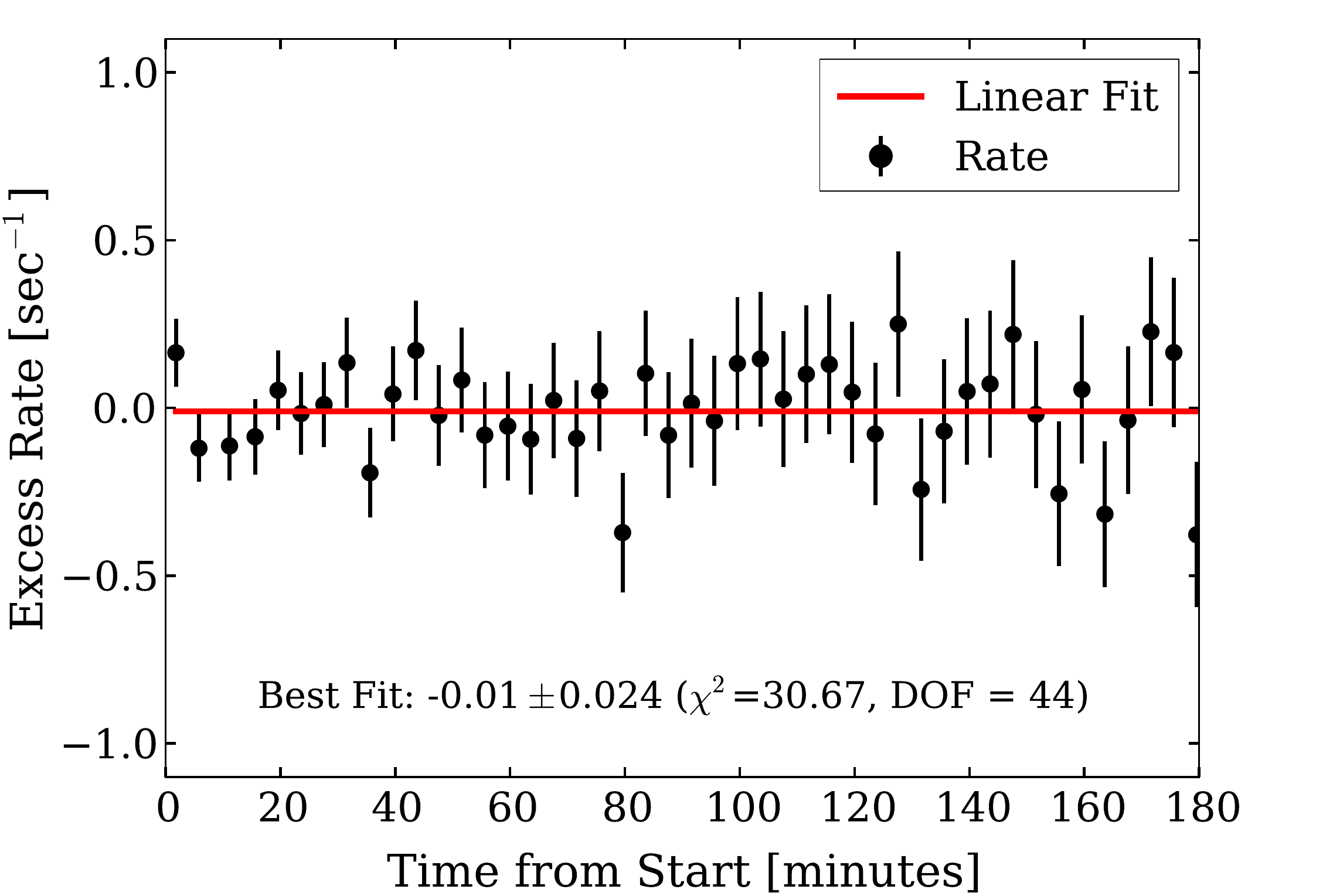} \caption{Lightcurve
     showing the excess counts per second for the source region described in
     Fig~\ref{sky_rate}. The overall excess count for the 3 hour period is found
     be consistent with zero.} 
     \label{light_curve} 
   \end{figure}

Figure~\ref{light_curve} shows the resulting lightcurve for the above example
within an arbitrary 3 hour window. The data points are binned at 6 minute intervals. As
expected, the excess count around the region is found to be consistent with zero
events (the red line shows a straight line fit the data points). 

\section*{Discussion}

Initially, an automated procedure is set up to routinely analyze the lightcurves
for increased gamma-ray emission from the locations of established TeV emitting
AGN. This list will be updated in the future to include the nearest of variable
AGN detected by Fermi to select promising candidates for TeV emission. To
estimate the variability in excess gamma-ray counts from a given candidate
object, each lightcurve is associated with a significance lightcurve.  The
latter measures the significance of the observed fluctuations above or below the
time-averaged gamma-ray counts,

  \begin{equation}
    {\rm V}_{\sigma} = \frac{{\rm Excess}_i - {\rm
    Excess}_{avg}}{\sqrt{\sigma_{i}^{2}+\sigma_{avg}^{2}}}
  \end{equation}

\noindent where the ${\rm Excess}_i$ and $\sigma_{i}$ denotes the excess and
error, respectively for an individual point in the lightcurve and ${\rm
Excess}_{avg}$, $\sigma_{avg}$ are the time-averaged excess counts and errors,
respectively.  The analysis also generates cumulative lightcurves on a daily and
weekly basis in order to detect day-scale or week-scale flares from an AGN.

Finally, we have established a preliminary flare criterion to alert the analysis
team if a single data point in the lightcurve exceeds 3$\sigma$ above the
time-averaged gamma-ray count. In the event of a possible detection, the
automated procedure will generate an email giving the position of the excess
along with necessary statistical details for immediate distribution to an
internal list. Pending thorough verifications, HAWC plans to release the alerts
to the general community (via e.g.,
ATEL\footnote{http://www.astronomerstelegram.org/} or
AMON\footnote{http://amon.gravity.psu.edu/index.shtml}) as early as possible.
Timely alerts will be particularly crucial for the ground-based ACTs in order to
plan the observing schedule for viable targets.

\section*{Conclusion} 

HAWC offers an unprecedented all-sky coverage to promptly detect increased
emission from AGN and trigger follow-up multiwavelength observations. This will
greatly enhance our exposure to flaring AGN, a crucial factor to understanding
the physics behind the ultra-relativistic jets. Beginning with the first 100
WCDs in August of 2013, HAWC plans to establish a dedicated program for alerting
the greater astrophysics community to flaring AGN. The present online flare
monitoring tools focuses on both known as well as predicted TeV emitting AGN.
Detections of flaring emission from unknown objects using non-targeted searches
will be the subject of future improvement to the existing analysis tools.

\section*{Acknowledgments}

We acknowledge the support from: US National Science Foundation (NSF); US
Department of Energy Office of High-Energy Physics; The Laboratory Directed
Research and Development (LDRD) program of Los Alamos National Laboratory;
Consejo Nacional de Ciencia y Tecnolog\'{\i}a (CONACyT), M\'exico; Red de
F\'{\i}sica de Altas Energ\'{\i}as, M\'exico; DGAPA-UNAM, M\'exico; and the
University of Wisconsin Alumni Research Foundation, and the Institute of
Geophysics and Planetary Physics at Los Alamos National Lab.

\clearpage


\newpage
\setcounter{section}{6}
\nosection{HAWC Observations of the Crab Nebula\\
{\footnotesize\sc Brian Baughman, James Braun, Jordan Goodman, Asif Imran,
Barbara Patricelli, John Pretz}}
\setcounter{section}{0}
\setcounter{figure}{0}
\setcounter{table}{0}
\setcounter{equation}{0}
%
%
%
%
\title{HAWC Observations of the Crab Nebula}

\shorttitle{HAWC Observations of the Crab}

\authors{
B.~M.~Baughman$^{1}$,
J.~Braun$^{1*}$,
J.~A.~Goodman$^{1}$,
A.~Imran$^{2}$,
B. Patricelli$^{3}$,
J.~Pretz$^{4}$,
for the HAWC Collaboration$^{5}$
}

\afiliations{
$^1$Department of Physics, University of Maryland, College Park, MD, USA\\
$^2$Department of Physics, University of Wisconsin-Madison, Madison, WI, USA\\
$^3$Instituto de Astronom\'ia, Universidad Nacional Aut\'onoma de M\'exico, Mexico D.F., Mexico\\
$^4$Physics Division, Los Alamos National Laboratory, Los Alamos, NM, USA\\
$^5$For a complete author list, see the special section of these proceedings\\
}

\email{jbraun@umdgrb.umd.edu}

\abstract{
We present observations of the Crab Nebula with the HAWC Observatory.
HAWC is a gamma-ray shower array under construction at 4100 m on the
volcano Sierra Negra in Mexico.
When complete, HAWC will consist of 300 water-Cherenkov detectors
(WCDs) measuring 7.3 m (diameter) by 4.5 m (height) and instrumented with four
upward-facing large-photocathode PMTs.
The modular design of HAWC allows for the operation of completed WCDs
during construction.
HAWC has been recording data with at least 28 WCDs since late 2012. 
}


\maketitle

\section*{Flares from the Crab Nebula}

The Crab nebula is arguably the most well-observed gamma ray source in the sky.
Observations at MeV to GeV energies are obtained from satellite-borne instruments
including the Fermi Large Area Telescope (LAT) and AGILE.
Observations at GeV to TeV energies are obtained by ground-based experiments
in two categories.
Imaging air- Cherenkov telescopes (IACTs), including VERITAS, MAGIC, and H.E.S.S.,
have a low duty cycle ($\sim$10\%) but excellent sensitivity over a narrow
field-of-view ($<$5$^\circ$).
Ground- based air shower observatories, including Milagro, ARGO-YBJ, and HAWC,
have a continuous duty cycle and wide field-of-view ($\sim$2 steradians) of the
overhead sky, but a comparatively lower instantaneous sensitivity at any given
point in the sky.
The Crab, at declination 22$^\circ$, transits near zenith for these ground-based
observatories and is observed once per day for five to six hours.
The Crab nebula has historically been viewed by the gamma-ray community as a
standard reference source because of its relative stability.

During September 2010, AGILE and Fermi LAT recorded a gamma ray flare from the
Crab nebula of four times above the baseline flux at $>$100 MeV with a duration
of 2 to 3 days \cite{or1,or2} and variability at time scales of 12 hours \cite{or3}.
In response to these observations, ARGO-YBJ reported a 4$\sigma$ excess from the
Crab at $\sim$TeV energies during the time of the flare \cite{or4}, corresponding to a
flux of 3 to 4 times the expected value.
MAGIC and VERITAS monitored the Crab for several $\sim$20-minute windows during
this period, and each reported no excess above expectation from the Crab
\cite{or5, or6}.

Following the September 2010 flare, AGILE and Fermi published observations of
previous flares from the Crab Nebula in 2007 \cite{or1} and 2009 \cite{or2},
respectively.
Milagro, which ceased operation in 2008, reported no increase in flux from the
Crab Nebula during the 2007 flare \cite{or7}, and ARGO-YBJ did not observe an excess
during the 2009 flare \cite{or8}.

In April 2011, the Crab Nebula produced an extreme flare that was observed by
Fermi \cite{or9, or10} and AGILE \cite{or11}, with fluxes at $>$100 MeV increasing
by a factor of 30, with a doubling time of eight hours \cite{or10} and a new
component appearing in the Crab nebula spectral energy distribution.
ARGO-YBJ reported an excess during this flare, observing 3.2$\sigma$ to 3.4$\sigma$,
on a 0.53$\sigma$ to 0.62$\sigma$ expectation from the Crab \cite{or8}.
No IACT results were published for this flare.

In July 2012, the Crab produced a fifth flare that was detected by LAT \cite{or12}.
Again, ARGO-YBJ reported a 4$\sigma$ excess, corresponding to a flux of about eight
times the expected value \cite{or13}; however, IACT observations were not possible,
as the flare occurred during the daytime.

\begin{figure}[t!!]\begin{center}
\includegraphics[width=3.2in]{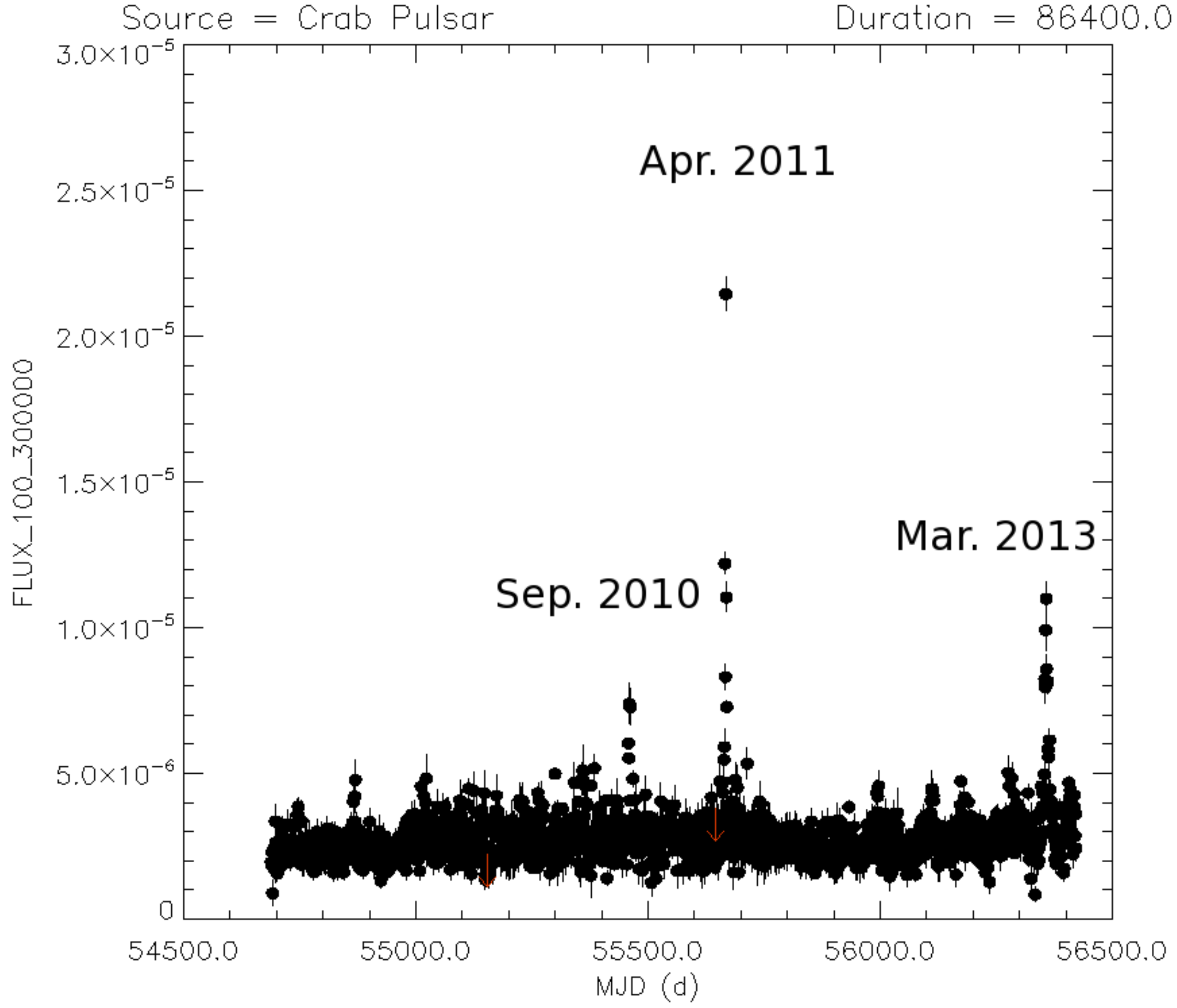}
\caption{Fermi-LAT light curve of the Crab at $>$100 MeV showing the September
2010, April 2011, and March 2013 flares \cite{fermicurve}.}\label{fermic}
\end{center}\end{figure}

Finally, in March 2013, the Crab produced another flare detected by Fermi-LAT
\cite{fermi2013} and AGILE \cite{agile2013}, the sixth flare in six years.
The Fermi-LAT data show flare lasted about a week, with the $>$100 MeV flux
peaking at a factor 4--5 above the quiescent flux.
No observations of the Crab by ARGO-YBJ or IACTs during the flare have
been published.
The HAWC Observatory was operational with 28 tanks during the majority of
this flare.  The Fermi-LAT light curve of Crab is shown in Fig \ref{fermic}.

Analysis of Crab nebula energy spectra during the April 2011 flare obtained by
Fermi LAT \cite{or10} and AGILE \cite{or11} strongly suggests that the emission
at MeV to GeV energies is synchrotron radiation from a population of freshly
accelerated PeV electrons that rapidly cool.
The mechanism responsible for this acceleration is not yet understood.
The accelerated electrons will also emit inverse Compton radiation at TeV to PeV
energies, with an intensity and spectrum determined by the properties of the
emission region. Measurements of this inverse Compton component therefore can
constrain the properties of the emission region \cite{or14}, including Lorentz factor,
and constrain acceleration models e.g. \cite{or15, or16}.
Current measurements in the $>$TeV energy range are unclear, but the marginal
ARGO-YBJ excesses suggest that these studies may be within the reach of HAWC.

\section*{The HAWC Observatory}

HAWC is a gamma-ray air-shower observatory under construction at 4100 m on the
volcano Sierra Negra, near Puebla, in Mexico.  HAWC consists of a densely-packed
array of WCDs sensitive to air showers.  Each WCD
consists of a 7.3 m (diameter) by 4.5 m (height) corrugated steel tank with a
plastic bladder inside that holds $\sim$200,000 L of purified water.  Four
photomultiplier tubes (PMTs) are deployed in each tank: One central 10-inch
Hamamatsu R7081 HQE PMT and three outer 8-inch Hamamatsu R5912 PMTs.  Signals
from the PMTs are processed by scalers and TDCs.  PMT timing and charge
information from the TDCs is used to reconstruct the core location and axis
of air showers landing on the array.  At the time of the flare, HAWC recorded
air showers at a rate of $\sim$8 kHz.  The vast majority of these events
are hadronic showers from cosmic rays.

HAWC is designed to be completely modular.  Tanks are added to the data stream
soon after they are constructed, increasing the sensitivity of HAWC over time.
HAWC initially operated with 28 tanks from late 2012 through mid-April 2013.
HAWC then operated with 43 tanks until mid-May, and has since been operating
with 77 tanks.  HAWC will operate with 111 tanks by the end of summer 2013 and
is scheduled to be complete with 300 tanks before the end of 2014.

To maximize the statistical significance of gamma-ray sources, HAWC will reject
hadronic cosmic ray events based on event morphology.  Hadronic showers produce
pions and muons which may have large transverse momentum and carry energy away
from the shower core, whereas electromagnetic showers produced by gamma-ray
events are more compact.  Using the largest-amplitude PMT signal outside
an exclusion radius of 40 meters as a statistic, in the 300-tank configuration
HAWC will reject 99\% of hadronic cosmic ray events above a few TeV, with negligible
loss of gamma-ray events.  In this configuration, HAWC will observe more than 7$\sigma$
on the Crab with each transit.  With 28 tanks, hadronic cosmic ray rejection
is not possible due to the small spatial extent of the detector, and only $\sim$0.3$\sigma$
is expected from each transit of the Crab.  With 77 tanks, HAWC is
large enough to reject a fraction of hadronic events.  We expect approximately 1.5$\sigma$
per Crab transit in this configuration.

\section*{HAWC Observations of the Crab}

\begin{figure}\begin{center}
\includegraphics[width=3.2in]{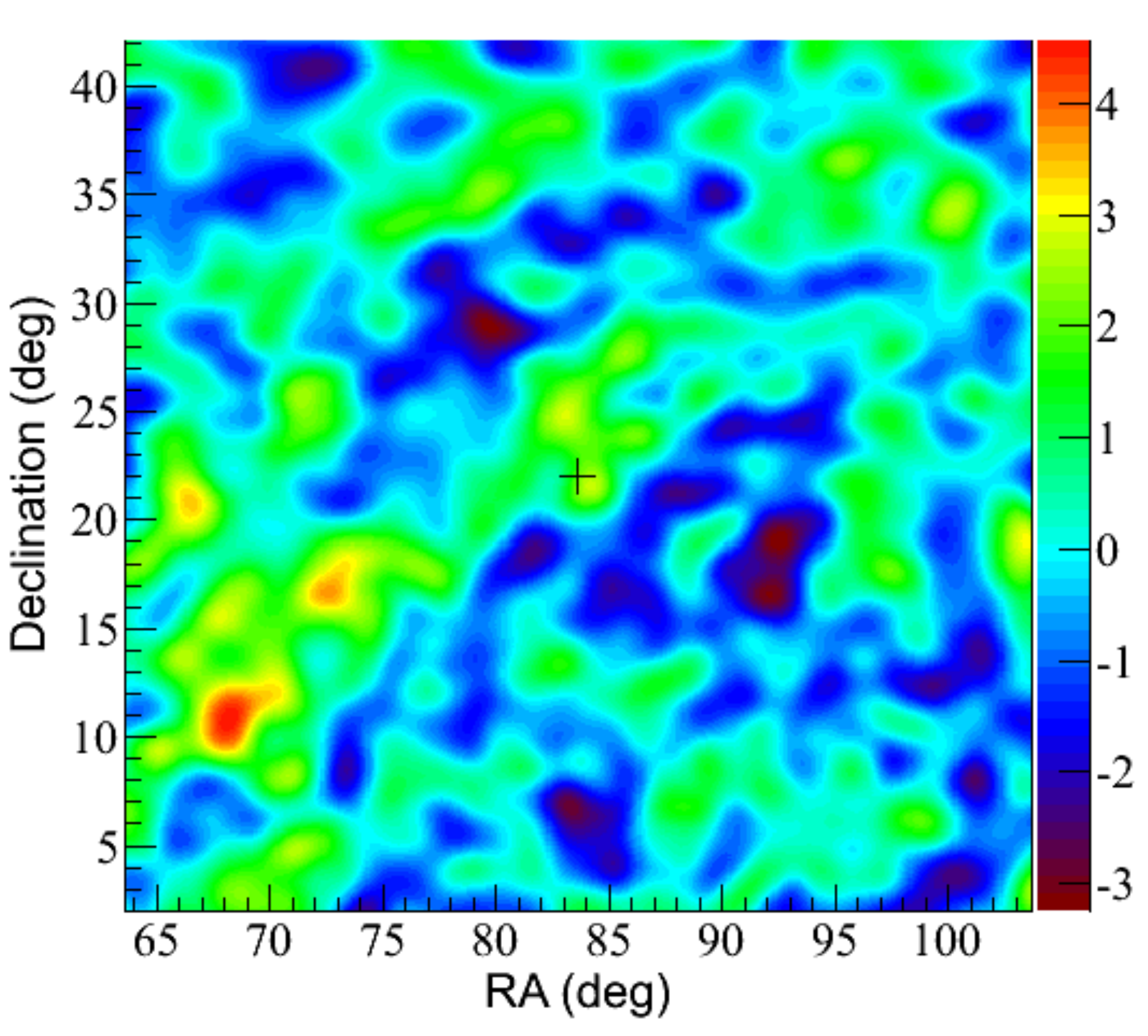}
\includegraphics[width=3.2in]{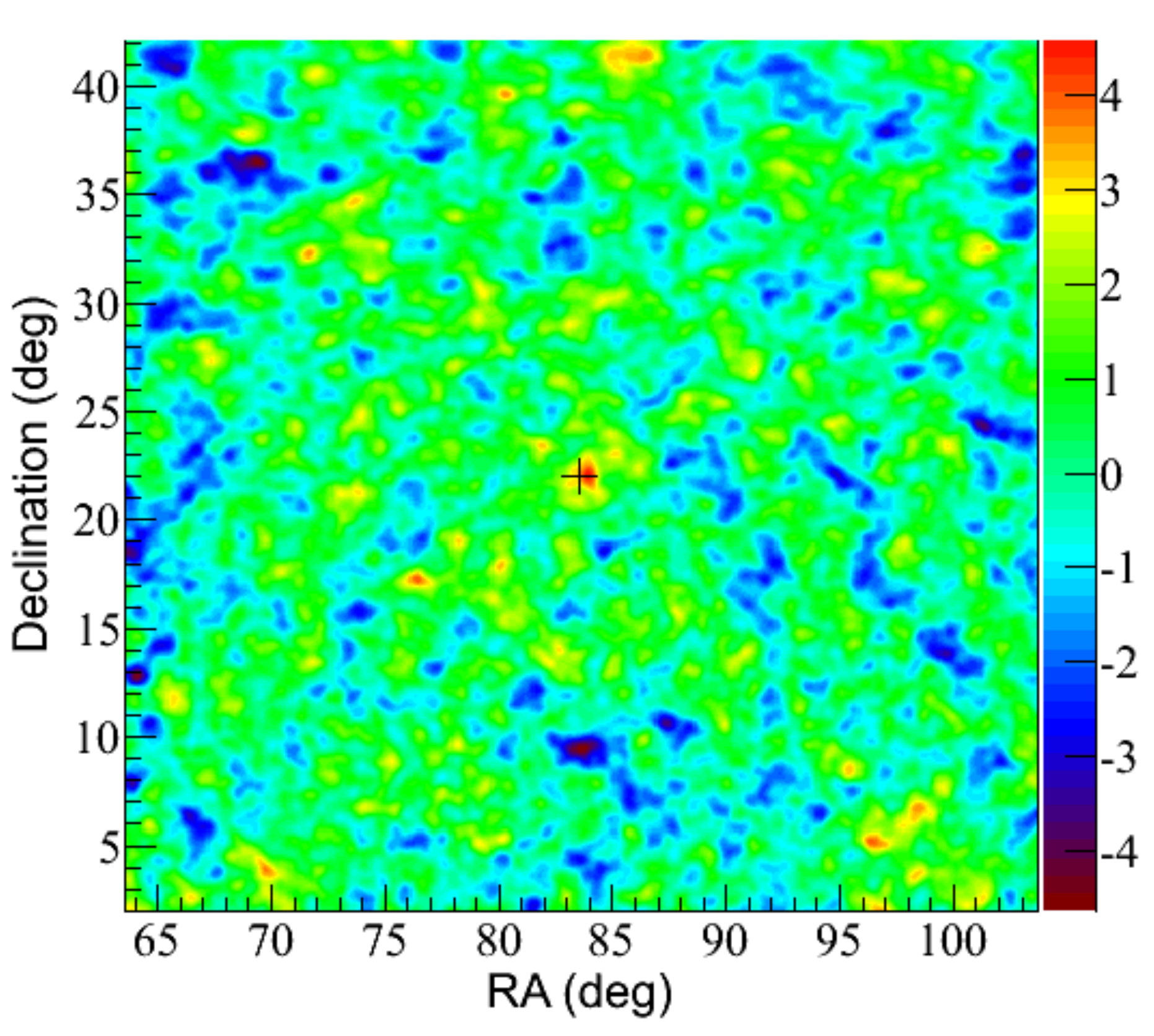}
\caption{135-day HAWC 28-tank (top) and 14-day HAWC 77-tank (bottom) significance skymaps for the region of the Crab. The color scale units are standard deviations.  2$\sigma$ and 3.5$\sigma$ were observed at the position of the Crab in the 28-tank data and 77-tank data, respectively.}\label{crabInt}
\end{center}\end{figure}

From October 2012 to May 2013, HAWC observed the Crab for 135 days.
HAWC observed a 2$\sigma$ excess at the location of
the Crab during this period, shown in figure \ref{crabInt}, roughly in-line with expectations from Monte Carlo.
HAWC has now recorded approximately 14 days of data in the 77-tank configuration and
observes a $\sim$3.5$\sigma$ excess at the location of the Crab, shown also
in figure \ref{crabInt}.

\section*{HAWC Observations during the March 2013 Flare}

\begin{figure}[t]\begin{center}
\includegraphics[width=3.2in]{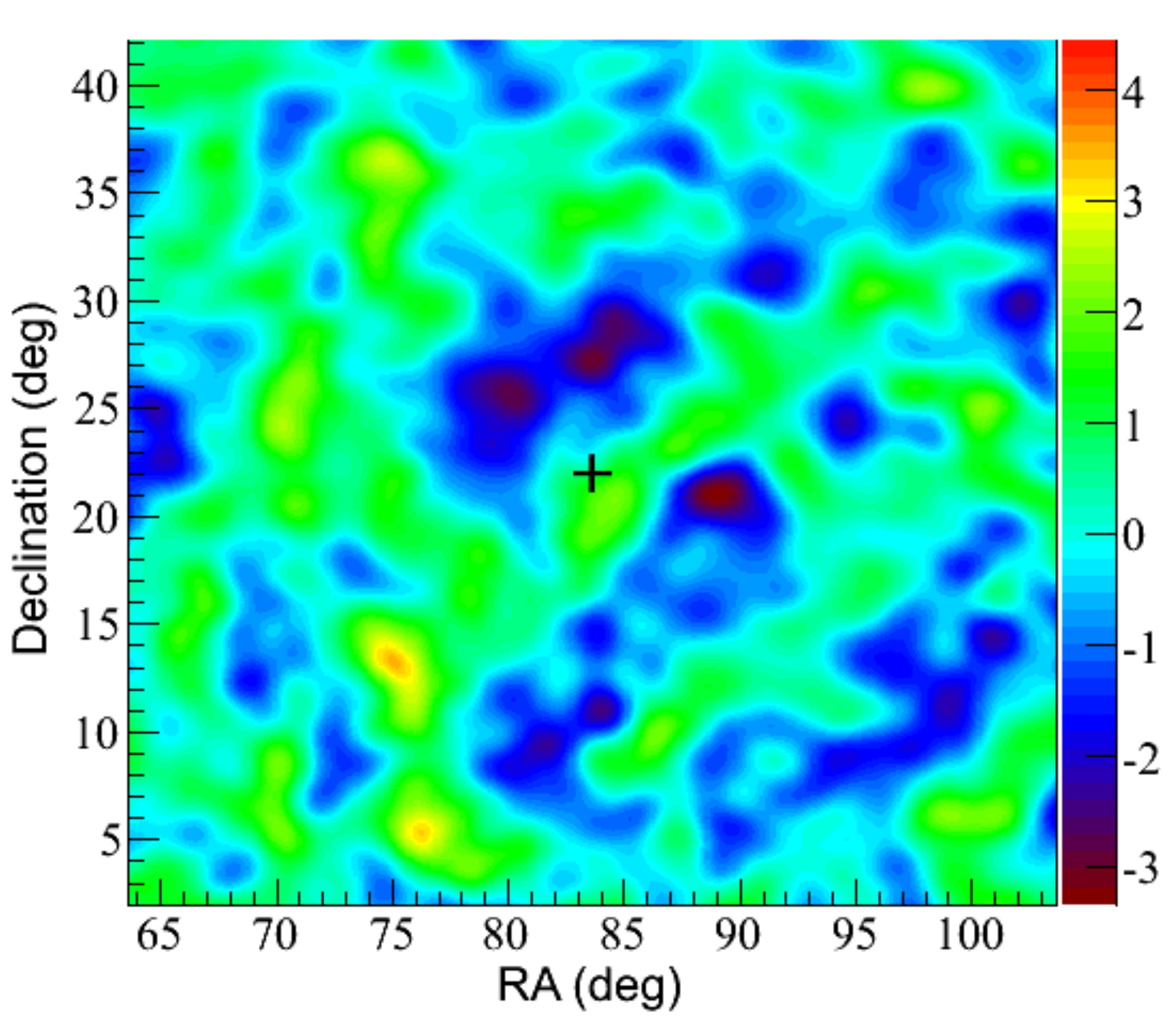}
\caption{5-day HAWC 28-tank significance skymap for the region of the Crab from March 4--8, 2013.
 The color scale units are standard deviations. 0.7$\sigma$ was observed at the position of the Crab.}\label{crabFlare}
\end{center}\end{figure}

The March 2013 flare from the Crab began on March 4 and faded slowly after
March 8.  HAWC was in the 28-tank configuration during this flare.
HAWC observed the Crab for a majority of the transit on each of
these five days.  HAWC observations are shown
in figure \ref{crabFlare}.  The measurement at the position of the Crab
was consistent with background, with an excess of 0.7$\sigma$.  This
measurement rules out a very large TeV component from this flare; however,
our data is preliminary, so we do not currently set a limit on the flux.

\section*{Performance of HAWC for Future Flares From the Crab}

HAWC will operate with 111 tanks beginning in August 2013.  In this configuration,
HAWC expects to record 2$\sigma$ with each Crab transit and approximately
5$\sigma$ for a 5-day period.  Therefore, for a similar five-day flare from the Crab,
HAWC will be capable of measuring changes in the TeV flux of the Crab at the level of
20\%.  When complete in 2014 with 300 tanks, HAWC
will observe 7$\sigma$ with each transit of the Crab.  At this significance level,
HAWC will measure the TeV flux of the Crab to better than 20\% for each day of the year,
enabling long-term studies of the Crab at TeV energies.

\section*{Acknowledgments}

We acknowledge the support from: US National Science Foundation (NSF); US
Department of Energy Office of High-Energy Physics; The Laboratory Directed
Research and Development (LDRD) program of Los Alamos National Laboratory;
Consejo Nacional de Ciencia y Tecnolog\'{\i}a (CONACyT), M\'exico; Red de
F\'{\i}sica de Altas Energ\'{\i}as, M\'exico; DGAPA-UNAM, M\'exico; and the
University of Wisconsin Alumni Research Foundation.

\clearpage


\newpage
\setcounter{section}{7}
\nosection{Enhancing the Response of HAWC to sub-TeV Transient Sources\\
{\footnotesize\sc Ian Wisher}}
\setcounter{section}{0}
\setcounter{figure}{0}
\setcounter{table}{0}
\setcounter{equation}{0}
%
%
%

\title{Enhancing the Response of HAWC to Sub-TeV Transient Sources}

\shorttitle{Enhancing HAWC to Transient Sources}

\authors{
Ian G. Wisher $^{1}$,
for the HAWC Collaboration.$^{2}$
}

\afiliations{
$^1$ WIPAC and Department of Physics, University of Wisconsin - Madison \\
$^2$ For a complete author list, see M. Mostafa in these proceedings \\
}

\email{ian.wisher@wipac.wisc.edu}

\abstract{The High Altitude Water Cherenkov (HAWC) Observatory, currently being built 
4,100 meters above sea level on Sierra Negra, Mexico, is a ground-based detector well-suited
for observing transient phenomena in the TeV energy range.  This is due to its large field 
of view (~2\,sr), high up time ($>$90\%), and high efficiency for triggering on showers 
above 1\,TeV.  However, sub-TeV transient events are of interest due to the overlap 
in energy with satellite experiments such as the Fermi gamma-ray space telescope.  
The standard HAWC reconstruction chain will achieve an effective area of $100\,{\mathrm{m}}^2$ 
at 100\,GeV while rejecting lower multiplicity events in the detector.  Triggering on small 
showers from primaries with energies below 100\,GeV is difficult due to the large 
number of uncorrelated signals in the detector during the large time window required 
to accept showers from all arrival directions.  To address this problem and augment 
the sensitivity of HAWC below 100\,GeV, we propose a method in which particle arrival 
directions are fit for triplets of triggered photomultipler tubes (PMTs) in a short 
sliding trigger window (100\,ns).  The resulting arrival directions are then summed 
in a coarsely-binned significance map of the sky with a time window of one to several 
seconds.  This fast algorithm will run online and will be able to localize the positions 
of transient sources to within $8^{\circ}$.  Applying this technique to data from the 
currently-operational 10$\%$ of the final array allows us to obtain actual noise 
rates and use them in conjunction with simulations to calculate the sensitivity to 
transients.}


\maketitle

\section*{Introduction}

The Gamma-Ray Coordinates Network (GCN) \cite{bib:gcn} has made possible rapid observations 
between multiple experiments to be triggered on transient phenomena.  Satellite experiments 
such as Fermi and Swift have so far contributed thousands of alerts in the MeV and GeV bands. 
The size of the orbital experiments limits the observations of the highest energy gamma rays, 
but the alerts allow imaging air cherenkov telescopes to follow up with observations and probe 
for TeV emission.  Simultaneous observations of a transient source like a gamma-ray burst (GRB) 
\cite{bib:hawc} by multiple experiments with a combined sensitivity spanning keV to TeV would 
be of particular interest .  Such an observation would give insight not only into the emission mechanism 
of the GRB but also other interesting physics, {\it e.g.}, limits on Lorentz invariance violation 
\cite{bib:lor,bib:lorenz} and bulk Lorentz factors of jets \cite{bib:blor}.  

The High Altitude Water Cherenkov (HAWC) observatory is an extensive air shower array currently 
being constructed at 4,100 meters above sea level on the Volcano Sierra Negra in Mexico. 
HAWC will have a wide field of view ($\sim$2\,sr) and a high duty cycle (\textgreater90\%).
It will cover an area of $\sim22,000\,{\mathrm{m}}^2$, giving the detector excellent sensitivity 
to gamma rays between 100\,GeV and 100\,TeV \cite{bib:hawcpro}.  These traits make HAWC ideal 
for unbiased surveys of the northern sky and searches for transient sources such as GRBs,
which so far have not been observed in the TeV band.

\section*{The HAWC Detector}

The final HAWC array will consist of 300 optically isolated water cherenkov detectors (WCDs). 
The WCDs are cylindrical tanks 7.3\,m in diameter and 4.5\,m tall that hold approximately 
200,000 liters of purified, clear water.  They are instrumented with four upward facing PMTs (three Hamamatsu 
R5912 PMTs and one central Hamamatsu R7081 high quantum efficiency PMT).  The array is scheduled 
to be completed in the summer of 2014 but has started taking data with WCDs as they are added 
in September 2012. 

The PMTs are read out using custom front-end board electronics that are reused from the Milagro 
experiment.  The boards amplify, shape, and discriminate the pulse across a low and a high threshold. 
Instead of digitizing the waveform, the time over these thresholds (ToT) is used as a proxy for pulse 
height and width.  These digital ToT signals are time-stamped by a CAEN VX1190 Time to Digital Converter (TDC), 
and read out using a VME single-board computer.  All the data are then sent to an online computer farm for processing. 
This allows the full data stream of all 1200 PMTs ($\sim$500\,MB/s ) to be seen by the online system 
which triggers in real time. 
                                                                            
\section*{Low-Energy Challenges}

Low energy showers produce challenges for extensive air shower detectors due to the limited 
amount of information that reaches the ground.  As the energy of the primary gamma ray 
decreases, the number of PMTs that detect secondary particles in a sliding time window ($nHit$) 
decreases.  Even though HAWC is built at a high altitude in order to get close to the shower maximum,
in sub-TeV showers, typically only a small fraction of the maximum number of secondary 
particles reaches the array.  The low number of hits causes the separation of gamma-ray and hadronic 
primaries to fail. 

An additional challenge comes from the noise rate of the PMTs, $\sim 20$\,kHz.  For events with
a low $nHit$, the noise hits become a significant fraction of the hits in the time window. 
This will significantly hinder the performance of standard N-point plane fits and skew them from 
their original direction.  Despite these challenges, the relative abundance of sub-TeV showers
makes them particularly interesting for transient searches. 

\section*{Online Reconstruction}

The online software trigger system gives the HAWC detector a great deal of flexibility to explore interesting 
trigger options.  The only caveat is that since the algorithms have to keep up with the full data 
rate of the detector, they cannot be processor intensive.  In addition, the trigger has to have 
sufficient compression or rejection capability to reduce the 500\,MB/s rate to a more manageable level. 
The default running mode is a simple multiplicity trigger that looks for events with more than $n$ 
hits in a short time window.  If the trigger condition is passed, all the hit information is saved.  

Due to the high rate, low multiplicity events will be reconstructed online, and only reconstructed 
track and energy parameters will be saved.  The remaining low-hit events are not usable by the 
standard reconstruction but still contain useful information that can be extracted using a different method. 

\subsection*{Low Multiplicity Reconstruction}

Extracting useful information from low-energy, low-hit showers is difficult, but the challenges 
described in Section 3 can be mitigated or overcome.  Our method (affectionally called the ``vector
telescope''), breaks down into two steps: a low multiplicity event fitter and a short time-window 
skymap search. 

A triplet fitter finds all combinations of 3 WCDs in a 100\,ns time window and analytically fits 
a plane to all triplets individually.  The normal vector to this plane then points back to the 
source.  This procedure has two advantages; it can be performed for a low number of hits very 
quickly; and it decouples the noise hits from those associated with the shower.  Though the noise 
hits will pull a fraction of the triplet normal vectors away from the source the remaining triplet 
normal vectors point back to the source.  Due to the rapid ($_nC_3$) growth of the number of 
triplets with event size, we only apply the algorithm to events with $n\le10$.

Once a fit to the triplets is calculated, it is pushed into a 1\,s time buffer.  Events in 
this buffer are placed in a HEALPix skymap, which is evaluated for significant excesses. 
The length of the buffer and the size of the HEALPix binning can be adjusted to search for 
different types of sources, {\it e.g.} GRBs of different duration. 

\section*{Performance and Results}

\begin{figure}[ht]
  \centering
  \includegraphics[width=0.4\textwidth]{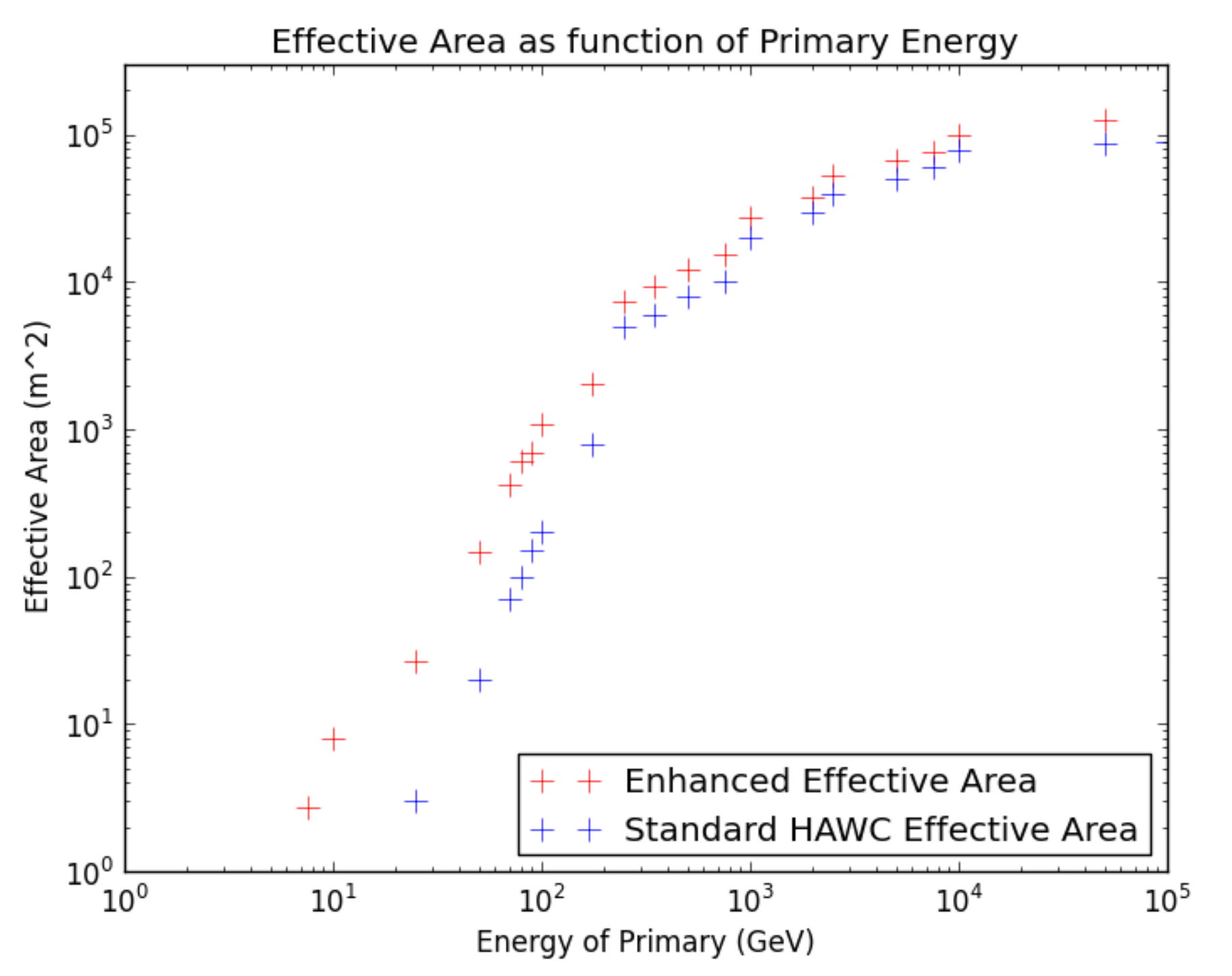}
  \caption{Effective area as a function of primary energy for the standard event reconstruction 
           (blue) and the vetor telescope algorithmR described here (red).  There is a considerable
           increase of effective area at low energies.}
  \label{area_fig}
 \end{figure}

To determine the performance, the method was applied to simulated data for the full detector.
Cuts were made to remove select events that would pass the normal trigger in HAWC. 
Since the method utilizes air showers that are normally discarded, the effective area increases 
with respect to the standard trigger (Fig. \ref{area_fig}). 

\begin{figure}[ht]
  \centering
  \includegraphics[width=0.4\textwidth]{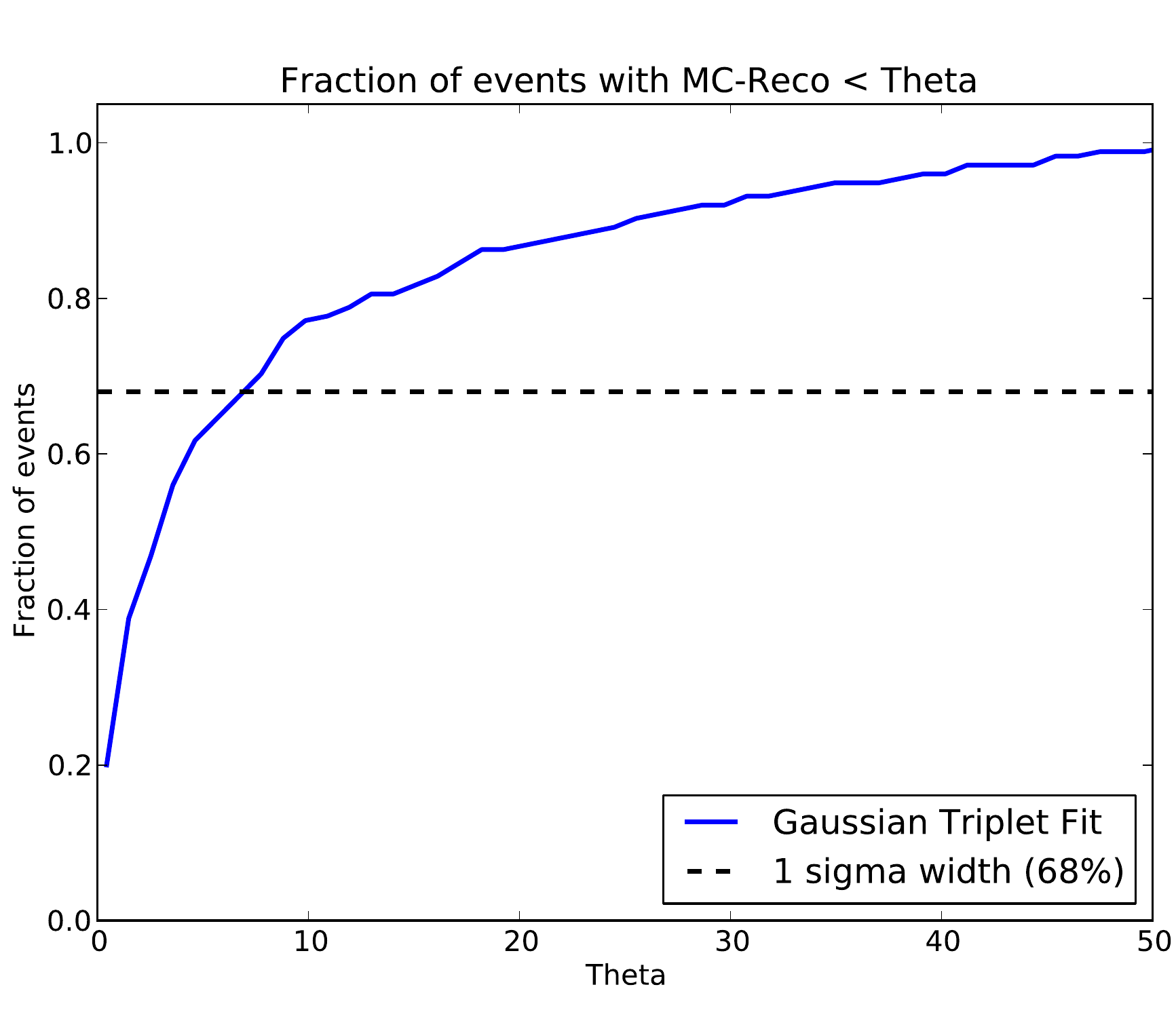}
  \caption{Angular resolution of the vector telescope algorithm.  The plot shows the fraction
           of events reconstructed with an angular distance of less than $\theta$ between the 
           simulated and reconstructed direction, as a function of $\theta$.}  
  \label{pointing_fig}
 \end{figure}

We also study the performance of the vector telescope reconstruction by calculating the 
detector's Point Spread Function (PSF) for the algorithm.  The PSF is derived by determining 
the difference between the simulated direction and the reconstructed direction.  While
the resulting PSF of $8^{\circ}$ (Fig. \ref{pointing_fig}) is worse than the PSF of the
standard reconstruction, it still provides a good angular resolution for low energy events
that would otherwise not be reconstructed.  The detection of a significant excess in the skymap
can trigger alerts to other detectors, for example air Cherenkov detectors.

The method cannot distinguish between hadronic and gamma-ray primaries.  This is not so much 
a limit of the method, but more likely a lack of information at the ground to distinguish the 
two types using the standard containment cuts.  Future work that could tag muons could provide 
the needed information.

\section*{Conclusions}

We are optimizing the skymap significance transient finder using data from the 30 tank deployment 
of HAWC. Results will be shown at the conference. 

\section*{Acknowledgments}

We acknowledge the support from: US National Science Foundation (NSF); US
Department of Energy Office of High-Energy Physics; The Laboratory Directed
Research and Development (LDRD) program of Los Alamos National Laboratory;
Consejo Nacional de Ciencia y Tecnolog\'{\i}a (CONACyT), M\'exico; Red de
F\'{\i}sica de Altas Energ\'{\i}as, M\'exico; DGAPA-UNAM, M\'exico; and the
University of Wisconsin Alumni Research Foundation; The Institute of
Geophysics, Planetary Physics and Signatures at Los Alamos National Lab.

\clearpage


\newpage
\setcounter{section}{8}
\nosection{An All-Sky Simulation of the Response of HAWC to Sources of Cosmic
Rays and Gamma Rays\\
{\footnotesize\sc Segev BenZvi}}
\setcounter{section}{0}
\setcounter{figure}{0}
\setcounter{table}{0}
\setcounter{equation}{0}
%
%
%

\title{An All-Sky Simulation of the Response of HAWC to Sources of Cosmic Rays
and Gamma Rays}

\shorttitle{All-Sky HAWC Simulation}

\authors{Segev BenZvi$^{1}$
for the HAWC Collaboration.
}

\afiliations{
$^1$ WIPAC and Department of Physics, University of Wisconsin-Madison, WI USA\\
}

\email{sybenzvi@icecube.wisc.edu}

\abstract{The High-Altitude Water Cherenkov Gamma-Ray Observatory (HAWC) is set
to become one of the world's most sensitive wide-field observatories of TeV
gamma rays and cosmic rays. In anticipation of commissioning the full
observatory, we have created simulated data sets which model the response of
the HAWC detector to TeV particles. These data sets contain the same time
dependence and statistical fluctuations as the data. They are then used to
optimize the data analysis, investigate artifacts introduced by our analysis
techniques, and estimate the sensitivity of HAWC. We present a simulation of
cosmic rays which includes the shadow of the moon and a $10^{-3}$ anisotropy in
the cosmic ray arrival directions. We also show how the simulation is used to
incorporate time-dependent models of gamma-ray sources.}


\maketitle

\section*{Introduction}

The HAWC Observatory is a water Cherenkov array currently under construction in
Sierra Negra, Mexico.  The HAWC ``data challenge'' is an all-sky simulation
designed to model the response of the detector to sources of TeV cosmic rays
and gamma rays.  The purpose of the data challenge is to produce simulated data
sets with realistic distributions of event energies, local arrival directions,
and arrival times.  The data also contain a small number of reconstruction
observables, such as the number of PMT hits per event or gamma/hadron
separators, which are commonly used in analysis cuts.

The HAWC detector Monte Carlo is based on simulations of air showers with
CORSIKA~\cite{CORSIKA} and of the water Cherenkov detectors with
Geant4~\cite{Geant4}.  The Monte Carlo is useful for characterizing the
detector response, but the computational cost of event production with
CORSIKA+Geant4 makes the full simulation of many different signal and
background hypotheses impractical.

One workaround for this problem is to reweight simulated events to match a
desired source hypothesis.  In the data challenge, we use a
complementary approach.  Rather than reweight existing simulations, we produce
new fake events using a model of the detector effective area
$A_\text{eff}(\theta,E,t)$ which depends on time $t$, energy $E$, zenith angle
$\theta$, and mass composition.  Given $A_\text{eff}$ it is possible to produce
$N$ events from $m$ particle species with differential fluxes $f$ from a solid
angle $\Omega$ by integrating the expression
\begin{equation}\label{eq:rate}
  N(t)=\sum_{i}^{m}
    \int dt\int dE\int d\Omega\
    f_i(\theta,E,t)\cdot A_{\text{eff},i}(\theta,E,t).
\end{equation}
From the detector Monte Carlo we produce tables $A_{\text{eff},i}$ for each
particle type $i$, and for given source fluxes $f_i$ we numerically integrate
eq.~\eqref{eq:rate} to create new events.  To produce reconstruction
observables we also construct multi-dimensional energy- and zenith-dependent
tables of the observables of interest and sample quantities from the tables
using a linear interpolation algorithm~\cite{Rovatti:1998}.

\section*{Simulation of Cosmic Rays}

The earliest experimental observations accessible to HAWC are the anisotropy of
the cosmic rays~\cite{BenZvi:2013} and the shadow of the
moon~\cite{Fiorino:2013}.  To estimate the sensitivity of the detector in the
data challenge we start with simulations of these cosmic-ray signals.  We
simulate eight cosmic ray nuclear species in the HAWC Monte Carlo: H, $^4$He,
$^{12}$C, $^{16}$O, $^{20}$Ne, $^{24}$Mg, $^{28}$Si, and $^{56}$Fe.  With these
simulated events we have constructed effective area and reconstruction tables
for each particle type.  To produce fake events we use broken power law
parameterizations of recent measurements of the cosmic-ray flux between 10~GeV
and 100~TeV \cite{Ave:2008,Panov:2009,Ahn:2010gv,Adriani:2011cu}.

\subsection*{Cosmic Ray Anisotropy}

\begin{figure}[t]
  \centering
    \includegraphics[width=0.49\textwidth]{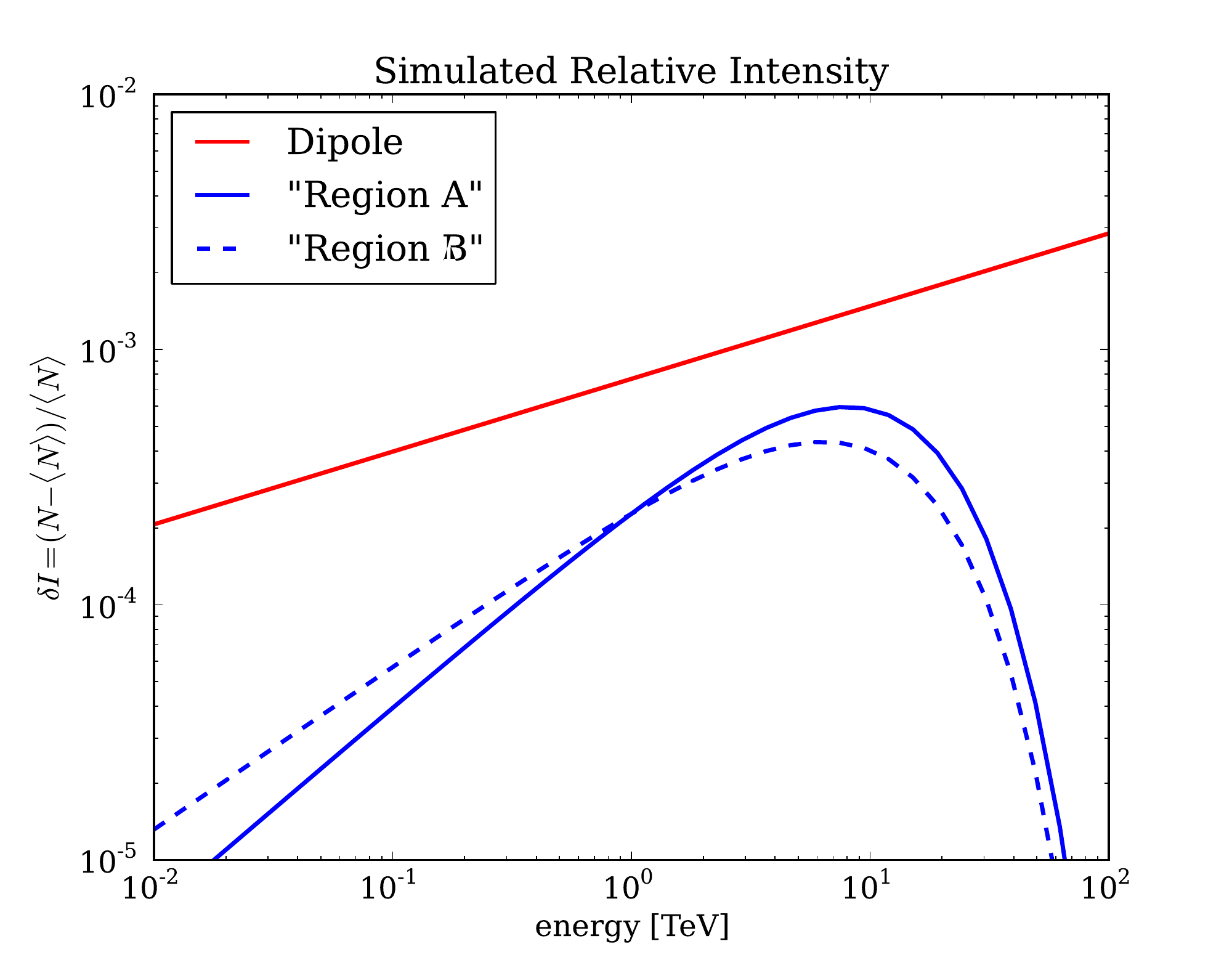}
    \caption{Residual intensity of the dipole (red) and small-scale
    structures (blue) injected into the isotropic cosmic-ray background.}
    \label{fig:cr_dc_spectrum}
\end{figure}

\begin{figure}[t]
  \centering
    \includegraphics[width=0.49\textwidth]{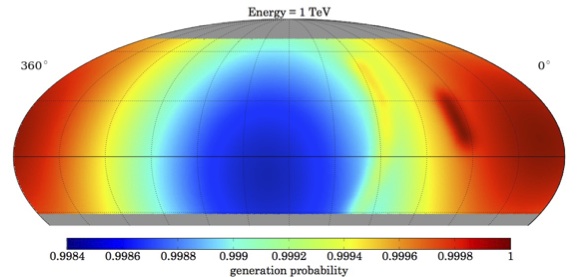}
    \includegraphics[width=0.49\textwidth]{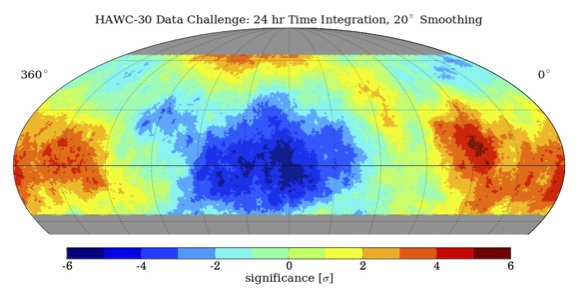}
    \caption{\textsl{Top}: Cosmic ray angular probability density at 1~TeV,
    showing the dipole and small structure atop a large isotropic background.
    \textsl{Bottom}: Significance of the anisotropy in 30 days of simulated
    data, recovered using the time integration algorithm with
    $\Delta t=24$~hr (see below).}
    \label{fig:cr_anisotropy}
\end{figure}

Isotropic cosmic rays are produced by drawing random energies from the
cosmic-ray spectrum and then uniformly generating arrival directions on the
unit sphere.  The detector response is applied using the species-dependent
$A_\text{eff}$ tables with a simple acceptance/rejection calculation.  Once an
event of energy $E$ and zenith angle $\theta$ is accepted, a tuple of
observables such as the number of hits, the gamma/hadron separator, and the
opening angle between the true and reconstructed shower track are resampled
from the observable tables.  The opening angle is then used to scatter the
arrival direction about the initial random draw, automatically introducing the
smearing effects of the detector reconstruction into the simulated data.

To produce an anisotropy in the cosmic rays, we alter the algorithm slightly by
using an energy- and direction-dependent probability density function (PDF)
describing the anisotropy we wish to model.  Arrival directions and energies
are generated from this PDF.

The energy spectrum and morphology of the anisotropy used in the data challenge
are shown in Fig.~\ref{fig:cr_dc_spectrum} and the top panel of
Fig.~\ref{fig:cr_anisotropy}.  We model a large-scale anisotropy as a dipolar
structure with an excess oriented toward right ascension $\alpha=15^\circ$ and
declination $\delta=10^\circ$.  The dipole strength increases mildly as a
function of energy but does not change its orientation.  An additional
small-scale anisotropy is modeled after the elongated regions excess flux at
$\alpha=60^\circ$ (region A) and $\alpha=120^\circ$ (region B) reported by
Milagro and ARGO-YBJ \cite{Abdo:2008kr,DiSciascio:2013}.

The combined isotropic background, dipole, and small-scale structure at 1~TeV
is plotted in Fig.~\ref{fig:cr_anisotropy}.  In the bottom panel of the figure
we plot the significance of features in 30 days of simulated data, derived from
the same direct integration algorithm \cite{Atkins:2003} that we apply to the
real data~\cite{BenZvi:2013}.  In this case an integration interval of 24~hours
is chosen.  Both the large and small-scale structures are visible in the
significance map; region A is detected at $>5\sigma$ in 30 days, which we also
observe in the real data \cite{BenZvi:2013}.  Un-physical artifacts are also
visible at high northern declinations.  It is in this manner that we use the
data challenge to test our analysis algorithms and understand the artifacts
they may produce.

\subsection*{Shadow of the Moon}

\begin{figure}[t]
  \centering
    \includegraphics[width=0.49\textwidth]{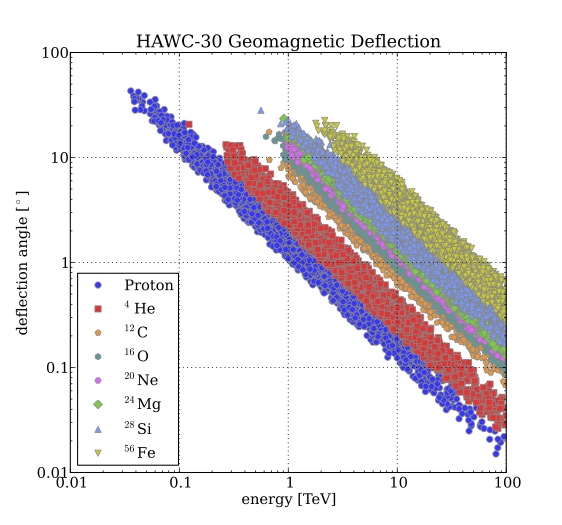}
    \caption{Simulated deflection angles of eight nuclear species in the
    geomagnetic field as a function of primary energy.  The cosmic rays are
    produced isotropically up to a zenith of $50^\circ$.  These deflections are
    used to determine the offset of the moon shadow observed in the 30-tank
    configuration of HAWC (HAWC-30).}
    \label{fig:geo_deflections}
\end{figure}

The moon is opaque to cosmic rays, so the point source analysis can be used to
search for a small deficit of cosmic rays from the direction of the moon.  HAWC
is sensitive to air showers with energies below 100~GeV, where the
deflection of cosmic rays in the geomagnetic field becomes significant.
Therefore, the apparent position of the moon in cosmic rays will differ from
its true position, and we must account for this effect if we wish to use the
moon shadow to verify the absolute pointing and angular resolution of the
detector.

We simulate the deficit of events from the moon as follows.  First we generate
a simulated set of cosmic rays as described in the last section.  Using the
cosmic-ray charges, energies, and arrival directions, we then trace each
particle back through the geomagnetic field \cite{Finlay:2010} to the radius of
the lunar orbit.  At this location we check to determine if the final position
of the particle is located inside the lunar surface.  If so, the particle is
removed from the simulated sample.  Any surviving events are smeared in
direction to account for the angular resolution of the detector.

The geomagnetic deflection of particles arriving at the HAWC site is shown in
Fig.~\ref{fig:geo_deflections}, and is fit by the power law
\begin{equation}
  \delta\theta\approx 1.6^\circ\cdot Z\left(\frac{E}{\text{TeV}}\right)^{-1}.
\end{equation}
Using the simulation of cosmic rays observed in the 30-tank configuration of
HAWC, we estimate that the moon shadow should be offset $-0.4^\circ\pm0.1^\circ$
in right ascension from the true position of the moon.  This prediction is
matched well by observations~\cite{Fiorino:2013}.

\section*{Simulation of Diffuse Emission}

\begin{figure*}[t]
  \centering
    \includegraphics[width=0.49\textwidth]{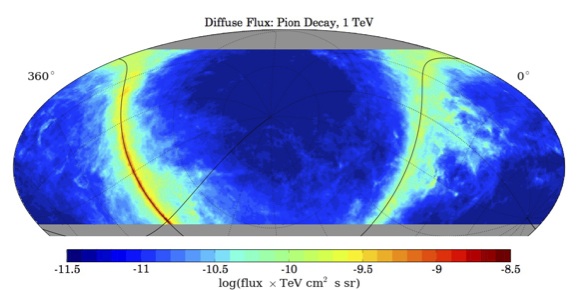}
    \includegraphics[width=0.49\textwidth]{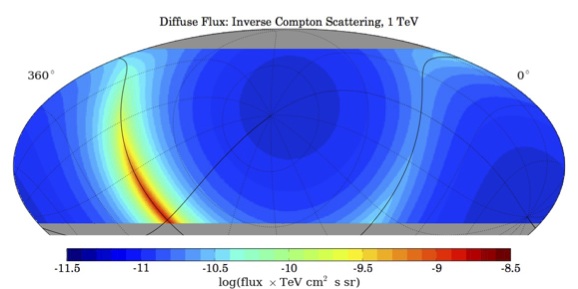}
    \includegraphics[width=0.49\textwidth]{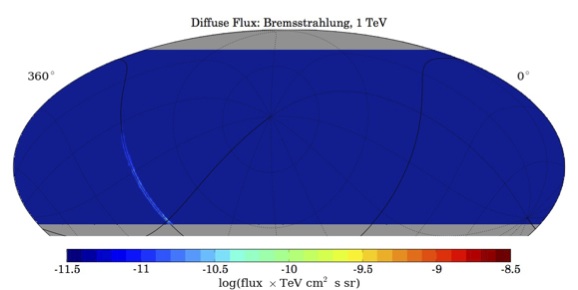}
    \includegraphics[width=0.49\textwidth]{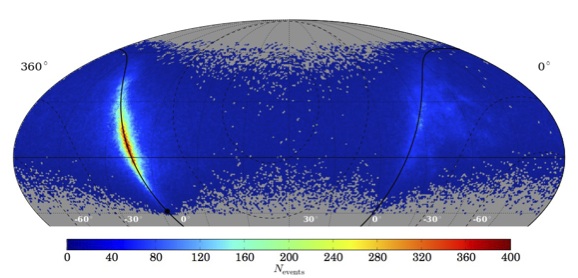}
    \caption{\textsl{Top Left}: Simulated galactic diffuse gamma-ray flux
    produced by the decay of $\pi^0$ particles, shown in the region of the sky
    visible to HAWC.
    \textsl{Top Right}: Diffuse flux produced by inverse Compton scattering.
    \textsl{Bottom Left}: Diffuse flux due to bremsstrahlung.
    \textsl{Bottom Right}: Six months of diffuse gamma rays observed with the
    HAWC-300 detector.}
    \label{fig:pidecay_diffuse_1TeV_map}
\end{figure*}

HAWC is the only TeV observatory currently able to observe the large-scale
diffuse emission of gamma rays, so the simulation of these sources is another
critical test of our analysis.  Diffuse emission in the data challenge refers
to galactic gamma rays from three sources.  The first population of diffuse
gamma rays we model is the decay of the $\pi^0$ particles produced when cosmic
rays interact with dust and gas in the interstellar medium.  A simulation of
$\pi^0$ emission at 1~TeV produced with GALPROP~\cite{Strong:2009} is shown in
the upper left panel of Fig.~\ref{fig:pidecay_diffuse_1TeV_map}.  A second
population is produced by the inverse Compton Scattering (ICS) of photons from
galactic and extragalactic radiation fields up to TeV energies by cosmic-ray
electrons (upper right panel of Fig.~\ref{fig:pidecay_diffuse_1TeV_map}).  A
third population is produced by bremsstrahlung as cosmic rays diffuse through
the interstellar medium (lower left panel of
Fig.~\ref{fig:pidecay_diffuse_1TeV_map}).  While diffuse emission can also
refer to extragalactic gamma-ray fluxes and unresolved point sources, these are
not separately modeled in the current data challenge.

Diffuse gamma rays are simulated much like the anisotropic background of cosmic
rays.  An energy- and position-dependent 3D sky map is used to sample the
source flux, which is then numerically integrated against the detector
effective area.  Since the all-sky diffuse gamma-ray flux has not been observed
in the TeV band, we use 3D tables giving flux as a function of energy and
galactic coordinates produced with GALPROP.  The GALPROP tables were created by
propagating galactic cosmic rays through molecular gas maps.  GALPROP uses
measurements of the cosmic-ray flux and nuclear abundance ratios, as well as
galactic and extragalactic photons, to generate flux predictions for the gamma
rays.

We have simulated the diffuse flux from $\pi^0$-decay, ICS, and bremsstrahlung
between 10~GeV and 100~TeV in the field of view of HAWC.  The number of diffuse
gamma-ray events observed in six months of simulated data is shown in the
bottom right panel of Fig.~\ref{fig:pidecay_diffuse_1TeV_map}.  This diffuse
map can be used to check analysis code and tune cuts to optimize the
sensitivity of HAWC to diffuse emission.

\section*{Simulation of Point Sources}

\begin{figure*}[t]
  \centering
    \includegraphics[width=0.395\textwidth]{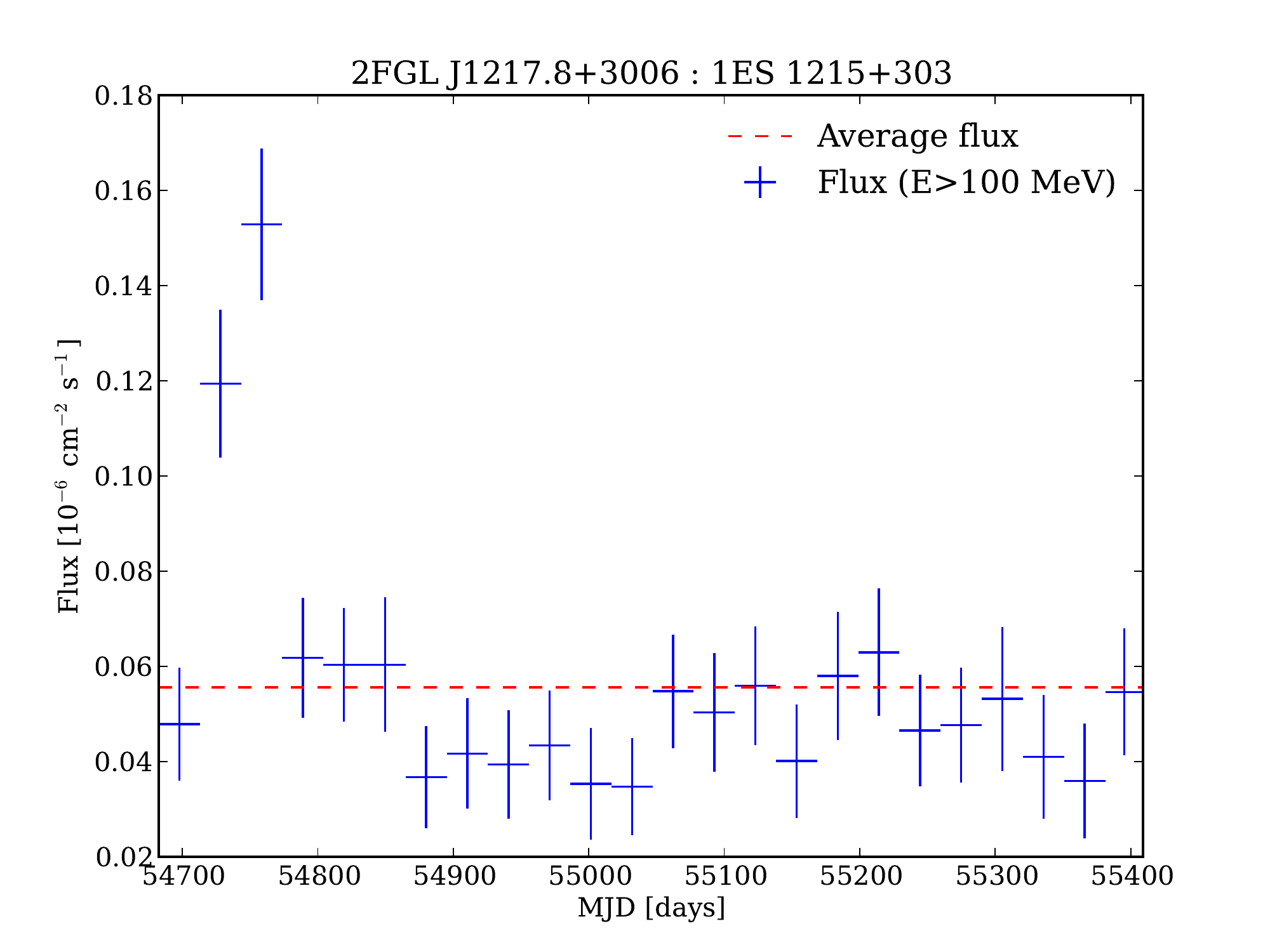}
    \includegraphics[width=0.595\textwidth]{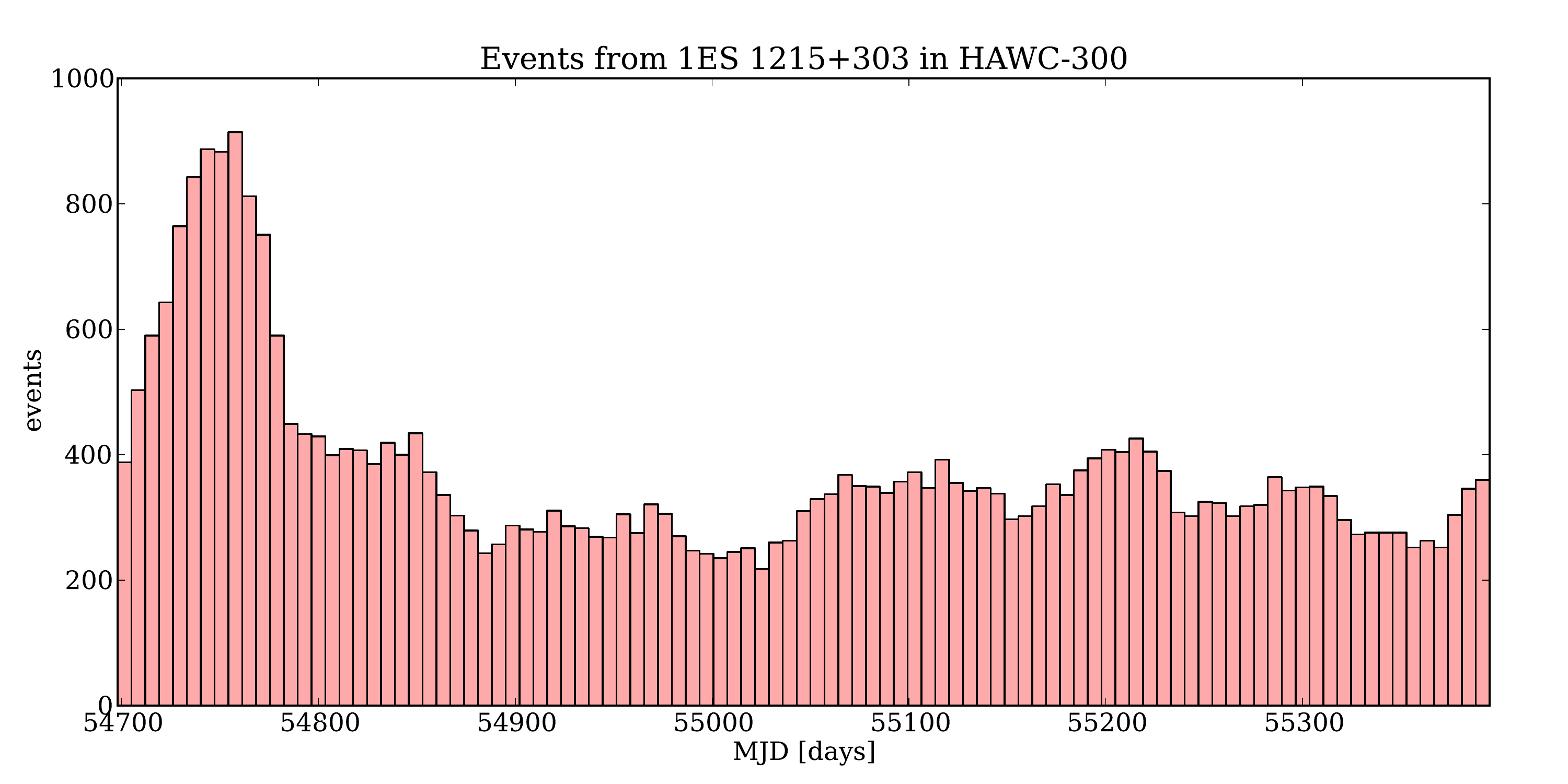}
    \caption{
      \textsl{Left}: Light curve of the AGN 1ES~1215+303 from the Fermi-LAT
      2FGL catalog~\cite{Nolan:2012}.
      \textsl{Right}: Simulated events in HAWC-300 from the source 1ES~1215+303.
    }
    \label{fig:1ES_1215_303}
\end{figure*}

The observation of transient emission from point sources is a major component
of the HAWC science program.  Time-dependent emission is a feature of many, if
not most, of the known sources of gamma rays in the sky.  Of particular
interest for HAWC are gamma-ray bursts (GRBs) and flaring active galactic
nuclei (AGN), because strong flares may be the only way that HAWC can be used
to observe objects with quiescent fluxes below 10~mCrab during the lifetime of
the observatory~\cite{Pretz:2013}.  Finally, HAWC may be the only facility that
can observe many transients (even those lasting days) because of the low uptime
of other TeV detectors.

We have written the data challenge software to support the simulation of
time-dependent fluxes of gamma rays from point sources.  It is only necessary
to provide a table of flux values versus modified Julian Date (MJD).  When
events are generated from a transient source, the flux is scaled up or down as
a function of time with respect to the average flux in the light curve table.
Once events are generated, their arrival directions are scattered about the
location of the source using the re-sampled opening angle distribution of the
HAWC detector Monte Carlo.

The light curve of 1ES~1215+303, an AGN which flared between August and November
2008, is shown in Fig.~\ref{fig:1ES_1215_303}.  The flare was observed by
Fermi-LAT and its light curve in the GeV band is provided in the 2FGL point
source catalog~\cite{Nolan:2012}.  Using the measurements from the LAT, we
simulated two years of observations of this source in the HAWC-300 detector.
The flux was extrapolated to TeV using the power-law spectral fit provided in
the 2FGL catalog.  Since 1ES~1215+303 is located at redshift $z=0.13$, we
attenuated the high-energy tail of the spectrum using a model of the infrared
component of the extragalactic background light~\cite{Gilmore:2009}.

The number of simulated events from 1ES~1215+303 as a function of time is
plotted in the right panel of Fig.~\ref{fig:1ES_1215_303}.  The flux during the
flare, $>2\times$ the quiescent average, is reproduced in the simulation.  In
order to define a flare with more time resolution or a different shape, it is
only necessary to provide the desired light curve table.

\section*{Conclusions}

We have developed a suite of programs to produce simulated data sets in which
we model sources of cosmic rays and gamma rays.  The software has been used to
generate anisotropic cosmic-ray backgrounds, sky maps of the moon shadow in
cosmic rays, and data sets with TeV gamma rays from point-like and diffuse
sources.  The simulation of point sources includes real and simulated light
curves.  We have verified the software and used the results to test the
analysis of data from the 30-tank configuration of HAWC, in particular the
study of the cosmic-ray anisotropy and the moon shadow.

\clearpage



\begin{thebibliography}{}
\bibitem{2FGL} Nolan, P.~L., et al. 2012, ApJS, 199, 31
\bibitem{cocoon} Ackermann, M., et al. 2011, Science, 334, 1103
\bibitem{MGRO07} Abdo, A. A., et al. 2007, ApJL, 658, L33
\bibitem{MGRO12} Abdo, A. A., et al. 2012, ApJ, 753, 159
\bibitem{MGRO-BSL} Abdo, A. A., et al. 2009, ApJL, 700, L127
\bibitem{MGRO-diffuse} Abdo, A. A., et al. 2008, ApJ, 688, 1078
\bibitem{Fermi-diffuse} Ackermann, M., et al. 2012, A\&A, 538, A71
\bibitem{HAWC-diffuse} Huentemeyer, P. 2013, 33rd International Cosmic
  Ray Conference, \textit{HAWC Sensitivity to Diffuse Emission}
\bibitem{HAWC-HL} Mostafa, M. 2013, 33rd International Cosmic Ray
  Conference, \textit{The HAWC Observatory}
\bibitem{HAWC-status} Pretz, J. 2013, 33rd International Cosmic Ray
  Conference,  \textit{Status and Sensitivity of HAWC}
\bibitem{Aliu-Gamma12}
Aliu, E., and the VERITAS Collaboration 2012, 5th International
Meeting on High Energy Gamma-Ray Astronomy
\bibitem{Aliu-ICRC11} 
Aliu, E. 2011, 32nd International Cosmic Ray Conference, 7, 227 
\bibitem{radio}
Paredes, J. M., et al. 2009, A\&A 507, 241
\bibitem{xmm}
Zabalza, V., \& Pardedes, J.~M. 2010, International Journal of Modern Physics D, 19, 811
\bibitem{Whipple}
Lang, M. J., et al. 2004, A\&A, 423, 415
\bibitem{HEGRA}
Aharonian, F., et al. 2005, A\&A, 431, 197
\bibitem{MAGIC}
Albert, J., et al. 2008, ApJL, 675, L25
\bibitem{radioLAT}
Camilo, F., et al. 2009, ApJ 705, 1
\bibitem{suzaku}
Murakami, H., et al. 2011, PASJ, 63, 873

\end{thebibliography}

\begin{thebibliography}{}

\bibitem{bib:Aharonian2006} Aharonian {\it et al.}, A\&A, 449, 223 (2006).
\bibitem{bib:Su2010} Su {\it et al.}, ApJ 724, 1044, (2010).
\bibitem{bib:Aharonian1996} Aharonian, and Atoyan, A\&A, 309, 917 (1996).
\bibitem{bib:Aharonian2000} Aharonian, and Atoyan, A\&A, 362, 937 (2000).
\bibitem{bib:casanova2010} Casanova {\it et al.}, PASJ 62, No.5, 1127 (2010).
\bibitem{bib:Gabici2009} Gabici, Aharonian, and Casanova, MNRAS 396, 3, 1629 (2009).
\bibitem{bib:TeVCat} http://tevcat.uchicago.edu/
\bibitem{bib:Aharonian2006Nature} Aharonian {\it et al.}, Nature, 439, 695 (2006).
\bibitem{bib:Atkins2005} Atkins {\it et al.}, Phys. Rev. Let., 95, 251103 (2005).
\bibitem{bib:Abdo2008} Abdo {\it et al.}, ApJ 688, 1078 (2008).
\bibitem{bib:Abdo2007} Abdo {\it et al.}, ApJ, 658, L33 (2007).
\bibitem{bib:Strong2000} Strong, Moskalenko, and Reimer, ApJ, 537, 763 (2000).
\bibitem{bib:Strong2004_1} Strong, Moskalenko, and Reimer, ApJ, 613, 956 (2004).
\bibitem{bib:Strong2004_2} Strong, Moskalenko, Reimer, Digel, and Diehl, A\&A 422, L47 (2004).
\bibitem{bib:Bi2009} Bi, Chen, Wang, and Yuan, ApJ, 695, 883 (2009).
\bibitem{bib:Crocker2010} Crocker, Bell, Balazs, and Jones, Phys.~Rev.~D 81, 063516 (2010).
\bibitem{bib:Bertone2009} Bertone, Cirelli, Strumia, and Taoso, JCAP, 3, 9 (2009).
\bibitem{bib:Meade2010} Meade, Papucci, Strumia, and Volansky, Nuclear Physics B, 831, 178 (2010).
\bibitem{bib:Casanova2008} Casanova, and Dingus, Astropart. Phys., 29, 63 (2008).
\bibitem{bib:Ackermann2012} Ackermann {\it et al.}, A\&A 538, A71 (2012).
\bibitem{bib:Abdo2009} Abdo {\it et al.}, ApJ 703, 1249 (2009).
\bibitem{bib:Sinnis2010} Sinnis  {\it et al.}, Astro2010: The Astronomy and Astrophysics Decadal Survey, Science White Papers, no. 275 (2010).
\bibitem{bib:Gabici2007} Gabici, and Aharonian, ApJ, 665, L131 (2007).
\bibitem{bib:BenZiv2013} BenZvi for the HAWC Collaboration, these proceedings, contribution 706.
\end{thebibliography}

\begin{thebibliography}{}
\bibitem{Jim}Observations of the Crab Nebula with HAWC, J. Braun (this conference)
\bibitem{Petra} HAWC Sensitivity to Diffuse Emission, P. Huentemeyer et. al. (this conference)
\bibitem{Asif} Real-time AGN Flare Monitor for the HAWC Observatory, A. Imran et. al. (this conference)
\bibitem{Sparks} Search for high energy emission from GRBs with the HAWC Observatory, K. Sparks et. al. (this conference)
\bibitem{Brian}Limits on Indirect Detection of WIMPs with the HAWC Observatory, B. Baughman et. al.  (this conference)
\bibitem{Tilan}HAWC Sensitivity to Primordial Black Holes T. Ukwatta et. al. (this conference)
\bibitem{moonpaper} Observation of the Moon Shadow and Characterization of the Point Response of HAWC-30, D. Fiorino et. al.  (this conference) 


\end{thebibliography}

\begin{thebibliography}{}


\bibitem{bib:ruffert} M. Ruffert et al., A\&A 344 (1999) 573-606

\bibitem{bib:rosswog} S. Rosswoget al., MNRAS 345 (2003) 1077-1090

\bibitem{bib:woosley} S. E. Woosley, ApJ 405 (1993) 273-277

\bibitem{bib:macfadyen} A. MacFadyen et al., ApJ 524 (1999) 262

\bibitem{bib:meszaros} P. M\'{e}sz\'{a}ros, Science 291 (2001) 79-84

\bibitem{bib:meszaros2} P. M\'{e}sz\'{a}ros, Rept. Prog. Phys. 69 (2006) 2259

\bibitem{bib:piran} T. Piran, Physics Reports 314 (1999) 575-667

\bibitem{bib:gilmore} R. C. Gilmore et al., MNRAS 399 (2009) 1694-1708

\bibitem{bib:band} D. Band et al., ApJ 413 (1993) 281

\bibitem{bib:grb090510} M. Ackermann et al., ApJ 716 (2010) 1178-1190

\bibitem{bib:grb090902b} A.Abdo et al.,  ApJ 706 (2009) L138

\bibitem{bib:fermilatcatalog} Fermi-LAT Collaboration, arXiv:1303.2908 [astro-ph.HE].

\bibitem{bib:hinton} J. Hinton, New J. Phys. 11 (2009) 055005

\bibitem{bib:sinnis} G. Sinnis, New J. Phys. 11 (2009) 055007 

\bibitem{bib:vernetto} S. Vernetto, GRB Coordinates Network (GCN): A Status Report

\bibitem{bib:icrcscalers} D. Lennarz et. al., paper 1160 (these proceedings)

\bibitem{bib:grbsensitivity} HAWC Collaboration, Astropart. Phys. 35, 641 (2012)

\bibitem{bib:icrchawc} J. Goodman et. al., paper 0702 (these proceedings)


\bibitem{bib:grb130427a} Yi-Zhong Fan et al., arXiv:1305.1261 [astro-ph.HE] 

\bibitem{bib:feldman} G. J. Feldman, R. D. Cousins, Phys. Rev. D 57 (1998), 3873-3889 

\bibitem{bib:hurley} K. Hurley, private communication

\bibitem{bib:williams} Williams D. A. et. al., AIP Conf. Proc. 921, p476 (2007)

\bibitem{bib:zhu} S. Zhu, GRB Coordinates Network (GCN) Circular 14508

\end{thebibliography}

\begin{thebibliography}{}

\bibitem{bib:review_Gehrels} N. Gehrels et al., ARA\&A 47 (2009) 567

\bibitem{bib:band_function} D. Band et al., ApJ 413 (1993) 281

\bibitem{bib:Fermi_LAT_GRB090902B} A.~A. Abdo et al., ApJ 706 (2009) L138

\bibitem{bib:Fermi_LAT_GRB090510} M. Ackermann et al., ApJ 716 (2010) 1178

\bibitem{bib:Fermi_LAT_GRB090926A} M. Ackermann et al., ApJ 729 (2011) 114

\bibitem{bib:Fermi_LAT_GRB_catalogue} Fermi-LAT Coll., ArXiv e-prints (2013) 1303.2908

\bibitem{bib:HAWC} HAWC Coll., HAWC highlight talk these proceedings

\bibitem{bib:HAWC_GRB_sensitivity} A.~U. Abeysekara et al., Astropart. Phys. 35 (2012) 641

\bibitem{bib:HAWC_GRB_ICRC} HAWC Coll., paper 783 these proceedings

\bibitem{bib:scaler_method} S. Vernetto, Astropart. Phys. 13 (2000) 75

\bibitem{bib:CORSIKA} D. Heck et al., Report FZKA 6019 (1998)

\bibitem{bib:Geant4} S. Agostinelli et al., Nuclear Instruments and Methods in Physics Research A 506 (2003) 250

\bibitem{bib:ATIC} A.~D. Panov et al., Bull. Rus. Acad. Sci. Phys. 73 (2009) 564

\bibitem{bib:LAT_refined} S. Zhu et al., GCN Circ. 14508 (2013)

\bibitem{bib:GBM_detection} A. von Kienlin, GCN Circ. 14473 (2013)

\bibitem{bib:LAT_detection} S. Zhu et al., GCN Circ. 14471 (2013)

\bibitem{bib:BAT_detection} A. Maselli et al., GCN Circ. 14448 (2013)

\bibitem{bib:MAXI_detection} T. Kawamuro et al., GCN Circ. 14462 (2013)

\bibitem{bib:INTEGRAL_detection} A. Pozanenkoet et al., GCN Circ. 14484 (2013)

\bibitem{bib:Konus_Wind_detection} S. Golenetskii et al., GCN Circ. 14487 (2013)

\bibitem{bib:AGILE_detection} F. Verrecchia et al., GCN Circ. 14515 (2013)

\bibitem{bib:RHESSI_detection} D.~M. Smith et al., GCN Circ. 14590 (2013)

\bibitem{bib:redshift_gemini} A.~J. Levanet al., GCN Circ. 14455 (2013)

\bibitem{bib:redshift_NOT} D. Xu, et al., GCN Circ. 14478 (2013)

\bibitem{bib:redshift_VLT} H. Flores et al., GCN Circ. 14491 (2013)

\bibitem{bib:BAT_refined} S.~D. Barthelmy et al., GCN Circ. 14470 (2013)

\bibitem{bib:GRB130504C_LAT_detection} D. Kocevski et al., GCN Circ. 14574 (2013)

\bibitem{bib:GRB130504C_GBM_detection} J.~M. Burgess et al., GCN Circ. 14583 (2013)

\bibitem{bib:GRB130504C_Wind_detection} S. Golenetskii et al., GCN Circ. 14587 (2013)

\bibitem{bib:GRB130504C_Suzaku_detection} T. Yasuda et al., GCN Circ. 14601 (2013)

\bibitem{bib:HAWC_GCN} D. Lennarz et al., GCN Circ. 14549 (2013)

\end{thebibliography}

\begin{thebibliography}{}

  \bibitem{bib:abdo} Abdo, A. \etal, ApJS, \textbf{199} (2011) 31

  \bibitem{bib:wagner} Wagner, M., MNRAS,  \textbf{385} (2008) 119

  \bibitem{bib:maraschi} Maraschi, L. \etal, Astrophys. J., \textbf{397}
    (1992) L5

  \bibitem{bib:dermer} Dermer, C. \& Schlickkeiser, R., Astrophys. J., 
    \textbf{416} (1993) 458

  \bibitem{bib:tavecchio} Tavecchio, F. \etal, A\&A,   \textbf{534} (2011) 86

  \bibitem{bib:aliu} Aliu, E. \etal, Astrophys. J.,  \textbf{755} (2012) 118

  \bibitem{bib:punch} Punch, M. \etal,  Nature,  \textbf{358} (1992) 477

  \bibitem{bib:cerutti} Cerutti, B. \etal,  Astrophys. J., \textbf{754}
    (2012) 33

  \bibitem{bib:kraw} Krawczynski, H. \etal, Astrophys. J., \textbf{601}
    (2004) 151

  \bibitem{bib:ackermann} Ackermann, M. \etal, Astrophys. J., \textbf{741}
    (2011) 30

  \bibitem{bib:abramo13} Abramowski, A. \etal, A\&A,  \textbf{550} (2013) A4

  \bibitem{bib:abramo12} Abramowski, A. \etal, Astropart.~Phys., 
    \textbf{738} (2011) 34

  \bibitem{bib:tavecchio12} Tavecchio, F. \etal, Phys.~Rev.~D, \textbf{86}
    (2012) 5036

  \bibitem{bib:murase} Murase, K. \etal, Astrophys. J., \textbf{749}
    (2012) 63

  \bibitem{bib:abey} Abeysekara, A. \etal, Astropart.~Phys., \textbf{35}
    (2012) 641

  \bibitem{bib:mostafa} Mostafa, M., for the HAWC Collaboration, Brazilian
    Journal of Physics, Proceedings of the 33$^{{ \rm rd}}$ International Cosmic Ray
    Conference

  \bibitem{bib:lauer} Lauer, R., for the HAWC Collaboration, Brazilian
    Journal of Physics, Proceedings of the 33$^{{ \rm rd}}$ International Cosmic Ray
    Conference

  \bibitem{bib:atkins} Atkins, R. \etal, Astrophys. J., \textbf{595}
    (2003) 803

  \bibitem{bib:gorski} Gorski, K. M. \etal, Astrophys. J., \textbf{622}
    (2005) 759


\end{thebibliography}

\begin{thebibliography}{}

\bibitem{or1}  M. Tavani {\it et al.}, Science {\bf 331}, 736 (2011).
\bibitem{or2}  A. A. Abdo {\it et al.}, Science {\bf 331}, 739 (2011).
\bibitem{or3}  M. Balbo {\it et al.}, A\&A {\bf 527}, L4 (2011).
\bibitem{or4}  G. Aielli {\it et al.}, The Astronomer's Telegram \#2921 (2010).
\bibitem{or5}  M. Mariotti {\it et al.}, The Astronomer's Telegram \#2967 (2010).
\bibitem{or6}  R. Ong {\it et al.}, The Astronomer's Telegram \#2968 (2010).
\bibitem{or7}  J. Braun {\it et al.}, Proc. 32nd ICRC, Beijing, (2011).
\bibitem{or8}  S. Vernetto {\it et al.}, Proc. 32nd ICRC, Beijing, (2011).
\bibitem{or9}  R. Buehler {\it et al.}, The Astronomer's Telegram \#3276 (2011).
\bibitem{or10} R. Buehler {\it et al.}, ApJ {\bf 749}, 26 (2012).
\bibitem{or11} E. Striani {\it et al.}, ApJL {\bf 741}, L5 (2011).
\bibitem{or12} R. Ojha {\it et al.}, The Astronomer's Telegram \#4239 (2012).
\bibitem{or13} B. Bartoli {\it et al.}, The Astronomer's Telegram \#4258 (2012).
\bibitem{fermi2013} R.~Ojha {\it et al.}, The Astronomer's Telegram \#4855 (2013).
\bibitem{agile2013} E.~Striani {\it et al.}, The Astronomer's Telegram \#4856 (2013).
\bibitem{fermicurve} http://fermi.gsfc.nasa.gov/FTP/glast/data/
lat/catalogs/asp/current/lightcurves/CrabPulsar\_86400.png
\bibitem{or14} K. Korhi, Y. Ohira, and K. Ioka, MNRAS {\bf 424}, 2249 (2012).
\bibitem{or15} S. S. Komissarov and M. Lyutikov, MNRAS {\bf 414}, 2017 (2011).
\bibitem{or16} W. Bednarek and W. Idec, MNRAS {\bf 414}, 2229 (2011).

\end{thebibliography}

\begin{thebibliography}{}

\bibitem{bib:gcn} http://gcn.gsfc.nasa.gov

\bibitem{bib:hawc} A.U. Abeysekara {\it et al.} (HAWC Collaboration), 
Astropart. Phys. 35\,(2012)\,641 (arXiv:1108.6034).

\bibitem{bib:lor} A.A. Abdo {\it et al.}, Nature 462\,(2009)\,331.

\bibitem{bib:lorenz} T. Piran and U. Jacob, Proceedings of the Spanish Relativity Meeting - 
Encuentros Relativistas Espanoles (ERE2007): Relativistic Astrophysics and Cosmology (2008). 
GRBs and Lorentz Invariance Violation.  EAS Publications Series 30\,(2008)\,27.

\bibitem{bib:blor} X. Zhao, Z. Li, and J. Bai, Astrophys. J. 726\,(2011)\,89 (arXiv:1005.5229).

\bibitem{bib:hawcpro} M. Mostafa {\it et al.} (HAWC Collaboration), these proceedings.

\end{thebibliography}

\begin{thebibliography}{}

\bibitem{CORSIKA}
CORSIKA: \url{www-ik.fzk.de/~corsika}

\bibitem{Geant4}
Geant4: \url{geant4.cern.ch}

\bibitem{Rovatti:1998}
R.~Rovatti et al., IEEE Trans. Comp., 1998, {\bf 47}, 894.

\bibitem{BenZvi:2013}
S.~BenZvi et al., Proc. ICRC 2013: 0710.

\bibitem{Fiorino:2013}
D.~Fiorino et al., Proc. ICRC 2013: 0784.

\bibitem{Ave:2008}
M.~Ave et al., Astrophys. J., 2008, {\bf 678}: 255

\bibitem{Panov:2009}
A.~Panov et al., Bull. Russ. Acad. Sci, 2009, {\bf 73}: 564

\bibitem{Ahn:2010gv}
H.S.~Ahn et al., Astrophys. J., 2010, {\bf 714}: L89

\bibitem{Adriani:2011cu}
O.~Adriani et al., Science, 2011, {\bf 332}: 69

\bibitem{Abdo:2008kr}
A.A.~Abdo {et~al.} Phys. Rev. Lett., 2008, {\bf 101}: 221101

\bibitem{DiSciascio:2013}
G.~Di~Sciascio, Proc. ISVHECRI 2012, Berlin, Germany

\bibitem{Atkins:2003}
R.~Atkins et al., Astrophys. J., 2003, {\bf 595}: 803

\bibitem{Finlay:2010}
C.~Finlay et al., Geophys. J. Int., 2010, {\bf 183}: 1216

\bibitem{Strong:2009}
A.~Strong, private communication, 2009.

\bibitem{Pretz:2013}
J.~Pretz et al., Proc. ICRC 2013: 0702.

\bibitem{Nolan:2012}
P.~Nolan et al., Astrophys. J. Suppl., 2012, {\bf 199}: 31

\bibitem{Gilmore:2009}
R.C.~Gilmore et al., MNRAS, 2009, {\bf 399}: 1694

\end{thebibliography}
\end{document}